\documentclass[12pt,a4paper,color]{article}
\usepackage{epsfig}
\usepackage{fullpage}
\usepackage{algorithm}
\usepackage{graphicx}
\usepackage{amssymb}
 \usepackage{amsmath}
 \usepackage{color}

\newcommand{\U}{\Upsilon}
\newcommand{\beq}{\begin{equation}}
\newcommand{\eeq}{\end{equation}}

\title{Static properties of 2D spin-ice as a sixteen-vertex model}
\author{Laura Foini$^1$, Demian Levis$^1$, Marco Tarzia$^2$ and Leticia F. Cugliandolo$^1$\\
 $^1$ {\small Laboratoire de Physique Th\'eorique et Hautes Energies} \\
{\small Universit\'e Pierre et Marie Curie, Paris, France} \\
$^2$ {\small Laboratoire de Physique de la Mati\`ere Condens\'ee} \\
{\small Universit\'e Pierre et Marie Curie, Paris, France} }
\date{\today}

\begin{document}
\maketitle

\begin{abstract}
We present a thorough study of the static properties of $2D$ models of spin-ice type on the square
lattice or, in 
other words, the sixteen-vertex model. We use extensive Monte Carlo simulations to 
determine the phase diagram and critical properties of the finite dimensional system.
We put forward a suitable mean-field approximation, by defining the model on carefully chosen trees.
We employ the cavity (Bethe-Peierls) method to derive self-consistent equations, the fixed points of which
yield the equilibrium properties of the model on the tree-like graph. 
We compare mean-field and finite dimensional results. 
We discuss our findings in the context of experiments in 
artificial two dimensional spin-ice.
\end{abstract}

\newpage

\tableofcontents

\newpage

\section{Introduction}

Many interesting classes of classical and quantum magnetic systems are highly constrained. In particular, geometric
constraints lead to frustration and the impossibility of satisfying all competing interactions
simultaneously. In most cases this phenomenon gives rise to the existence  of highly degenerate ground
states~\cite{DiepBookCH7short,Balents2010}, often associated with an excess entropy at zero
temperature. Anti-ferromagnets on a triangular lattice and spin-ice samples are materials of this kind.

In conventional spin-ice~\cite{Harris1997} (see~\cite{Bramwell2001a,Gingras2010} for reviews)
magnetic ions form a tetrahedral structure in $3D$, {\it i.e.} a pyrochlore lattice.
This is the case, for instance, of the
Dy$^{+3}$  ions in the Dy$_2$Ti$_2$O$_{7}$ compound. Since the $f$-electron spins have a large magnetic moment, they
can be taken as classical variables at low enough temperature, say, $T<10$ K, behaving as
Ising doublets, forced to point along the axes joining the centers of the tetrahedra shared by the considered spin.
The origin of geometric frustration in these systems
is twofold: it arises from the non-colinearity of the crystal field and the effective ferromagnetic
exchange, and from long-range couplings between the spins~\cite{Gingras2010}.
Since within this Ising formulation the long ranged dipolar interactions are almost  perfectly screened at low
temperature~\cite{Gingras2000,Isakov2005},
in a simplified description only short-range ferromagnetic exchanges
are retained~\cite{Harris1997}. Thus, topological frustration arises from the fact that
the Ising axes in the unit cell are fixed and different, forced to point towards the centers of neighboring
tetrahedra. The configurations that 
minimize the energy of each tetrahedron are the six states with two-in and two-out pointing spins.

The ground state of the 
system is more easily visualized by realizing that each tetrahedron in $3D$ can be considered as
a vertex taking one out of six possible configurations on a lattice of coordination four.
With this mapping the magnetic problem described above becomes equivalent to the model of
water ice introduced in~\cite{Bernal1933,Pauling1935}. In this context,
the entropy of the ground state satisfying all the local (two-in - two-out) ice constraints,
with all six vertices taken with the same statistical weight, was estimated by 
Pauling with a simple counting argument~\cite{Pauling1935}. Pauling's result is very close to the
earlier measurements performed by Giauque and Stout~\cite{Giauque1936} on water ice
and to the ground-state 
entropy of the magnetic spin-ice sample measured in the late 90s~\cite{Ramirez1999}.
Experimentally, the Boltzmann weights of the six vertices can be tuned by applying 
pressure or magnetic fields along different crystallographic axes. Indeed,
the extensions of Pauling's ice model to describe more general ferroelectric systems lead to `ice-type models'~\cite{Slater1941}.
We will come back to this issue below.

The local constraint leads to many peculiar features that have been studied experimentally, numerically,
and analytically. In the ground state, the total spin surrounding a lattice point is conserved and
constrained to vanish according to the two-in -- two-out rule. This fact has been interpreted
as a zero-divergence condition on an emergent vector field~\cite{Isakov2005,Isakov2004}.
Spins are regarded as fluxes and, quite naturally, 
an effective fluctuating electromagnetism emerges, where each equilibrium configuration
is made of closed loops of flux. This analogy  can be used to derive power-law decaying spatial correlations
of the spins~\cite{Youngblood1980,Youngblood1981}, with a parameter dependent exponent,
 that were recently observed experimentally via neutron 
scattering measurements~\cite{Fennel2009}. 
Such criticality of the disordered ground state, also called spin-liquid or Coulomb phase,
had been observed numerically first~\cite{Stillinger1973}
(see~\cite{Huse2004} for a more general discussion).
A detailed description of the Coulomb phase can be found in~\cite{Henley2010}.

Thermal (or other) fluctuations are expected to generate defects, 
in the form of ice-breaking rule vertices.  In the electromagnetic analogy described above
a defect corresponds to a charge,  defined as the number of outgoing arrows minus the number of ingoing arrows.
As such, a tetrahedra with three-in and one-out spins contains a negative charge of magnitude two
and the reversed configuration a positive charge also of magnitude two. In turn, the 
four-in cells carry a charge minus four and the four-out ones a charge plus four.
Such vertices  should be present in the samples under adequate conditions. 
The possibility of observing magnetic monopoles and Dirac strings 
as being associated to defects has been proposed by
Castelnovo {\it et al.}~\cite{Castelnovo2008} and investigated experimentally
by a number of groups~\cite{Fennel2009,Morris2009}.

Spin-ice can be projected onto $2D$ Kagome planes by applying  specially chosen magnetic fields
or pressure. Recently, the interest in spin-ice physics  has been boosted by the advent of artificial
samples~\cite{Wang2006} on simpler square lattices.
These are stable at room temperature and have magnetic moments that are large enough to be
easily observed in the lab, thus  giving access to the micro-states that can be directly visualized with
microscopy.  Following the same line of reasoning exposed in the previous paragraphs,
such $2D$  ice-type systems should be modeled by the complete sixteen-vertex model on a square lattice, where
all kind of vertices are allowed.

The exact solution of the $2D$ ice model ({\it i.e.}, the six-vertex model)~\cite{Lieb1967a},
and some generalizations of it in which
a different statistical weight is given to the six allowed vertices~\cite{Lieb1967b,Lieb1967c,Sutherland1967},
was obtained by Lieb and Sutherland in the late 60s
using the transfer matrix technique with the Bethe {\it Ansatz}. Soon after, Baxter
developed a more powerful method to treat the generic six- and eight- vertex models~\cite{BaxterBook}
and founded in this way the theory of integrable systems (in the eight-vertex model vertices with four in-going
or four out-going arrows are allowed). The equilibrium phase diagrams of the six- and eight- vertex models are very rich:
depending on the Boltzmann weight of the vertices the systems can be found in a variety of different
phases, such as a quasi long-range ordered spin liquid phase (SL), two ferromagnetic (FM) and one (or two) antiferromagnetic
(AF) phases, separated by different types of transition surfaces in the phase diagram. In the six-vertex case the SL
phase is critical similarly to what is observed in the $3D$ Coulomb phase.

From a theoretical perspective integrable vertex models are of notable interest. 
Their static properties can be mapped into spin models with many-body interactions, 
loop models, three-coloring problems, random tilings, surface growth,  {alternating} 
sign matrices and quantum spin chains. 
{The Coulomb gas method~\cite{Nienhuis,diFrancesco}
and conformal field theory techniques~\cite{diFrancesco,Klumper} have added significant insight into the 
phase transition and critical properties of these systems as well.}
A comprehensive discussion of these mappings goes beyond the scope of the present work. For a
review the interested reader may consult Refs.~{\cite{Zinn-Justin,Wuphasetransitions}}. 
Some comments on these alternative representations,
when useful, will be made in the text.

Much less is known about the equilibrium (and dynamic~{\cite{Levis2012,Wysin12,Bud12,CepCan12}}) 
properties of the sixteen-vertex model in two and
three dimensions. 
{At present, the experimental interest in classical frustrated magnets of spin-ice type is  focused on
understanding 
the nature and the role of defects (magnetic monopoles) and their effects on the 
samples' macroscopic properties~\cite{Castelnovo2008,Fennel2009,Slobinsky2010}. 
This motivates us to revisit the generic vertex model with defects, i.e. the sixteen-vertex model, and complete its analysis.} 
The special experimental simplicity of two dimensional
samples suggests to start from the $2D$ case. Moreover, it is worth trying to extend at least part of the very powerful analytic
toolbox developed for the integrable six- and eight- vertex models to the (non-integrable)
complete sixteen-vertex models, where all defects are allowed. 

In this paper we describe the equilibrium properties of  spin-ice systems with defects.
We proceed in two directions. On the one hand,
we study the static properties of the sixteen-vertex model on a square lattice by using 
Monte Carlo {(MC)} simulations in equilibrium. For simplicity, and in order to 
keep the model close to experimental realizations, we assume that the statistical weight of
the defects remains relatively small (as will be explained carefully in the text). 
We establish its phase diagram and its critical properties, {which} we compare
to the ones of the integrable models. On the other hand, we introduce a suitable mean-field
approximation, defined on a well-chosen tree of vertices, and we adapt the cavity  (Bethe-Peierls)
method to derive self-consistent equations on such trees, the 
fixed points of which yield the exact solution of the mean-field model. This method  
allows us to describe all expected phases and to unveil some of their properties, such
as the presence of anisotropic equilibrium fluctuations in the symmetry broken phases.
We discuss the range of validity of the mean-field approximation, and we compare the analytical solution
to the numerical results for the finite dimensional system. In the conclusions we summarize our results and 
we briefly discuss some experimental features that we can describe with our methods~\cite{Levis12a}.

\section{Vertex models}
\label{sec:vertex-models}

In vertex models the degrees of freedom (Ising spins,
$q-$valued Potts variables, etc.) sit on the edges of a lattice. The interactions take place on the vertices
and involve the spins of the neighboring edges. 
In this Section we recall the definition and main equilibrium properties of 
Ising-like vertex models in two dimensions defined on a square lattice.

\subsection{Definitions}

We focus on an $L\times L$ square lattice {$\mathcal{L}$} with periodic boundary conditions. We label the coordinates of the lattice
sites by $(m,n)$. This lattice is bipartite, namely, it can be partitioned in two sub-lattices $A_1$ and $A_2$ such
that the sites having  
$m+n$ even belong to $A_1$, those having $m+n$ odd belong to $A_2$, and 
each edge connects a site in $A_1$ to one in $A_2$. The degrees 
of freedom sit on the links between two sites or, in other words, on the ``medial'' lattice whose sites are placed on the midpoints of the     
links of the original lattice. The midpoints are hence labeled by $(m+1/2, n)$ and $(m, n+1/2)$.
Thus, in the models we consider the degrees of freedom are arrows aligned along the edges of a square lattice,
which can be naturally mapped into Ising spins, say $s_{m+1/2,n}=\pm 1$. 
Without loss of generality, we choose a convention such that $s = + 1$          
corresponds to an arrow pointing in the right or up direction, depending on the orientation of the link, and conversely 
$s= - 1$ corresponds to arrows pointing down or left.

\subsection{The six-vertex model}

\begin{figure}[h]
\begin{center}
\includegraphics[scale=1.8]{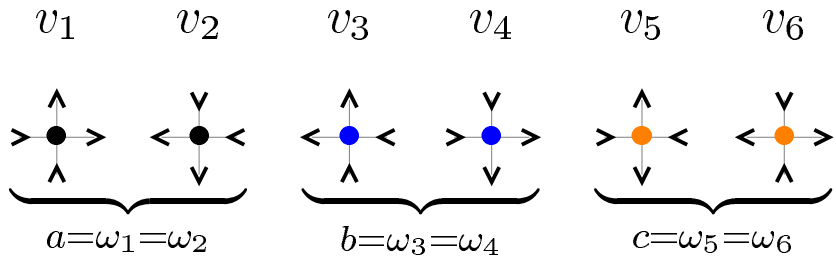}
\end{center}
\caption{\small (Color online.) The zero-charge six-vertices of the six-vertex model.
$v_1$, $v_2$, $v_3$ and $v_4$ are FM vertices while $v_5$ and $v_6$ are 
AF ones. Vertices are grouped in spin-reversal symmetric pairs.}
\label{fig:six-vertex}
\end{figure}

In the six-vertex model ({\it i.e.}, $2D$ spin-ice) arrows (or Ising spins) sit on the edges of a 
coordination four square lattice and they are constrained to satisfy the  two-in two-out 
rule~\cite{Bernal1933,Pauling1935}. 
Each node on the lattice has four spins attached to it with two possible directions, as 
shown in Fig.~\ref{fig:six-vertex}. A {\it charge} can be attributed to each single vertex configuration, simply defined as
the number of out-going minus the number of in-coming arrows.
Accordingly, the six-vertex model vertices have zero charge. (Note that the charge is not 
the sum of the spins attached to a vertex{.} Such a {\it total spin} will be defined and used in 
Sec.~\ref{sec:model-on-the-tree}.)

Although in the initial modeling of spin-ice  all such vertices 
were equivalent, the model was later generalized to describe ferroelectric systems by giving different statistical weights 
to different vertices: $w_k \propto e^{-\beta \epsilon_k}$ with  $\epsilon_k$ the energy of each of the $k=1,\dots, 6$
vertices.  $\beta=1/(k_BT)$ is the inverse temperature.
Spin reversal symmetry naturally imposes that $w_1=w_2=a$ for the first pair of ferromagnetic (FM) vertices, $w_3=w_4=b$ 
for the second pair of FM vertices, and $w_5=w_6=c$ for the anti-ferromagnetic (AF) ones, see Fig.~\ref{fig:six-vertex}.
We have here introduced the conventional names $a$, $b$, and $c$ of the three fugacities corresponding to the Boltzmann weight 
of the three kinds of vertices. In the literature it is customary 
to parametrize the phase diagram and equilibrium properties in terms of $a/c$ and $b/c$.  This is the choice we also
make in this paper. Particular cases include: the 
F model of anti-ferroelectrics, in which the energy of the AF vertices is set to zero and all other ones are taken to be 
equal and positive, {\it i.e.} $c> a=b$~\cite{Rys1963}; the KDP model of 
ferroelectrics, in which  the energy of a pair of FM vertices is set to zero and all other ones are equal and positive, 
e.g. $a>b=c$~\cite{Slater1941}; the Ice model, in which the energies of all vertices are equal, 
{\it i.e.} $a=b=c$~\cite{Lieb1967a}. It is important to note, however, 
that in the context of experiments in artificial spin-ice type 
samples, vertex energies are fixed and the control parameter is the temperature. In~\cite{Levis12a} we used this 
alternative parametrization and we compared the model predictions to experimental observations~\cite{Nisoli10,Marrows12}.

The six-vertex model is integrable.
The free-energy density of the model with $a=b=c$ and periodic boundary conditions 
was computed by Lieb in the late 60s with  the transfer matrix technique and the Bethe {\it Ansatz} to solve the 
eigenvalue problem~\cite{Lieb1967a}. The method was then extended to 
calculate the free-energy density of the F model~\cite{Lieb1967b}, the  KDP model~\cite{Lieb1967c}, and also models with generic 
values of $a$, $b$, and $c$, and periodic boundary conditions~\cite{Sutherland1967}, and the same general case with 
antisymmetric~\cite{Batchelor1995} and domain wall boundary conditions~\cite{Korepin1982}. The effect of 
the boundary conditions 
is indeed very subtle in these systems as some thermodynamic properties depend upon them~\cite{Korepin2000} contrary to 
what happens in usual short-range statistical physics models. Very powerful analytic methods such as the 
Yang-Baxter equations have been employed in the analysis of these integrable systems~\cite{BaxterBook}.

{An order parameter that allows one to characterize the different phases is the total {\it direct} and {\it staggered} polarization per spin\footnote{{A different order parameter $\Phi$ can also be introduced in the eight-vertex model. The latter can be reformulated as an Ising model with nearest, next-to-nearest and plaquette interactions between spins sitting on the sites of the dual lattice $\mathcal{L}^*$~\cite{Kadanoff1971}. The usual spontaneous magnetization of the corresponding Ising model defines the order parameter $\Phi$~\cite{BaxterBook}. } }}
\begin{equation}
\langle M_{\pm} \rangle = \frac{1}{2} (\langle | m_{\pm}^{x}| \rangle + \langle | m_{\pm}^{y}| \rangle)
\label{eq:magn}
\end{equation}
with the horizontal and vertical fluctuating components given by 
\begin{eqnarray}
L^2m^{x}_{\pm}=\sum_{(m,n)\in
  A_1}s_ {m+1/2,n} \ \pm \sum_{(m,n)\in
  A_2}s_{m+1/2,n}
  \; , 
  \label{eq:magnx}
  \\
  L^2m^{y}_{\pm}=\sum_{(m,n)\in
  A_1}s_ {m,n+1/2} \ \pm \sum_{(m,n)\in
  A_2}s_{m,n+1/2}
  \; .
  \label{eq:magny}
\end{eqnarray}
The angular brackets $\langle \dots \rangle$ denote here, and throughtout the rest of this article, the statistical average.
A finite value of $\langle M_{+} \rangle$ indicates the spontaneous breaking of $Z_2$ symmetry, while 
a finite value of $\langle M_{-} \rangle$ corresponds to the spontaneous breaking of $Z_2$ and translational symmetry.

\begin{figure}[h]
\begin{center}
\includegraphics[scale=0.5,angle=-90]{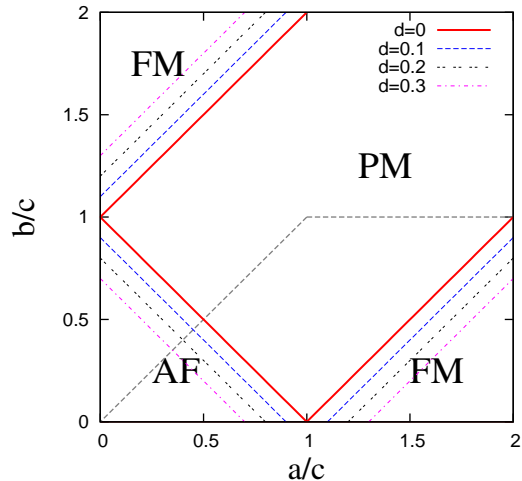}
\end{center}
\caption{\small (Color online.) The phase diagram of the six- (red solid lines) and eight- (dashed lines) 
vertex models. For the eight-vertex model the curves correspond to the projection on the $d=0$ plane.
Only for $d=0$ (six-vertex model) the PM region becomes an SL (Coulomb) phase.}
\label{fig:phase-diagram-8vertex}
\end{figure}

The four equilibrium phases  are classified by the anisotropy parameter
\begin{equation}\label{delta6}
\Delta_6 = \frac{a^2+b^2-c^2}{2ab} 
\; ,
\end{equation}
and they are the following.

\vspace{0.2cm}

\noindent
$a$-{\it Ferromagnetic} ($a$-FM) {\it  phase}: 
$\Delta_6 > 1$; {\it i.e.} $a > b + c$. Vertices $v_1$ and $v_2$ are favored. Spin reversal symmetry is broken.
The lowest energy state in the full FM phase is
doubly degenerate: either all arrows point up and right or down and left [{\it i.e.} $M_+=1$, with $M_+$ the 
magnetization density defined in eq.~(\ref{eq:magn})]. In this
phase the system is frozen as the only possible excitations involve a number of degrees of freedom of
the order of $L$.  In all this phase the exact free-energy per vertex is given
by~\cite{BaxterBook}
\begin{equation}
f_{\rm FM} = \epsilon_1
\; . 
\end{equation}
At $a = b + c$ ($\Delta_6=1$) the system experiences a frozen-to-critical phase transition from
the frozen FM to a disordered (D) or spin liquid (SL) phase that we 
discuss below.

\vspace{0.2cm}

\noindent
$b$-{\it Ferromagnetic} ($b$-FM) {\it phase}:
$\Delta_6>1$; {\it i.e.} $b > a+c$. This phase is equivalent to the previous one by replacing
$a$- by $b$-vertices. The free-energy is $f_{\rm FM}=\epsilon_3$ and the phase transition 
towards the SL phase is also of frozen-to-critical type. 

\vspace{0.2cm}

\noindent
{\it Spin liquid} (SL) {\it phase}: 
 $-1 <\Delta_6 <1$; {\it i.e.} $a < b+c$, $b < a+c$ and $c < a+b$. In this phase the averaged magnetization
is zero, $\langle M_\pm \rangle =0$,  and one could expect the system to be a conventional paramagnet
(PM). However, the ice constraints are strong
enough to prevent the full decorrelation of the spins even at finite temperature.
The system is in a quasi long-range ordered phase with an infinite correlation length. At $c = a+b$
there is a Kosterlitz-Thouless (KT) phase transition between this critical phase and an AF phase with 
staggered order that is discussed below. Some particular points in parameter space belong to the SL phase as the 
spin-ice point $a = b = c$ for which  $\Delta_6= 1/2$. At this special point 
the ground state is macroscopically degenerate
giving rise to the residual entropy~\cite{Lieb1967a}
\begin{equation}
S/N=3/2 \ \ln 4/3 
\label{eq:entropy-spin-ice}
\end{equation} 
with $N=L^2$ the number of vertices in the sample.

The exact solution found by Baxter yields the free-energy density  as a function of the 
parameters through a number of integral equations~\cite{BaxterBook}. In Sec.~\ref{sec:model-on-the-tree} we evaluate it numerically 
and we compare it to the outcome of the Bethe approximation. Close to the FM transitions the free-energy density can be approximated by
\begin{equation}
f_{\rm SL} \simeq \max (\epsilon_1,\epsilon_3) - \frac{1}{2} k_BT \left( \frac{b+c}{a} -1\right)
= \max(\epsilon_1,\epsilon_3) - \frac{1}{2} k_B T \ t^{2-\alpha}
\; , 
\label{eq:free-energy-6vertex-PM}
\end{equation}
with $t$ being the reduced distance from criticality, $t=(b+c)/a-1$, and $\alpha$ an exponent that plays the role of the one of the 
heat-capacity and takes the value $\alpha=1$ here.
The first derivative of $f_{\rm SL}$ with respect to the distance from the transition $t$
shows a step discontinuity at the SL-FM transition as it would in a first-order phase transition, even though  
the FM phase is frozen. This corresponds to a critical-to-frozen phase transition.

\vspace{0.2cm}

\noindent{\it Antiferromagnetic} (AF) {\it phase}: 
$\Delta_6< -1$; {\it i.e.}  $c > a + b$. Vertices $v_5$ and $v_6$ are favored. The ground
state is doubly degenerate, corresponding to the configurations $M_- = \pm  1$. The staggered order 
is not frozen, due to thermal fluctuations. 
This is confirmed by the exact expression of the staggered magnetization found by Baxter~\cite{Baxter1971,Baxter1972}. 
The free-energy has an essential singularity at the critical temperature (towards the SL phase) 
\begin{equation}
f_{\rm AF} \simeq e^{-\mbox{cst}/\sqrt{t}}
\; , 
\label{eq:free-energy-6vertex-AF}
\end{equation}
with cst being a constant and $t=(a+b)/c-1$ the distance from criticality, as it is typical for
an infinite order phase transition.

{The transition lines are straight lines (given by $\Delta_6=1$ for the SL-FM  and $\Delta_6=-1$ for the SL-AF)
and they are shown in Fig.~\ref{fig:phase-diagram-8vertex} as solid (red) lines. 
The dashed line along the diagonal represents the range of variation of the 
parameters in the F model. The horizontal dashed line is the one of the KDP model. The intersection of this two lines corresponds to the ice-model.
Although the transitions are not of second order, critical 
exponents have been defined and are given in the first column of Table~\ref{table:exponents-th}
for the SL-FM transition.
The exponent $\alpha$ is taken from the expansion of the free-energy close to the transition, 
see eqs.~(\ref{eq:free-energy-6vertex-PM}) and (\ref{eq:free-energy-6vertex-AF}).  We denote by $\beta_e$ the critical exponent associated to the order parameter as defined in eq.~\ref{eq:magn}, i.e.  the polarization of the arrows on the edges of $\mathcal{L}$. The ratios
$\hat\gamma_e=\gamma_e/\nu$, $\hat\beta_e=\beta_e/\nu$ and $\hat\phi=(2-\alpha)/\nu$ are defined 
using 
the (divergent in the SL phase) correlation length $\xi$  instead of $t$ as the scaling variable~\cite{Suzuki1974}.}

\subsection{The eight-vertex model}

\begin{figure}[h]
\begin{center}
\includegraphics[scale=1.8]{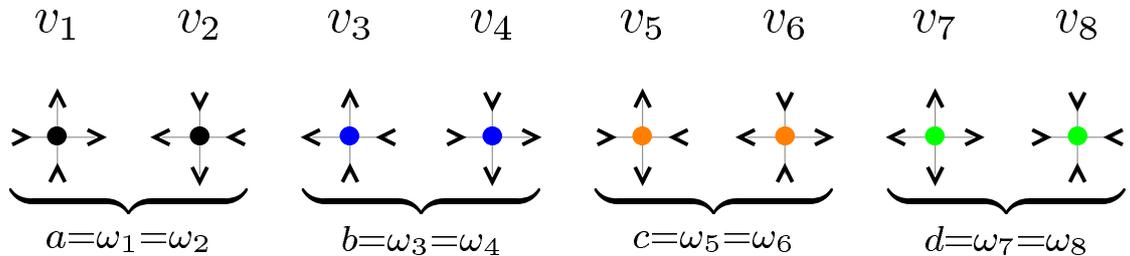}
\end{center}
\caption{\small (Color online.) The vertices in the eight-vertex model. The four-out and four-in  vertices, $v_7$ and $v_8$, 
with charge $+4$ and $-4$, 
respectively, have weight $d$.}
\label{fig:eight-vertex}
\end{figure}

The eight-vertex model is a generalization of the six-vertex model, first introduced to get rid of some of 
its very unconventional properties due to the 
hard ice-rule constraint (frozen FM state, quasi long-range order at infinite temperature, etc.)~\cite{Sutherland1970,Fan1970}. 
In this model the allowed local configurations are those for which each vertex is surrounded by 
an even number of arrows pointing in or out, resulting in the addittion of  the  two vertices with four ingoing and
four outgoing arrows shown in 
Fig.~\ref{fig:eight-vertex} to the ones in Fig.~\ref{fig:six-vertex}. The eight-vertex model can still be mapped into a loop model 
on the lattice in such a way that the integrability property is preserved. It was first
solved by Baxter in the zero-field case ({\it i.e.}, with $Z_2$ symmetry)~\cite{Baxter1971,Baxter1972}. In order to do that he 
introduced the celebrated Star-Triangle relations (now called Yang-Baxter equations). 

The phase diagram is 
characterized by the anisotropy parameter
\begin{equation}
\Delta_8=\frac{a^2+b^2-c^2-d^2}{2(ab+cd)}
\label{eq:Lambda8}
\end{equation}
which becomes the six-vertex one when $d=0$. This model sets into the following five phases depending on the weight of the 
vertices:

\vspace{0.2cm}

\noindent $a$-{\it Ferromagnetic phase} ($a$-FM): $\Delta_8>1 \,\, (a>b+c+d)$. 
Spin reversal symmetry is broken. This ordered phase is no longer frozen and $\langle M_+\rangle \leq 1$.

\vspace{0.2cm}

\noindent $b$-{\it Ferromagnetic phase} ($b$-FM): $\Delta_8>1 \,\, (b>a+c+d)$.
This phase is equivalent to the previous one replacing $a$- by $b$- vertices.

\vspace{0.2cm}

\noindent{\it Paramagnetic phase} (PM): $-1<\Delta_8<1 \,\, [a,b,c,d<(a+b+c+d)/2]$. As soon as $d>0$ this phase is truly 
disordered, with a finite correlation length. The averaged magnetization vanishes $\langle M_\pm \rangle =0$. 

\vspace{0.2cm}

\noindent $c$-{\it  Antiferromagnetic phase}  ($c$-AF):  
$\Delta_8<-1 \,\, (c>a+b+d)$. Translational symmetry is broken.
The configurations are dominated by $c$-vertices with an 
alternating pattern of vertices of type $v_5$ and $v_6$ with defects; $\langle M_- \rangle \leq 1$.

\vspace{0.2cm}

\noindent $d$-{\it  Antiferromagnetic phase} ($d$-AF): $\Delta_8<-1 \,\, (d>a+b+c)$. 
Translational symmetry is broken. The configurations are dominated by $d$-vertices, 
with an alternating pattern of vertices $v_7$ and $v_8$ with defects. $\langle M_-\rangle $ is also different from zero in this phase.  
This order parameter does not allow one to distinguish the $d$-AF from the $c$-AF.

\vspace{0.2cm}

The transition lines are given by $\Delta_8=1$ for the PM-FM ones and $\Delta_8=-1$ for the PM-AF ones.  The projection 
of the critical surfaces on the $d=0$ plane yields straight lines translated by $d/c$ with respect to the ones of the 
six-vertex model, in the direction of enlarging the PM phase, as shown by the dashed lines in 
Fig.~\ref{fig:phase-diagram-8vertex}\footnote{{Beyond the parameter $\Delta_8$ which is crucial in the classification of the phases of 
the eight-vertex model, a second parameter $\Gamma = \frac{a b - c d}{a b + c d} = \frac{1-r}{1+r} $ 
plays an important role~\cite{BaxterBook} and can be put in relation through the ratio $r = \frac{c d}{a b}$ with 
the values of the critical exponents (see Table~\ref{table:exponents-th}).}}.

The effect of the $d$-vertices on the order of the different phase transitions is very strong. 
As soon as $d>0$, the KT line between the $c$-AF and the SL phases
becomes `stronger', {\it i.e.} second order, except at the intersection with the $a=0$ or $b=0$ planes
when the transition is first order. On the contrary, the frozen-to-critical lines between the FMs and SL phases
get ``softer'', {\it i.e.} second order, and become KT transitions on the 
$a=0$ and $b=0$ planes. Finally, the separation between the $d$-AF and disordered phases
is second order for $a, \ b, \ d>0$ and it is of KT type on the $a=0$ and $b=0$ planes.
As we will show in Sec.~\ref{sec:numerical}, this is consistent with our numerical results
\footnote{{
In this work we will generically refer to the eight-vertex model as the model where {\it all} fugacities, $a,b,c$ and $d$ 
are different from zero, in contrast with the six-vertex model where $d=0$ (or other particular cases where any 
of the four Boltzmann weights is zero).}}.

{The critical exponents can be found from the analysis of the free-energy density close to the transition planes. In the $c$-AF regime (referred to as 'principal regime') they
depend explicitly on the weights of the vertices  via the parameter
$\tan(\mu/2) \equiv \sqrt{cd/ab}$~\cite{Baxter1971}. The critical behaviour close to the $a$-FM-PM transition is  obtained from the principal regime by replacing the parameter $c$ by $a$ and $d$ by $b$. 
The critical exponents for the PM-FM transitions in this model are given in the second column of 
Table~\ref{table:exponents-th}. 
The values of $\beta_e$, \ $\gamma_e$ and $\phi$  for the six-vertex model 
given in the first column are consistent with the eight-vertex model results as the limit $d\to 0$ (when $\mu=\pi$). }

\begin{table}[h]
\centering
 \begin{tabular}{|c|c|c|}
\hline
& six-vertex & eight-vertex \tabularnewline
\hline 
\hline 
$\gamma_e/\nu=\hat{\gamma_e}$ & 2 & $1+\mu/\pi$  \tabularnewline
\hline 
$\displaystyle{\beta_e/\nu=\hat{\beta_e}}$ & 0 & $(\pi-\mu)/(2\pi)$  \tabularnewline
\hline 
$(2-\alpha)/\nu=\hat{\phi}$ & 2 & $2$  \tabularnewline
\hline 
$\alpha$ & 1 &$2-\pi/ \mu$  \tabularnewline
\hline 
$\beta_e$ & 0 &$(\pi-\mu)/(4\mu)$  \tabularnewline
\hline 
$\gamma_e$ & 1 &$ (\pi+\mu)/(2\mu) $  \tabularnewline
\hline 
$\nu$ & 1/2 &$\pi/(2\mu)$  \tabularnewline
\hline 
 \end{tabular}
\caption{\small Exact critical exponents of the six- and eight- vertex model with {$\tan(\mu/2)=\sqrt{ab/cd}$}
along the SL/PM-FM transition.
In the limit $d\to 0$ the parameter $\mu\to\pi$ and the eight-vertex model exponents become 
the ones of the six-vertex model.}
\label{table:exponents-th}
\end{table}

\subsection{The sixteen-vertex model}

\begin{figure}[h]
\begin{center}
\includegraphics[scale=1.8]{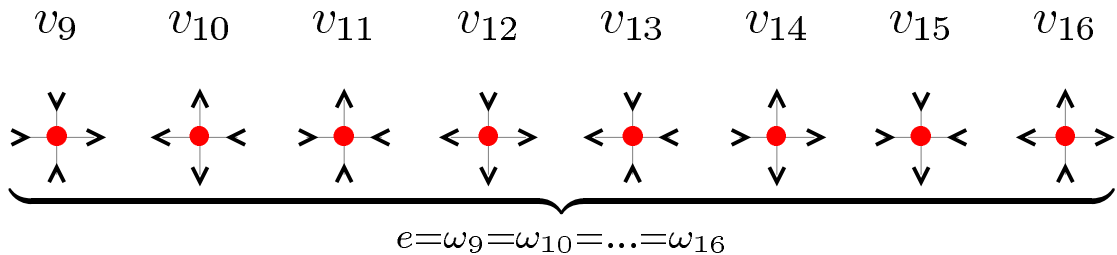}
\end{center}
\caption{\small (Color online.) The eight three-in  one-out  (with charge $-2$) or three-out one-in (with charge $+2$) 
vertices that are included in the sixteen-vertex model. We give them equal weight $e$. }
\label{fig:sixteen-vertex}
\end{figure}

The most general model obtained by removing the ice-rule is the sixteen-vertex model, in which no 
restriction is imposed on the value of the binary variables attached on each edge of the lattice, and all the 
$2^4=16$ vertex configurations 
can occur. The (eight) three-in one-out and three-out one-in vertices that are added to the ones already discussed are 
shown in Fig.~\ref{fig:sixteen-vertex}. In order to preserve the $Z_2$ symmetry  (in absence of an external 
magnetic field that would break the rotational symmetry), the same statistical weight $e$ is given
to all these `defects' with charge $2$ and $-2$. 
In the figure, vertices are ordered in pairs of spin-reversed 
couples ($v_{10}$ is the spin reversed of $v_9$ and so on) and the difference with the following 
couples is a rotation by {$\pi/2$} ($v_{11}$ is equal to $v_9$ apart from a $\pi/2$-rotation and so on).

The new vertices naturally entail the existence of new phases. 
One can envisage the existence of a critical SL phase 
for $a=b=c=d=\omega_{10,12,14,16}=0$ and $\omega_{9,11,13,15}>0$ as this new four-vertex model is equivalent to the dimer model 
solved by Kasteleyn~\cite{Kasteleyn1963}. It is quite easy to see that 
$e$-AF stripe order is also possible. For instance, one can build an ordered configuration 
with alternating lines of $v_9$ and $v_{10}$ vertices, or another one with alternating columns of 
$v_{11}$ and $v_{12}$ vertices.  Phases of this kind should appear if one favors one 
pair of spin-reversed related vertices by giving them a higher weight than the others, and the transition 
to this phase should be continuous, 
since local fluctuations made by elementary loops around a square plaquette are allowed.

The sixteen-vertex model loses the integrability properties~\cite{Wuphasetransitions,WuPRL1969,BelMaiVia92}. 
Nonetheless, some exact results are available for a few 
special sets of parameters.  The equivalent classical Ising model has only nearest and next-nearest neighbor 
two-body interactions when $e^4=abcd$~\cite{Wuphasetransitions,WuPRL1969}.  
In the $c$-AF sector  this condition leads to the generalised F model defined by $c=1$, 
$a=b<1$,  $d=a^u$ and $e=a^v$, with the constraint $4v=u+2$. The model has been solved 
for the special cases: (i) $v=1$ and $u=2$ ({\it i.e.} $e=a$ and $d=a^2$), 
the associated spin model simplifies into an AF Ising model with only nearest neighbor interactions. 
This model is known to exhibit a second-order phase 
transition\footnote{The transition occurs at $\epsilon/k_B T_c=2 \ln (\sqrt{2}+1)\approx 0.567$. Using our conjectured $\Delta_{16}$ 
(defined below) we get a critical temperature  
$\epsilon/k_B T_c\approx 0.607$. } with a logarithmic divergence of the specific heat 
($\alpha=0$). (ii) Using a different approach it has been {shown} that for  
 $v\rightarrow \infty$ and $u=2$ ({\it i.e.} $e=0$ and $d=a^2$ ) 
the system also exhibits a second-order phase transition in the same 
universality class as (i). Notice that the exactly solvable F model is recovered in the limit $v\rightarrow \infty$ 
and $u\rightarrow \infty$. In the same way, in the $a$-FM sector this leads to the generalised KDP 
model~\cite{WuPRL1970} by setting $a=1$,  $b=c<1$, $d=b^u$, $e=b^v$ and again 
$4v=u+2$.  For $v=1$ and 
$u=2$ ({\it i.e.} $e=b$ and $d=b^2$) the system exhibits a second-order phase transition with the same 
properties of its $c$-AF analog discussed above. For $v\rightarrow \infty$ and 
$u=2$ ({\it i.e.} $e=0$ and $d=b^2$ ) the system also exhibits a second-order 
phase transition in the same universality class as the previous case.

Since the phase diagram of the generic sixteen-vertex model is rather complex, and our aim is to 
consider $e$ and $d$ vertices as defects having a relative small statistical weight, 
in this work we will focus on the effect of the presence of $e$ vertices on the 
phases and the phase transition lines described in Fig.~\ref{fig:phase-diagram-8vertex}.

\subsection{Numerical simulations}

The numerical analysis of the equilibrium properties of $2D$ vertex models has been restricted, 
so far, to the study of the six and eight-vertex cases. Single spin-flip updates break the six- 
and eight- vertex model constraints and cannot be used to generate new allowed configurations. 
Instead, as each spin configuration 
can be viewed as a non-intersecting (six-vertex) or intersecting (eight-vertex) loop configuration, stochastic 
non-local updates of the loops have been used to sample phase space~\cite{Barkema1998,Syljuasen2004}.
By imposing the correct probabilities all along the construction of  non-local moves, cluster algorithms can be 
designed~\cite{Syljuasen2004,Evertz1993a}. Non-trivial issues {such} as the effect of boundary conditions have 
been explored in this way~\cite{Syljuasen2004,Allison2005}. 
  
Loop-algorithms, as usually presented in the context of Quantum Monte Carlo methods, exploit the world-line representation 
of the partition function of a given quantum lattice model~\cite{Suzuki1985}. It is well known that the $2D$ six- and eight-vertex 
models are equivalent to the Heisenberg XXZ and XYZ  quantum spin-$1/2$ chains, respectively~\cite{Sutherland1970,Suzuki1976}. 
It is then not surprising to find the same kind of loop-algorithms in the vertex models literature. A configuration in terms of bosonic 
world lines of the quantum spin chain in imaginary time can be one-to-one mapped into a vertex configuration on the square lattice, 
so that the same loop algorithm samples equivalently the configurations of both models.

The loop algorithms could be modified to include three-in -- one-out 
 and one-in -- three-out defects for the study of spin-ice systems~\cite{Jaubert2008,Melko2004}. However, the simultaneous 
 inclusion of  four-in and four-out defects makes this algorithm inefficient compared to a MC algorithm with local updates.  
For this reason, we will use local moves in our numerical studies, as we explain in the next Section. 
 
\section{Numerical study of the sixteen-vertex model}
\label{sec:numerical}

In this Section we summarize the results obtained for the sixteen-vertex model 
using MC simulations.
We first explain the numerical algorithm. Then, we discuss the phase diagram and the critical 
properties of the model. 
All our results are for a square lattice with linear size $L$ and periodic boundary conditions. 

\subsection{Numerical methods}

We used two numerical methods to explore the equilibrium 
properties of the generic model; the Continuous time Monte Carlo (CTMC) method that that 
we briefly explain in Sec.~\ref{subsubsec:CTMC} and in App.~\ref{app1} and the Non-equilibrium 
relaxation method (NERM)  that we equally briefly explain in Sec.~\ref{subsubsec:STR}.

\subsubsection{Continuous time Monte Carlo}
\label{subsubsec:CTMC}

The usual Metropolis algorithm, from 
now on called Fixed Step Monte Carlo algorithm (FSMC), is very inefficient to study the equilibrium 
properties of frustrated magnets. The dynamics freeze when $d, e\ll\min(a,b,c)$ 
and the acceptance probability for most of the single-spin flip updates is extremely small. We 
implemented an algorithm which overcomes this difficulty, the CTMC 
algorithm~\cite{Bortz1975,Barkema-Newman_Book}. This method 
is also known with different names: Bortz-Kalos-Lebowitz \cite{Bortz1975}, 
n-fold way or kinetic MC. The 
basic aim of this algorithm is to get rid of the time wasted  due to a large number of rejections when the physics of the problem 
imposes a very small acceptance ratio.  It is extremely useful for the study of the long time behavior of systems with complicated 
energy landscapes and a large number of metastable states. The main idea behind the method is to sample stochastically the time 
needed to update the system and then do it without rejections. In our case, 
we chose to use single spin updates such that the two vertices connected by the spin are updated in the move.
Details on this algorithm are given in App.~\ref{app1}. Equilibrium can be achieved for relatively large samples. 
The finite size scaling analysis of the equilibrium MC data yield the 
thermodynamic properties of the generic model.

\subsubsection{Non-equilibrium relaxation method}
\label{subsubsec:STR}

The fact that dynamic scaling applies during relaxation at a 
critical point~\cite{Janssen89} suggested to use {\it short-time}
dynamic measurements to extract equilibrium critical 
exponents with numerical methods~\cite{Barkema-Newman_Book,Jaster1999,Albano11}.
{With this method it is not needed to equilibrate the systems and 
only short-time scales are evaluated.  Therefore, it is not necessary to use the CTMC version but a plain MC is 
sufficient. We parametrize the consecutive steps of the MC simulation with a
parameter $t$ measured in MC step units, each of these corresponding to a sweep of the 
single spin flip MC algorithm over $N$ spins taken at random on the sample.}
Magnetized, $M^0_{\pm} \equiv M_{\pm}(t=0) \neq 0$, and non-magnetized, $M_{\pm}^0=0$, 
configurations can be used as starting conditions
and the critical relaxation
\begin{equation}
M_{\pm}(t) \simeq t^{-\beta_e/(\nu z)} \ F(t^{x_0/z} M^0_{\pm})
\end{equation}
where $z$ is the dynamic critical exponent, and $F(x) \simeq x$ for $x\ll1$ and $F(x) \to \mbox{cst}$ for 
$x\to\infty$, can be used to extract either the critical parameters or the critical exponents. This expression is expected to hold 
for $t^{1/z} \ll L$ and $t^{1/z} \ll \xi_{eq}$ with $\xi_{eq}$ 
the equilibrium correlation length. 

\subsection{Phase diagram and critical singularities}

In this subsection we present a selected set of results from our simulations, and we describe the 
kind of phases and critical properties found. The strategy to study the different phase transitions is 
the following.
We first chose the relevant order parameter, $\langle M_+\rangle$ or $\langle M_- \rangle$, to study FM or AF phases.
From the finite size scaling analysis of the corresponding fourth-order reduced Binder's cumulant
\begin{equation}
K_{M_{\pm}}=1-\frac{\langle M_{\pm}^4\rangle}{3\langle M_{\pm}^2\rangle^2}
\simeq \Phi_K(t L^{1/\nu})
\label{eq:K-def}
\end{equation}
where $t$ is the distance from the critical point, we extracted the critical exponent $\nu$.
From the maximum of the magnetic susceptibility 
\begin{equation}
\chi_{\pm} = L^2 \left[ \langle M_{\pm}^2 \rangle - \langle M_{\pm} \rangle^2  \right]
\simeq  L^{\gamma_e/\nu} \Phi_\chi(t L^{1/\nu})
\label{eq:suscp-magn}
\end{equation}
we extracted $\gamma_e/\nu$, then $\gamma_e$
as $\nu$ was already known. From the maximum of the specific heat 
\begin{equation}
C_E=L^{-2} \left[ \langle E^2\rangle - \langle E\rangle^2 \right]
\simeq L^{\alpha/\nu} \Phi_C(t L^{1/\nu})
\end{equation}
we extracted 
$\alpha/\nu$, then $\alpha$. Here, $E=\sum_k n_k \epsilon_k$ with $n_k$ the number of vertices of type $k$ and 
$\epsilon_k$ their energy.
The direct measurement of $\beta_e$ is difficult, we thus deduced it from the scaling 
relation $\beta_e=\frac{1}{2}(2-\alpha-\gamma_e)$.
Finally, we checked hyper-scaling, {\it i.e.} whether $d \nu=2-\alpha$ is satisfied by the exponent values 
obtained, that we summarize in Table~\ref{table:num-exponents} for the SL/PM-FM transition and two choices 
of parameters. 

\subsubsection{The PM-FM transition}

In order to reduce the number of parameters in the problem we studied the PM-FM transition for the 
special choice $d=e$.

\begin{figure}[h]
\begin{center}
\includegraphics[scale=1,angle=0]{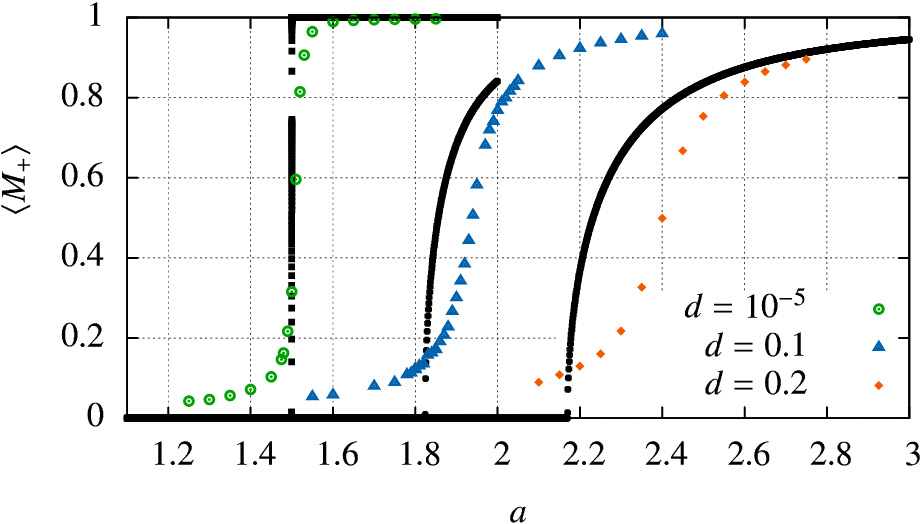}
\end{center}
\caption{\small (Color online.) Equilibrium magnetization density, $\langle M_+\rangle$, of the sixteen-vertex model 
for three different values of $d=e$ (given in the key) and $b=0.5$ as a function of $a$ (vertices fugacities are normalized by 
$c$). The colored data points are the result of the numerical simulations of the $2D$ model for $L=40$ while the
black dots corresponds to the analytic solution of the model defined on the tree of plaquettes, as explained in 
Sec.~\ref{sec:model-on-the-tree}.} 
\label{fig:magn-PM-FM}
\end{figure}

As the direct magnetization density $\langle M_+ \rangle$ defined in eqs.~(\ref{eq:magn})-(\ref{eq:magny})
is the order parameter for the PM-FM transition in the six-vertex model, we study this quantity 
to investigate the fate of the FM phase in the sixteen-vertex model. In Fig.~\ref{fig:magn-PM-FM} we       show $\langle M_+ \rangle$ 
as a function of $a$ for $b=0.5$ and three values of the fugacity $d=e$ (all normalized by $c$). The data 
for the $2D$ model (shown with colored points)
demonstrate that the presence of defects tends to disorder the system and, therefore, the extent of the PM phase is enlarged 
for increasing values of $d=e$. Moreover, the variation of the curves gets smoother for increasing values of $d=e$ suggesting
that the transitions are second order (instead of frozen-to-critical) in presence of defects. The data displayed with black 
points for the same parameters are the result of the mean-field analysis of the model, which will be 
discussed in Sec.~\ref{sec:model-on-the-tree}. 
In~\cite{Levis2012} we suggested that the equilibrium phases of the sixteen-vertex model with $d=e$ could be characterized 
by a generalization of the anisotropy parameter of the eight-vertex model recalled in eq.~(\ref{eq:Lambda8}):
\begin{equation}\label{Delta16_2D_model}
\Delta_{16}=\frac{a^2+b^2-c^2-(d+3e)^2}{2[ab+c(d+3e)]}
\; . 
\end{equation}
In the same way as in the integrable cases, the proposal is that the PM phase corresponds to the region of the 
parameters' space where $|\Delta_{16}|<1$, the FM phases corresponds to  $\Delta_{16}>1$, 
and the AF ones to $\Delta_{16}<-1$. It follows that 
the projection of the FM-transition hyper-planes onto the $(a/c,b/c)$ plane should then be parallel to the ones 
of the six- and eight-vertex models and given by $a_c=b+c+d+3e$ (or equivalently $b_c=a+c+d+3e$). 
As shown in Fig.~\ref{fig:magn-PM-FM}, the numerical results are well fitted by Eq.~(\ref{Delta16_2D_model}) which
is, however, not exact.
As we will explain in detail in Sec.~\ref{sec:model-on-the-tree}, our mean-field treatment 
of the model defined on a tree of single vertices
predicts a similar shift of the transition lines given by $a_c=b+c+d+2e$. The parameter capturing all transition lines is,
within this analytic approach, 
\begin{equation}
\Delta^{sv}_{16}=\frac{a^2+b^2-c^2-d^2+2(a+b-c-d)e}{2(cd+ab+e(a+b+c+d+2e))}
\end{equation} 
(see Sec.~\ref{sec:model-on-the-tree} for the technical details). For the more sophisticated 
tree made of `plaquette' units the transition lines are not parallel to the ones of the six- and eight-vertex models 
and an analytic form of the anisotropy parameter, $\Delta_{16}^{pl}$, has not been found. Nevertheless, 
the  transition lines can be computed numerically and their evaluation leads to the phase diagram depicted 
in Fig.~\ref{fig:phase_diagram-16vertex}.  

\vspace{0.25cm}
\vspace{0.25cm}

\begin{figure}[h]
\begin{center}
\includegraphics[scale=0.9,angle=0]{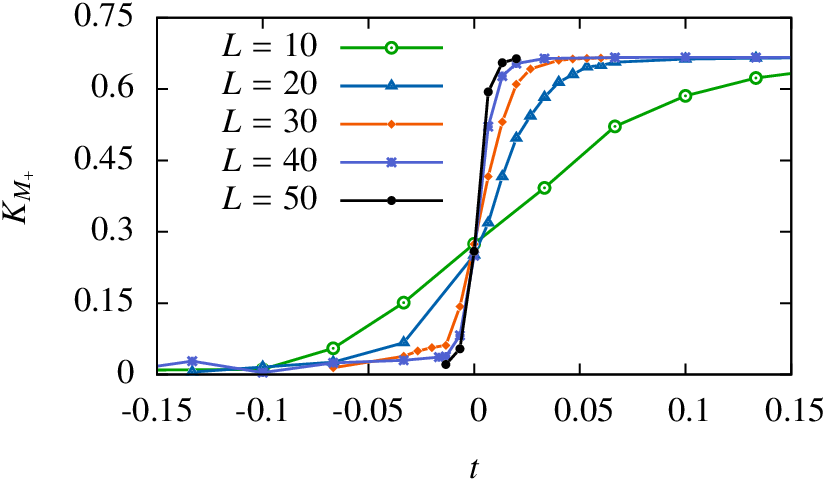}
\hspace{1.5cm}
\includegraphics[scale=0.9,angle=0]{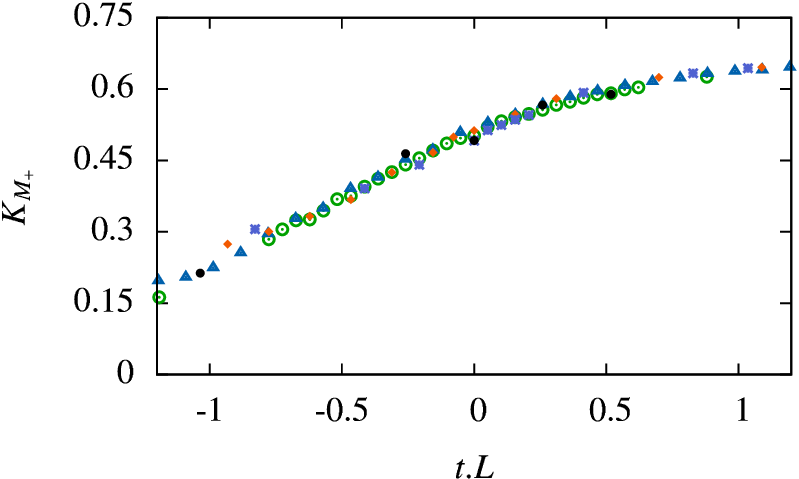}
\end{center}
\caption{\small (Color online.) Analysis of the Binder fourth-order  cumulant defined in eq.~(\ref{eq:K-def}) across the FM-PM 
transition in the sixteen-vertex model.
(a) Raw data for $d=e=10^{-5}$. 
(b) Scaling plot for $d=e=0.1$.
One extracts $1/\nu = 1.65 \pm 0.05$ in case (a) and $1/\nu = 1\pm 0.1$ in case (b)
from this analysis.}
\label{fig:K-PM-FM}
\end{figure}

Further evidence for the transition becoming second order comes from 
the analysis of the fourth-order cumulant defined in eq.~(\ref{eq:K-def}). Raw data on such Binder's cumulant across the 
FM-PM transition were shown in~\cite{Levis2012} where we showed that they  
intersect at a single point, as expected in a second order phase transition. 
In Fig.~\ref{fig:K-PM-FM} we display raw data for $d=e=10^{-5}$ (a) and scaled data for $d=e=0.1$ (b)
as a function of $t=(a-a_c)/a_c$. 
In both cases $b=0.5$ and, as above, we normalize all fugacities by $c$. From the analysis of the scaling properties we extract  
$a_c = 1.5$ for $d = e= 10^{-5}$ and $a_c=1.93$ for $d=e=0.1$. Sets of 
data for linear system sizes $L = 10, \ 20, \ 30, \ 40, \ 50$ are scaled quite satisfactorily by using $1/\nu = 1.65 \pm 0.05$ for the small 
$d$ and $1/\nu = 1\pm 0.1$ for the large value of $d$.

\begin{figure}[h]
\includegraphics[scale=0.85, angle=0]{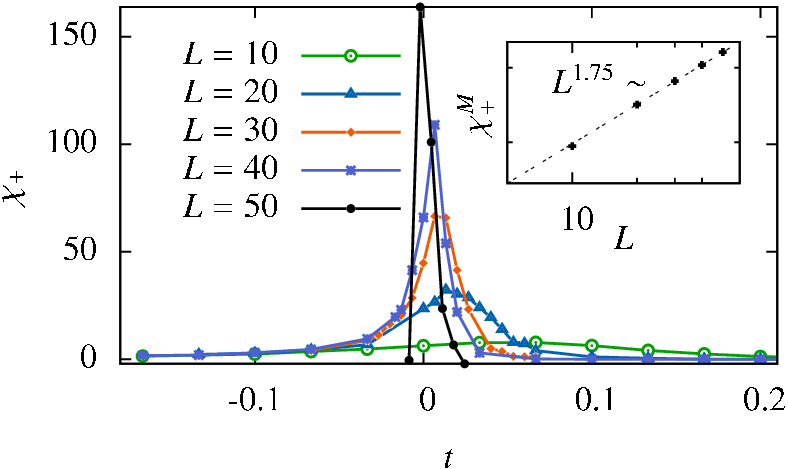}
\hspace{1.5cm}
\includegraphics[scale=0.85, angle=0]{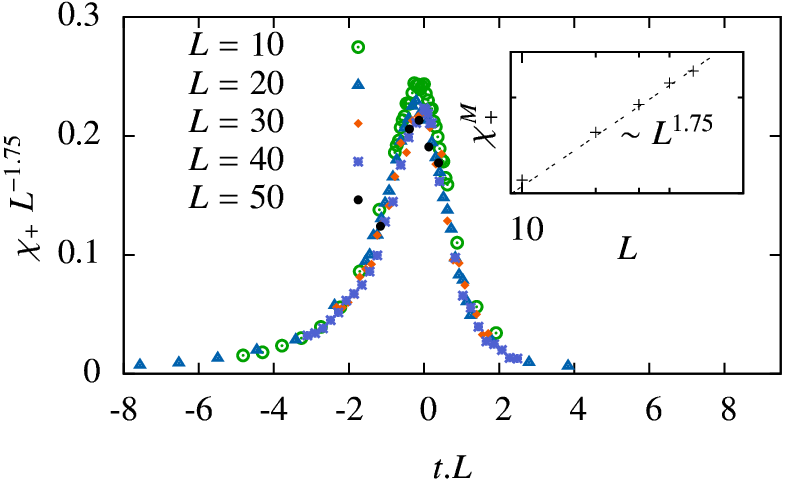}
\caption{\small (Color online.) The magnetic susceptibility across the PM-FM transition
for $b=0.5$, $d=e=10^{-5}$ (a) and $d=e=0.1$ (b). Where 
$t$ is the distance from the critical point measured as 
$t=(a-a_c)/a_c$ with $a_c=1.5$ (a) and $a_c=1.93$ (b). From the finite size scaling of the maximum shown in 
the insets one extracts $\gamma_e/\nu\simeq 1.75 \pm 0.02$ in both cases. }
\label{fig:chi-PM-FM}
\end{figure}

In order to complete the analysis of this transition we studied the magnetic susceptibility  
(\ref{eq:suscp-magn}) associated to the direct magnetization $M_+$ and its finite size scaling. Figure~\ref{fig:chi-PM-FM} 
displays $\chi_+$ for $b = 0.5$, $d = e = 10^{-5}$ (a) and $d=e=0.1$ (b), 
and five linear sizes, $L=10, \ 20, \ 30, \ 40, \ 50$. 
The data 
collapse is very accurate and it allows us to extract the exponent $\gamma_e/\nu \simeq 1.75 \pm 0.02$
in both cases. The study of the maximum of $\chi_+$ displayed in the insets confirms this estimate for 
$\gamma_e/\nu$. We repeated this analysis for other values of $d=e$ and we found that in all cases
critical scaling is rather well obeyed and, interestingly enough, $\gamma_e/\nu$ 
is, within numerical accuracy, independent of $d=e$. This is similar to what happens in the eight-vertex model
where, as shown in Table~\ref{table:exponents-th}, this ratio is independent of the parameters.

The ratio $\alpha/\nu$  is obtained from the  finite size analysis of  the 
specific heat (not shown) that is consistent with $C^{max} \simeq L^{\alpha/\nu}$ (instead of $L^D$ for a first 
order phase transition). We found $\alpha/\nu \simeq 1.30 \pm 0.06$ for $b=0.5$ and $d=e=10^{-5}$
and a logarithmic divergence of the heat capacity, {\it i.e.} $\alpha/\nu \approx 0$ for $d=e=0.1$ 
(cf. Table~\ref{table:num-exponents}), although it is very difficult to distinguish numerically a logarithmic divergence
from a power-law one with a very small exponent.

The critical exponents extracted numerically at the second order phase transition with $d=e>0$ are 
compared to the ones of the six-vertex model and the $2D$ Ising model in Table~\ref{table:num-exponents}.
It is interesting to note that for very small value of $d=e$ the exponents are rather close to the ones of the 
six-vertex model while for large value of $d=e$ they approach the ones of the $2D$ Ising model.
In terms of a Renormalization Group 
{(RG)} approach, this behavior suggests the existence of two
fixed points, one in the $d=e=0$ plane describing the critical behavior of the six-vertex model, and another
for $d,e>0$, belonging to the $2D$ Ising universality class. The first one might be unstable as soon as an infinitesimal
amount of defects is allowed. A RG treatment of the model is required to confirm this guess. 

{The critical exponents obtained from the numerical analysis depend on the fugacity $d=e$, namely, they vary along the transition
lines, as it happens for the eight-vertex model. 
However, as shown in Table~\ref{table:num-exponents}, the ratios of critical exponents $\hat{\gamma_e}$, $\hat{\beta}_e$ 
and $\hat{\phi}$ do not depend qualitatively 
on the choice of the parameters and are equal to the ones of the
$2D$ Ising model.}

The NERM yields values of the critical $a$ that are in agreement (within 
numerical accuracy) with the ones found with the conventional analysis. We do not show 
this analysis here. 

\begin{table}[h]
\centering
\begin{tabular}{|c|c|c|c|c|}
\hline 
 & six-vertex &  MC ($d=e=10^{-5}$) & MC ($d=e=0.1$) & $2D$ Ising \tabularnewline
\hline 
\hline 
$\gamma_e/\nu=\hat{\gamma}_e$  & $2$&  $1.75 \pm 0.02$ &  $1.75 \pm 0.02$ & 7/4 \tabularnewline
\hline 
$\beta_e/\nu=\hat{\beta}_e$ & $0$ & {$0.14\pm0.05$} & $\approx 0.125$& 1/8 \tabularnewline
\hline 
$(2-\alpha)/\nu=\hat{\phi}$ & $2$ &  
$2.03\pm0.15$ & $\approx 2$ & $2$ \tabularnewline
\hline 
$\alpha$ & $1$  & $0.78\pm0.23$ & $\approx 0$ & 0 \tabularnewline
\hline 
$\beta_e$ & $0$  & $0.085\pm0.014$ & $\approx 0.125$ & 1/8 \tabularnewline
\hline 
$\gamma_e$ & $1$  & $1.05\pm0.03$ & $1.75\pm0.18$ & 7/4  \tabularnewline
\hline 
$\nu$ & $1/2$ & $0.60 \pm 0.02$ & $1.0\pm0.1$ & 1 \tabularnewline
\hline 
$2\nu=2-\alpha$ ? & yes & yes & yes & yes \tabularnewline   
\hline 
\end{tabular}
\caption{\small Numerical values of the critical exponents at the FM-PM  transition in the sixteen-vertex model as compared to the 
ones in the six-vertex model (first column) and $2D$ Ising model (fourth column). 
We did not include errorbars in the column corresponding to $d=e=0.1$ as our determination of $\alpha$ 
is not precise enough to distinguish between $\alpha=0$ (the value used to extract the remaining exponents) 
and a very small but non-vanishing value.}
\label{table:num-exponents}
\end{table}

\subsubsection{The $c$-AF-PM transition}

We now focus on the transition between the $c$-AF and PM phases. For the six-vertex model this is a KT transition 
while for the eight-vertex model it is of second order as soon as $d>0$. In this case we chose to work with 
$d= 10^{-5} \neq e=10^{-3}$ and with $d=e=10^{-5}$. We present data obtained with the NERM.

\begin{figure}[h]
\hspace{2cm}
\includegraphics[scale=1,angle=0]{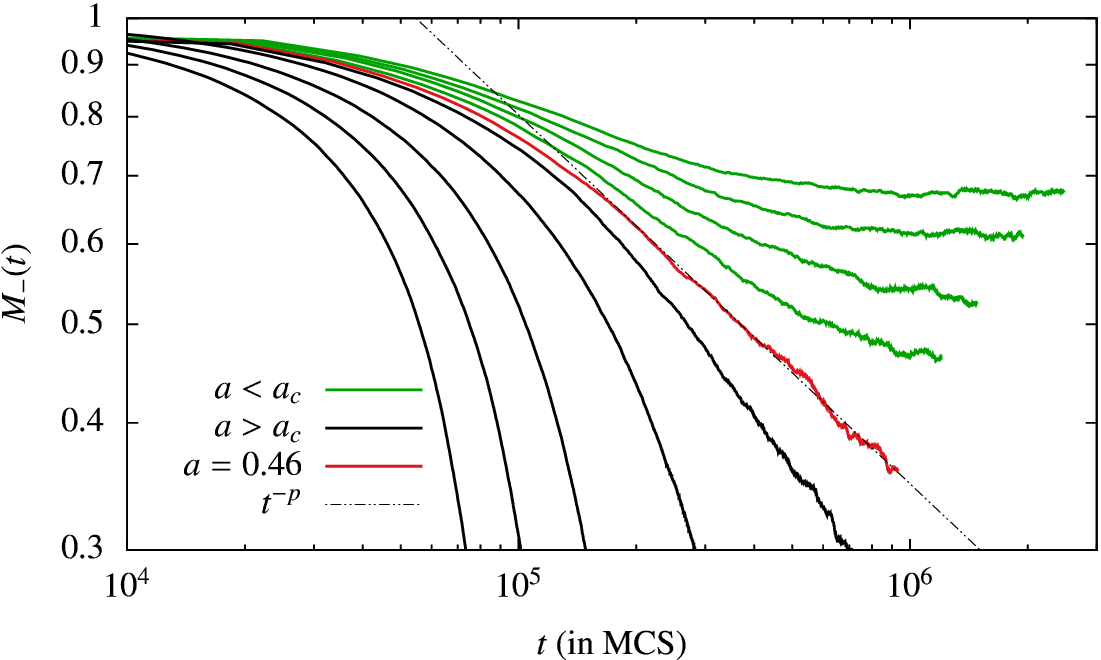}
\caption{\small (Color online.) Non-equilibrium relaxation of the staggered  magnetization from a fully ordered initial 
condition $M_-^0=1$ at different values of $a=b$, for $c=1$, $e=10^{-3}$ and $d=10^{-5}$. After a short transient the 
relaxation at the critical point follows a power law $t^{-p}$ with $p=\beta_e/(\nu z)$. We identify such critical relaxation at $a_c=0.46\pm0.01$.}
\label{NERMAF}
\end{figure}

Figure~\ref{NERMAF} shows the relaxation of the staggered averaged magnetization at $b=c=1$ and 
different values of $a$ given in the caption. The power-law relaxation, typical of the 
critical point, is clearly identifiable from the figure. We extract 
the critical value $a_c=0.46\pm0.01$ for the $c$-AF-PM phase transition. Moreover, the data demonstrate 
that the PM phase is not of SL-type as soon as a finite density of defects is allowed. Indeed, the relaxation 
of $M_-$ does not follow a power law in the PM phase, the decay being exponential for $a>a_c$.
Although this strategy gives a rather precise determination of $a_c$, it is very hard to determine the 
value of the exponent $p$, and hence of $z$, with good precision as
$p$ is extremely sensitive to the choice of $a_c$. 

\begin{figure}[h]
\includegraphics[scale=0.9,angle=0]{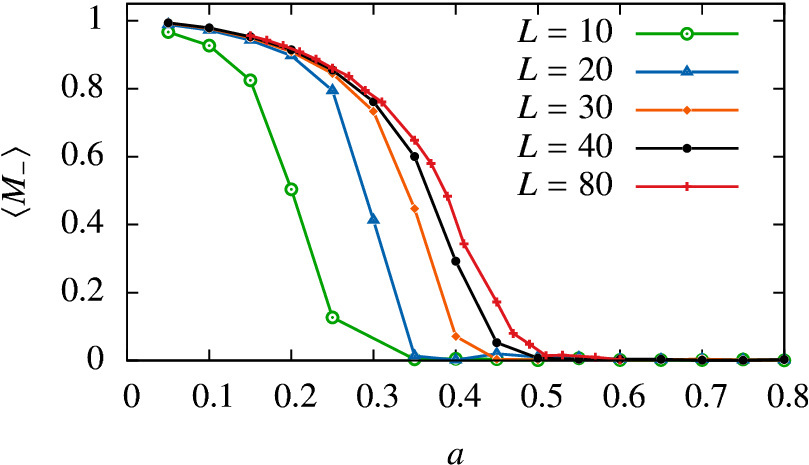}
\hspace{1.5cm}
\includegraphics[scale=0.9,angle=0]{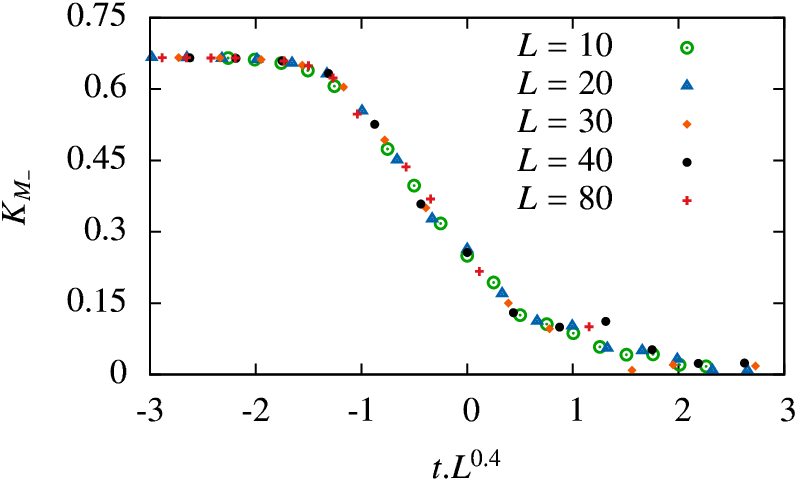} 
\caption{\small (Color online) Study of the $c$-AF-PM transition for $b=0.5$ and $d=e=10^{-5}$.
(a) The averaged staggered magnetization as a function of $a$ for several system sizes given in the key.
(b) Scaling plots across the 
$c$-AF-PM transition of the Binder cumulant of the staggered magnetization.
$t$ is the distance from the critical point $t=(a-a_c)/a_c$ with $a_c=0.5$.
The best scaling of data is obtained for $1/\nu = 0.4 \pm 0.05$. 
}
\label{fig:Binder-AF}
\end{figure}

The standard analysis of the $c$-AF-PM transition is not as clean as for the FM-PM one. Figure~\ref{fig:Binder-AF} (a)
shows the scaling plot of the Binder cumulant of the staggered magnetization for $b=0.5$ and, in 
this case, $d=e=10^{-5}$. From it one extracts $1/\nu=0.4 \pm 0.05$. The susceptibility fluctuates too much 
to draw any stringent conclusion about the exponent $\gamma_e$. The analysis of the specific heat (not 
shown) suggests a logarithmic divergence $\alpha \approx 0$. The dependence of the critical line on the 
fugacities is reasonably well described by $\Delta_{16}$. 

Both $e$ and $d$ vertices make the disordered phase be a conventional PM, and the transitions be second order. 
Although $d$ does not break integrability while $e$ does, their effect, in these respects, are similar.
The fact that the critical exponents in the eight-vertex model depend on the fugacities is known from
the exact solution. In the sixteen-vertex model, where integrability is lost, this is not the case.
Unfortunately, our numerical analysis does not allow us to draw stringent enough
conclusions for the $c$-AF-PM transition, since
it is very hard to get precise measurement of the observables close to the critical lines. 
Possibly, a {RG} approach would be useful in this respect.

\section{Bethe-Peierls mean-field approximation: Vertex models on a tree}
\label{sec:model-on-the-tree}

In this Section we study the properties of the six-, eight- and sixteen-vertex models
defined on the Bethe lattice by using the cavity method. 
First, we will consider a standard Bethe lattice of uniform connectivity $\mathcal{C} = 4$ (Sec.~\ref{sec:tree-ver}). 
In order to get more accurate results, we also 
study the model on a tree of plaquettes of $2\times2$ vertices (Sec.~\ref{sec:tree-plaq}), which
account for local 
excitations and  
fluctuations induced on short scales 
by the presence of small loops. 

\subsection{The cavity method}

The cavity method is a technique that allows one to calculate the average properties of 
statistical models
defined on tree-like graphs in the thermodynamic limit~\cite{Bethe35,MM09}. 
The method, which is equivalent to the Bethe-Peierls (BP) approximation,  
is based on the assumption that
due to the 
tree-like structure of the lattice, 
in absence of a given site (the cavity), the neighbors of that site 
 are not correlated and their marginal joint probability factorizes. 
 
 Removing one site from the graph creates ${\mathcal C}$  
 {\it rooted trees}, each one being a tree where all the sites of the bulk have the same connectivity ${\mathcal C}$,
apart from the root which has only ${\mathcal C}-1$ neighbors. 
The evaluation of physical observables  is based 
on the determination of the properties of the site at the {\it root} of a rooted tree. 
Thanks to the above-mentioned factorization property one can 
write relatively simple
recursion equations for the marginal probabilities of the rooted
sites.
Such equations have to be solved self-consistently, the fixed points of which
yield the free energy of the system along with all the thermodynamic observables. 

The BP method can be interpreted either as the {\it exact} solution of the model on the tree, or as 
an {\it approximate} solution of the original model on the Euclidean lattice.
In order to mimic an Euclidean lattice in $D$ dimensions the tree should have
connectivity ${\mathcal C} = 2D$. 
Such an approach takes into account
short-scale ($O(1)$) correlations and it is
expected to give more accurate quantitative results than the standard fully-connected mean-field approach.
It is well-known, for instance, that for the Ising model the BP method yields the mean-field critical exponents
with a better estimate of $T_c$ than the one obtained on  the fully-connected graph.
 
\subsection{The trees}
\label{sec:the-tree}

The definition of the vertices requires the selection of a particular
orientation of the edges  adjacent to a given site. We will define 
``horizontal" and ``vertical" edges, each one with two possible orientations. With this procedure 
we associate a statistical weight to each vertex configuration,
even in such non-Euclidean geometry. 
In the recursion equations this 
partition will translate into four different species of rooted trees, depending on the ``position'' 
(left, right, down or up) 
of the missing edge.

\subsubsection{A tree of vertices} \label{sec:tree-ver}

\begin{figure}[h]
\begin{center}
\includegraphics[scale=0.7]{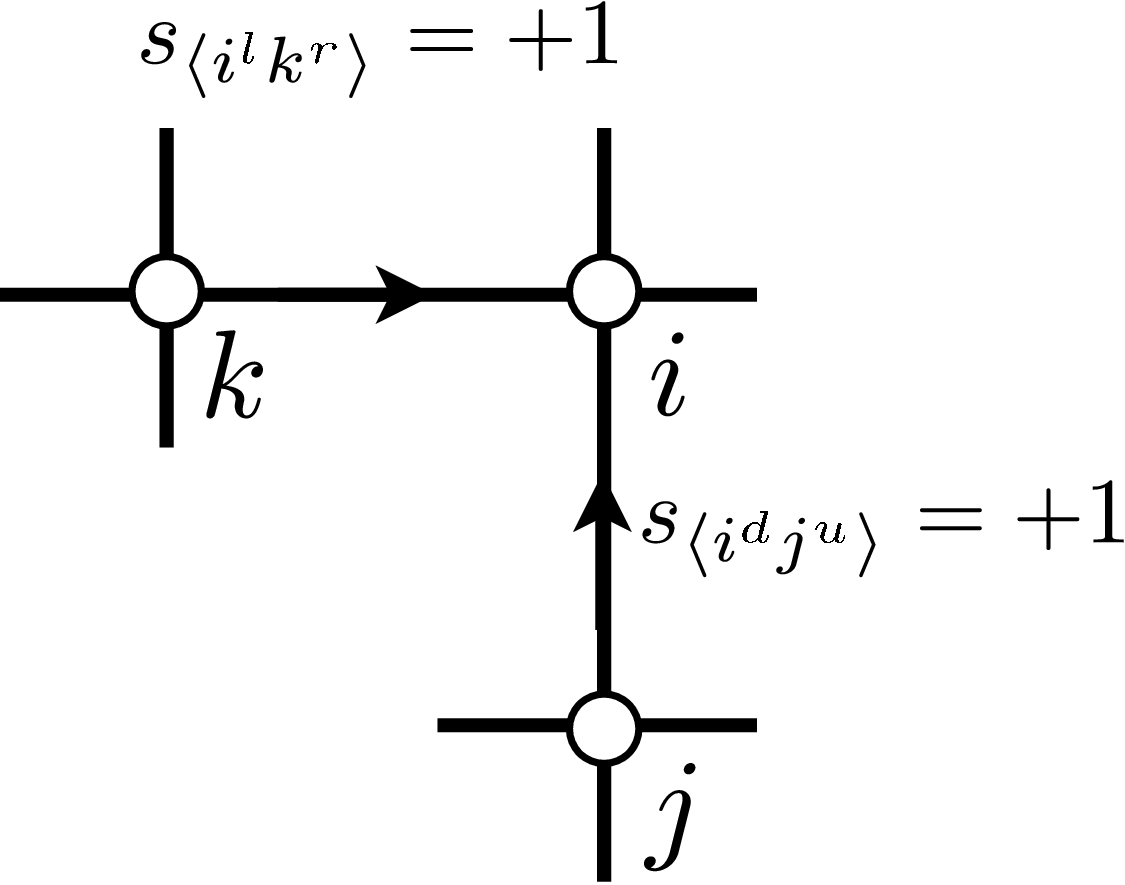}
\end{center}
\caption{\small (Color online.) Two spins living on an horizontal and a vertical
edge, both taking value $+1$. }
\label{fig:def_spins}
\end{figure}

In the models we are interested in, each site is a vertex and its coordination, which is fixed and equal to four, 
is the number of vertices connected to it. In order to distinguish one type of vertex from another, and 
to identify all possible phases, we define  the
analogue of the two orthogonal directions of the Euclidean square lattice:
each vertex has four terminals
that we call ``up" (u), ``down" (d), ``left" (l) and ``right" (r). So far,
the vertices were labeled by their positions. 
Here, for the sake of clarity, we label them with a single latin index, say $i$, $j$, $k$, $\ldots$.  
Vertices are connected through edges $\langle i^{d}j^{u}\rangle$ 
and $\langle i^{l}k^{r}\rangle$ that link respectively the down extremity of a vertex $i$ 
with the up terminal of a neighboring vertex $j$,  or the left end of $i$ with the right end of its neighbor $k$.
The symbols $\langle i^d j^u \rangle$  and $\langle i^l k^r \rangle$
denote undirected edges, so  that $\langle i^d j^u \rangle = \langle  j^u i^d \rangle$
and $\langle i^l k^r \rangle = \langle  k^r i^l \rangle$. 
In this way, one creates a bipartition of the edges into horizontal (left-right $\langle i^{l}k^{r}\rangle$) 
and vertical (up-down $\langle i^{d}j^{u}\rangle$) edges.  This notation gives a notion of which vertex is above ($i$) 
and which one is below ($j$) and similarly for the horizontal direction. See the sketch in Fig.~\ref{fig:def_spins}.

Each edge is occupied by an arrow shared by two vertices. An arrow defined on
the $\langle i^l k^r \rangle$ edge is the left arrow for the $i$ vertex and the 
right arrow for the neighboring $k$ vertex. A similar distinction holds for the vertical  $\langle i^d j^u \rangle$ edges. 
Each arrow, as any binary variable, can be identified with a spin degree of freedom, 
taking values in $\{-1,1\}$. In this construction there are two kinds
of spins, those living on horizontal edges, $s_{\langle i^l  k^r \rangle}$, and those sitting on 
vertical edges, $s_{\langle i^d j^u \rangle}$. 
Without {loss} of generality, we choose a convention such that 
$s_{\langle j^u i^d \rangle}=+1$ if the arrow points up and $-1$ otherwise. 
Similarly, $s_{\langle k^r i^l \rangle}=+1$ if the arrow points right and $-1$ otherwise, for the horizontal arrows
 (see Fig.~\ref{fig:def_spins}).
This is the analog of the convention 
 used for the spin sign assignment in the $2D$ model (of course, alternative choices of the signs of
the spins can be used). 

The local arrow configuration defines the state of the selected vertex. 
With the spin definition given above, we can assign a total spin $S_i$ to each vertex $i$, 
and define it as the sum 
of the spins attached to it, $S_i \equiv  \frac12 \sum_{j \in \partial i} s_{\langle i j \rangle}$,
where $\partial i$ indicates the neighborhood of $i$. 
With this assignment, the spin associated to each type of vertices is 
$S_i=\pm 2$ if the vertex is of type $a$, 
$S_i=0$ if it is a $b$ one, $S_i=0$ if the vertex is of kind $c$, $S_i=0$ for a $d$ vertex, 
and $S_i=\pm 1$ for the $e$ ones. 

\begin{figure}[h]
\begin{center}
\includegraphics[scale=0.5]{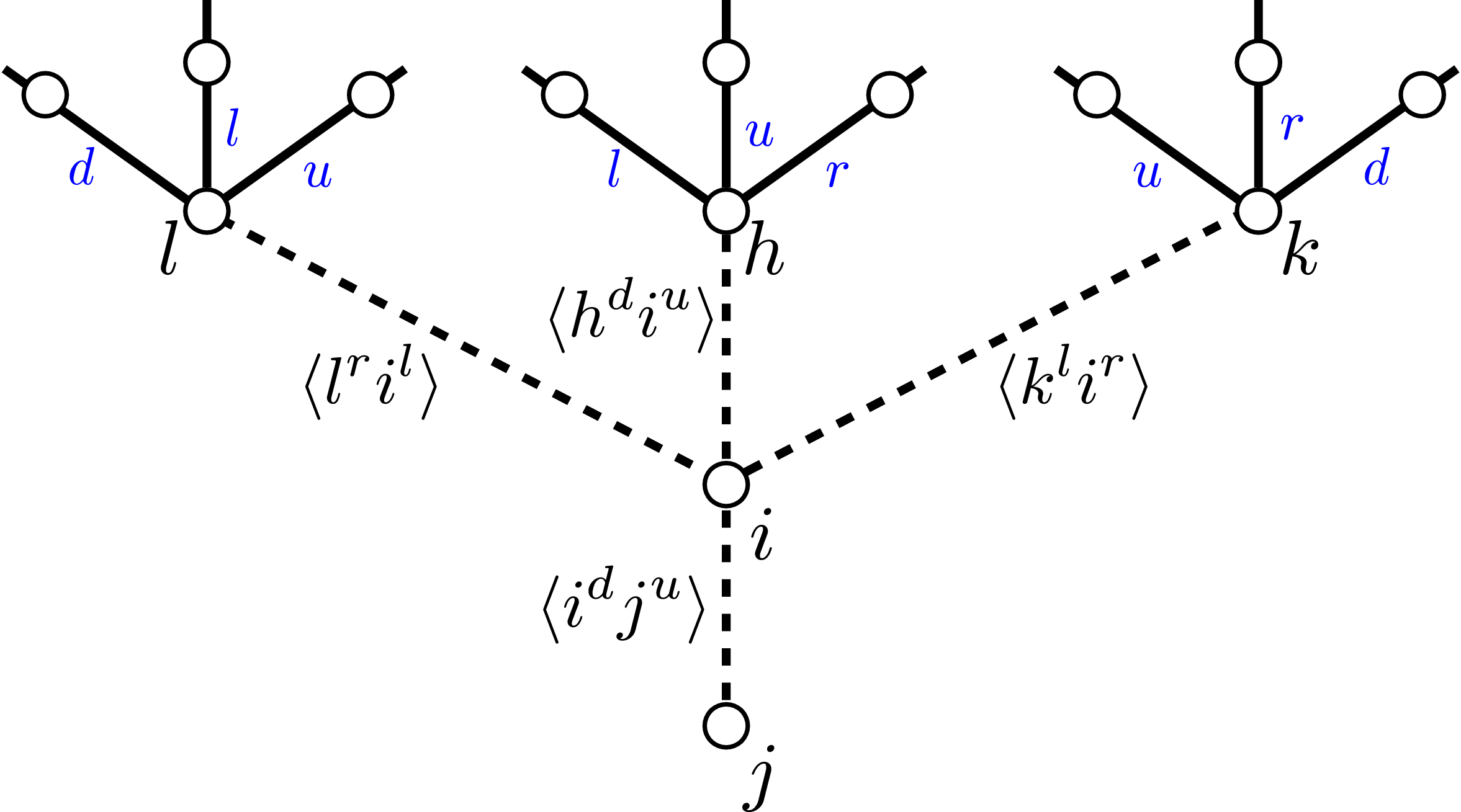}
\end{center}
\caption{\small (Color online.) Construction of un ``up rooted tree" from
the merging of a left, a right and an up rooted tree.}
\label{fig:rooted_up}
\end{figure}

Consider now a site (vertex) $i$ at the root of a rooted tree, in absence of an edge with
one of its neighbors, say site $j$.
There are four distinct rooted trees
depending on whether the  missing edge with vertex $j$ is the one on its 
left, right, up or down direction.  
By analogy with the $2D$ case, 
one could interpret these rooted trees as the result of the integration of a transfer 
matrix  in four possible directions.
This can be emphasized by taking into account the particular direction of the missing edge 
at the root that we will indicate as follows: $i^\alpha\to j^\beta$, with $\alpha,\beta \in \{u,d,l,r\}$.
For instance, an ``up rooted tree"  is the one in which the root $i$ has no connection
to the up terminal of the vertex $j$, {\it i.e.} the link $i^d\to j^u$ is absent.
As shown in Fig.~\ref{fig:rooted_up}, such a rooted tree is obtained by merging a left, 
an up and a right rooted tree (with root vertices  $l$, $h$ and $k$ respectively) 
with the addition of a new vertex $i$ through the links $l^r\to i^l$, $h^d\to i^u$, 
$k^l\to i^r$ (pictorially the transfer matrix is moving down). 
Similarly, a ``left rooted tree" is obtained by merging a down, a left
and an up rooted tree, and so on. 
The Bethe lattice is finally recovered by joining an up, a left, a down and a right
rooted tree with the insertion of the new vertex.
Equivalently, given a tree, one creates rooted (cavity) trees by removing an edge.

\subsubsection{A tree of plaquettes} \label{sec:tree-plaq}

\begin{figure}[h]
\begin{center}
\includegraphics[scale=0.8]{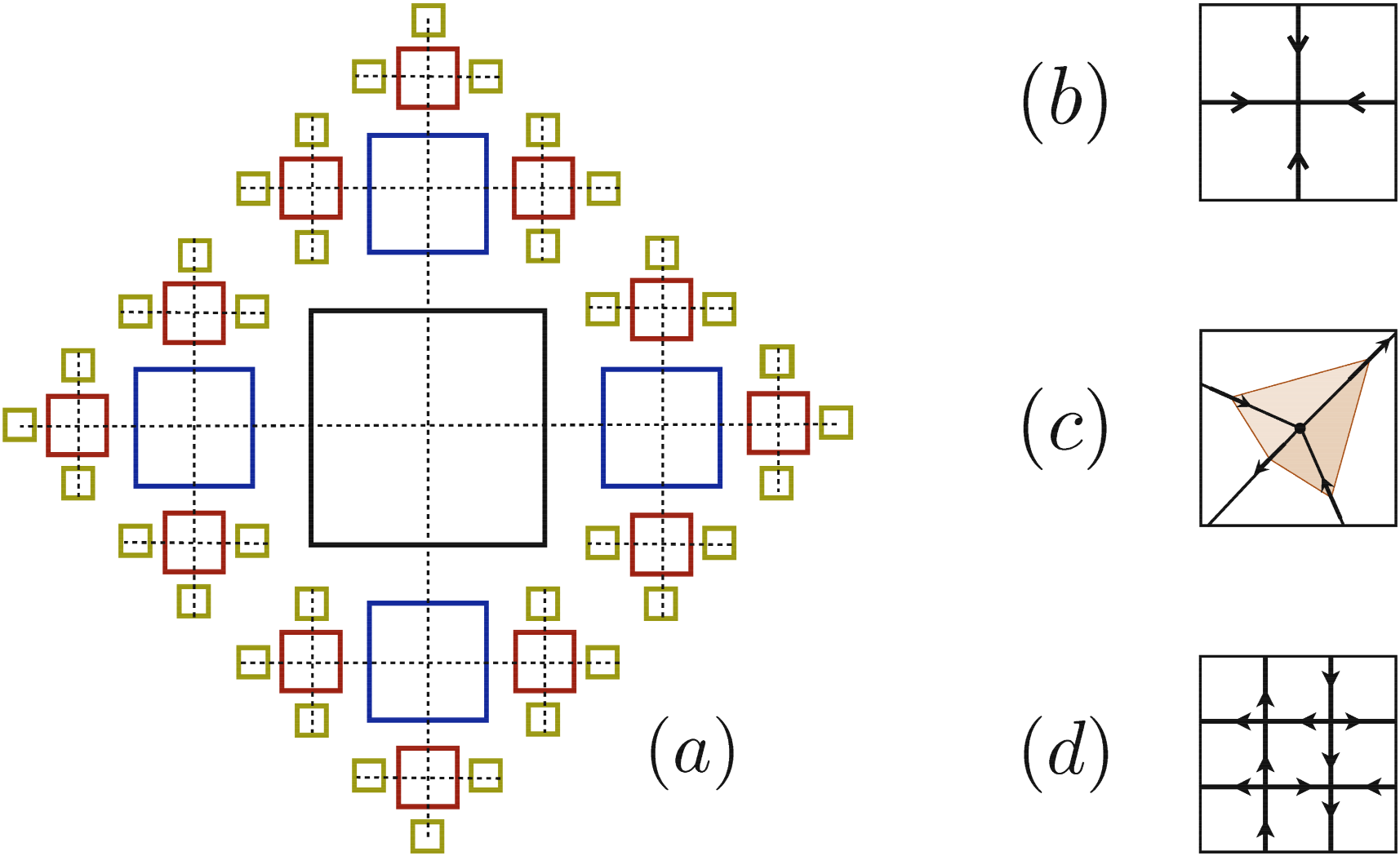}
\end{center}
\caption{\small (Color online.) The main panel  shows how to construct a Bethe lattice
of individual units that can be chosen at will. In the right panel we show
four different choices of such units. The image in (b) represents a single vertex that once
inserted in (a) builds the simplest Bethe lattice of vertices, described in Sec.~\ref{sec:tree-ver}. The tetrahedron in (c) 
is the individual unit used for the calculations of~\cite{Yoshida2004} and~\cite{Jaubert2008}, 
as discussed in Sec.~\ref{sec:discussionTree}. 
Finally panel (d) shows a plaquette of four vertices that is the individual unit of the plaquette model,
see Sec.~\ref{sec:tree-plaq}.}
\label{fig:tree_plaquette}
\end{figure}

As we will show in the following,  the results obtained by using the tree structure described above
compare extremely well to the ones in $2D$ in many respects. 
In order to further improve the approximation,  
in particular relatively to the nature of some transitions, 
 we introduce a Bethe lattice of  ``plaquettes'' (see Fig.~\ref{fig:tree_plaquette}),
where the basic unit cell is not a single vertex, but a {$2\times2$} square
of vertices. This tree is constructed by connecting each
plaquette to other four plaquettes (without forming loops of plaquettes), 
in the same way as we constructed the tree of 
single vertices. In this procedure one has to be careful 
with the orientation of 
 each plaquette and of the vertices on it, and with the order between the two 
 edges outgoing from each side of the unit. 

In the following we will refer to the first simpler geometry as the ``single vertex problem''
and to the second one as the ``plaquette model''.

\subsubsection{Discussion}\label{sec:discussionTree}

A mean-field approximation for the pyrochlore spin-ice system in $3D$, 
based on the same tree-like structure of single vertices described in Sec.~\ref{sec:tree-ver},  
has been already employed in~\cite{Yoshida2004} and~\cite{Jaubert2008}.
In these papers no distinction between the orientation (up, down, left, right) 
of the edges was made. This approach was apt to deal with the 
SL and FM phases~\cite{Jaubert2008} only. Conversely, our approach keeps track of the four different
directions and allows us to simultaneously study all possible phases, including the AF ones. 
Within our approach it is also easier to remove 
the degeneracy of the vertices by introducing  external magnetic fields or other kinds of perturbations. 

The single-vertex BP approximation proposed in~\cite{Yoshida2004,Jaubert2008}
provides a very good qualitative and quantitative description of the 
transition towards the frozen FM phase (KDP problem) in pyrochlore 
spin-ice in $3D$. This is due to the absence of fluctuations in the frozen FM phase.\footnote{This is also 
due to the fact that $D=3$ coincides with the upper critical dimension of the problem in the absence of defects
({\it i.e.}, the six-vertex model): 
the mean-field BP approximation is almost exact in $3D$ apart from logarithmic corrections.}
However, such a single-vertex approximation is not precise enough to describe the unfrozen
staggered AF order, due to thermal fluctuations.  
By constructing the tree of plaquettes we will manage to capture some of these fluctuations, and to obtain
a more accurate description of the unfrozen phases.

Let us finally mention that the ice-rule for the six-vertex model, or the ``parity" rule for
the eight-vertex might be viewed as a particular form of constraint that 
forces to flip all the variables on an entire loop in order to go from one allowed configuration
to another. Such form of constraints has in many respects a strong algorithmic impact,
in particular for algorithms that work with local updates. 
For this reason it has been investigated in the context of
combinatorial optimization, for a wide class of problems  mainly defined on tree-like graphs,  
with techniques similar to those adopted here for the tree of single vertices~\cite{ZM08}.

\subsection{The six- and eight-vertex models on a tree of vertices}
\label{Sec_6_and_8_vertex_single_vertex}

In this Subsection we study the six- and the eight-vertex models on a tree of single vertices.
We obtain the self-consistent recursive equations for the marginal probabilities along with their fixed points.
We perform the stability analysis of the solutions, compute the free-energy of the different phases, 
and present the resulting phase diagram.

\subsubsection{Recursion  equations}
 
We call $\psi^{i^\alpha\to j^\beta}_\zeta$, with  $\alpha,\beta \in \{u,d,l,r\}$, 
the probability that the root vertex  $i$ -- in a 
rooted tree where $\langle i^\alpha j^\beta\rangle$ is the missing edge -- be of type $\zeta \in \chi_v^8 = \{ v_1, \dots, v_{8} \}$. 
Such probabilities must satisfy the normalization condition 
\begin{equation}
\sum_{\zeta \in \chi_v^8}\psi^{i^\alpha\to j^\beta}_{\zeta} = 1
\ , \ \  \qquad \forall \ \langle i^\alpha j^\beta \rangle \ .
\end{equation}

In the recurrence procedure we will only be 
concerned with the state of the arrow on the missing edge.
As a consequence, on the root vertex $i$ of a rooted tree with a missing edge  $i^\alpha \to  j^\beta$,
we define $\psi^{\beta}_i \equiv\psi^{i^\alpha \to  j^\beta}(+1)$ 
as being the probability that the arrow that lies on the missing edge $\langle i^\alpha j^\beta\rangle$ 
takes the value $s_{\langle i^\alpha j^\beta\rangle} = +1$. 
Then,
\begin{align}
&\psi^{u}_i = \psi^{i^d \to  j^u}_{v_1} + \psi^{i^d \to  j^u}_{v_3} + \psi^{i^d \to  j^u}_{v_6} + \psi^{i^d \to  j^u}_{v_8} \ , \nonumber \\
&\psi^{d}_i = \psi^{i^u \to  j^d}_{v_1} + \psi^{i^u \to  j^d}_{v_3} + \psi^{i^u \to  j^d}_{v_5}+ \psi^{i^u \to  j^d}_{v_7} \ , \nonumber \\
&\psi^{l}_i = \psi^{i^r \to  j^l}_{v_1} + \psi^{i^r \to  j^l}_{v_4} + \psi^{i^r \to  j^l}_{v_6}+ \psi^{i^r \to  j^l}_{v_7}  \ , \label{eq:def_prob} \\
&\psi^{r}_i = \psi^{i^l \to  j^r}_{v_1} + \psi^{i^l \to  j^r}_{v_4} + \psi^{i^l \to  j^r}_{v_5} + \psi^{i^l \to  j^r}_{v_8} \ , \nonumber
\end{align}
and 
\begin{equation}
\psi^{i^\alpha \to  j^\beta}(-1) = 1 - \psi^{i^\alpha \to  j^\beta}(+1) = 1-\psi^\beta_{i} \ . \label{eq:psi}
\end{equation}
Clearly, other parameterizations are possible for these probabilities. 
For instance, one could use an effective field
acting on the spin $s_{\langle i^\alpha j^\beta\rangle}$, {\it i.e.} $\psi^{\beta}_i = e^{h^\beta_i}/(e^{h^\beta_i}+e^{- h^\beta_i})$,  
and the recursive equations would be equivalently written in terms of the
fields $h^\beta_i$, {as shown in Appendix~\ref{Appendix_Eq_cavity}}.

The operation of merging rooted trees allows us to obtain in a self-consistent fashion the set of probabilities 
associated to the new root in terms of those of the previous generation.
In the bulk of the tree one does not expect such quantities to depend on the precise site, 
since all expected phases are homogeneous. As a result, the  
explicit reference to the particular vertex index $i$ of the probabilities defined in eq.~(\ref{eq:def_prob}) 
can be dropped. By so doing, the following four coupled \emph{self-consistent equations} for the probabilities 
$\psi^{\alpha}$ are obtained:
\begin{equation}
 \begin{array}{ll}
& \psi^{u } 
  =  \hat{\Psi}^u[a, b, c, d, \psi^u , \psi^d , \psi^l , \psi^r ]=g^u(a, b, c, d, \psi^u , \psi^d , \psi^l , \psi^r )/z^u 
\\ \vspace{-0.2cm} \\
&  =    
\displaystyle{\frac{1}{z^{u }}} \Big[ a \psi^{l } \psi^{u } \psi^{r } +
b (1 - \psi^{l }) \psi^{u }(1-  \psi^{r }) 
+  c (1-\psi^{u })(1- \psi^{l }) \psi^{r }
+ d  \psi^{l }(1 - \psi^{u })(1 - \psi^{r })   \Big]
\\ \vspace{-0.2cm} \\
&  \psi^{l } 
 =  \hat{\Psi}^l[a, b, c, d, \psi^u , \psi^d , \psi^l , \psi^r ]  =g^l(a, b, c, d, \psi^u , \psi^d , \psi^l , \psi^r )/z^l 
\\ \vspace{-0.2cm} \\
&  =       \displaystyle \frac{1}{z^{l }} \Big[ a\psi^{d } \psi^{l }  \psi^{u } +
b(1-\psi^{d }) \psi^{l }(1 - \psi^{u } )
+  c \psi^{d }(1 - \psi^{l }) (1 - \psi^{u })  + d(1 - \psi^{d })(1 - \psi^{l })  \psi^{u } \Big]
\\ \vspace{-0.2cm} \\
&  \psi^{d } 
=  \hat{\Psi}^d[a, b, c, d, \psi^u , \psi^d , \psi^l , \psi^r ] =g^d(a, b, c, d, \psi^u , \psi^d , \psi^l , \psi^r )/z^d 
\\ \vspace{-0.2cm} \\
&  =     \displaystyle \frac{1}{z^{d }} \Big[ a\psi^{r } \psi^{d } \psi^{l }   +
b(1 - \psi^{r }) \psi^{d }(1 - \psi^{l })
+ c (1 - \psi^{r })(1 - \psi^{d }) \psi^{l }
+ d \psi^{r } (1 - \psi^{d })(1 - \psi^{l })   
 \Big]
\\ \vspace{-0.2cm} \\
& \psi^{r } =  \hat{\Psi}^r[a, b, c, d, \psi^u , \psi^d , \psi^l , \psi^r ]  =g^r(a, b, c, d, \psi^u , \psi^d , \psi^l , \psi^r )/z^r 
\\ \vspace{-0.2cm} \\
& 
    = \displaystyle \frac{1}{z^{r }} \Big[ a\psi^{u }  \psi^{r } \psi^{d }   +
b(1 - \psi^{u })  \psi^{r } (1 - \psi^{d })
+  c \psi^{u } (1 - \psi^{r } )(1 - \psi^{d } )  +  d (1 - \psi^{u })(1 - \psi^{r }) \psi^{d }
 \Big]
 \\
 &
 \vspace{-0.2cm}
    \end{array}
    \\
    \label{psiU_8vertex}
\end{equation}
where $a$, $b$, $c$, and $d$ are the fugacities of the vertices, as
defined in  Sec.~\ref{sec:vertex-models} (see
Figs.~\ref{fig:six-vertex} and~\ref{fig:eight-vertex}), 
and $z^\alpha$ are normalization constants that ensure the normalization of $\psi^{\alpha}$:
\begin{equation}
z_\alpha=g^\alpha(a, b, c, d, \psi^u , \psi^d , \psi^l , \psi^r )+g^\alpha(a, b, c, d,1- \psi^u ,1- \psi^d ,1- \psi^l ,1- \psi^r ) \ .
\end{equation}
The functions $g^\alpha$ have been defined in Eq.~(\ref{psiU_8vertex}).
The first term in this sum corresponds to the un-normalized contribution of a spin 
$(+1)$ while the second term is for a spin $(-1)$. For the sake
of simplicity in eq.~(\ref{psiU_8vertex}) we considered the argument in the
functions $\hat{\Psi}^{\alpha}$ to be the same for all the directions $\alpha$,
although in each equation the field defined along the opposite direction does not appear
 explicitly.

\subsubsection{Fixed points and free-energy}\label{FP_8vertex}

In order to allow for a fixed point solution associated to  $c$-AF and $d$-AF staggered order 
we study eqs.~(\ref{psiU_8vertex}) on a bipartite graph.
We partition the graph into two distinct subsets of vertices $A_1$ and $A_2$, such that each vertex belonging
to $A_1$ is only connected to vertices belonging to $A_2$ and {\it vice versa}.
This amounts to doubling the fields degrees of freedom $\{\psi^{\alpha}_1,\psi^{\alpha}_2\}$,
one for the sub-lattice $A_1$ and the other one for the sub-lattice $A_2$, and to 
{solving} the following set of coupled equations:
\begin{eqnarray}
\begin{array}{ll}
&\psi^{\alpha}_1 = \hat{\Psi}^{\alpha} [a, b, c, d, \psi^u_2 , \psi^d_2 , \psi^l_2 , \psi^r_2 ]
\; ,   \\
&\psi^{\alpha}_2 = \hat{\Psi}^{\alpha} [a, b, c, d, \psi^u_1 , \psi^d_1 , \psi^l_1 , \psi^r_1 ]
\ , 
\end{array}
\label{eq-cav}
\end{eqnarray}
with $\alpha=u,d,r, l$. The FM and PM phases are characterized by $\psi^{\alpha}_1 = \psi^{\alpha}_2$, while
the AF phases by $\psi^{\alpha}_1 \neq \psi^{\alpha}_2$ with $\psi^{\alpha}_1 = 1 - \psi^{\alpha}_2$. 

Considering  the solution associated to  $A_1$ only, the fixed points are
\begin{eqnarray}
\label{Fixed_points}
\begin{array}{ll}
& \qquad \boldsymbol{\psi}_{\rm PM} = (\psi^u=\frac12,\psi^l=\frac12,\psi^r=\frac12,\psi^d=\frac12) \ , 
\\ \vspace{-0.2cm} \\
& \qquad \boldsymbol{\psi}_{\text{$a$-FM}} = (\psi^u=1,\psi^l=1,\psi^r=1,\psi^d=1) \ , 
\\ \vspace{-0.2cm} \\
& \qquad \boldsymbol{\psi}_{\text{$b$-FM}} = (\psi^u=1,\psi^l=0,\psi^r=0,\psi^d=1) \ , 
\\ \vspace{-0.2cm} \\
& \qquad \boldsymbol{\psi}_{\text{$c$-AF}} = (\psi^u=1,\psi^l=1,\psi^r=0,\psi^d=0) \ , 
\\ \vspace{-0.2cm} \\
& \qquad \boldsymbol{\psi}_{\text{$d$-AF}} = (\psi^u=1,\psi^l=0,\psi^r=1,\psi^d=0) \ , 
\end{array}
\end{eqnarray}
as can be checked by inserting these values into eqs.~(\ref{psiU_8vertex}) and (\ref{eq-cav}).
The spin reversal symmetry 
implies that $\boldsymbol{\psi}' = 1 - \boldsymbol{\psi}$ is also a solution of the self-consistent equations. 
For the AF phase this corresponds to the excange of the two sublattices 
({\it i.e.}, the exchange of $\psi^{\alpha}_1 $ with $\psi^{\alpha}_2$).
These solutions exist for any value of the vertex weights $a, b, c$ and $d$.
Conversely, the stability of the solutions, that will be discussed in more detail in Sec.~\ref{Sec:Stabiliy_sv}, 
depends on the fugacities. The numerical iteration of the self-consistent equations, eqs.~(\ref{psiU_8vertex}) and 
(\ref{eq-cav}), confirms that the
fixed points given in eq.~(\ref{Fixed_points})
are the only possible attractive solutions in the different regions of the phase diagram.

In order to calculate the free-energy, 
it is useful to consider the contributions to the partition function 
coming from a vertex, an horizontal edge and a vertical edge. These quantities are defined as follows:
\beq\label{eq:Zv}
 \begin{array}{ll}
Z_{v}[\psi^l, \psi^r,\psi^u,\psi^d]  &  =  \;\; a~ \Big[ \psi^{l } \psi^{u } \psi^{r } \psi^{d} +
(1-\psi^{u })(1- \psi^{l })(1 - \psi^{r })(1 - \psi^{d }) \Big]  ~ 
\\ \vspace{-0.2cm} \\
&  \;\;\;\;\; + ~ b~ \Big[ (1 - \psi^{l }) \psi^{u }(1-  \psi^{r }) \psi^{d} +  \psi^{l }(1 - \psi^{u }) \psi^{r }(1 - \psi^{d}) \Big] ~
   \\ \vspace{-0.2cm} \\
&  \;\;\;\;\; + ~   c ~\Big[ (1-\psi^{u })(1- \psi^{l }) \psi^{r }\psi^{d } + \psi^{u } \psi^{l }(1 - \psi^{r })(1 - \psi^{d } )\Big] ~
\\ \vspace{-0.2cm} \\
&  \;\;\;\;\; + ~ d ~\big[ \psi^{l }(1 - \psi^{u })(1 - \psi^{r })\psi^{d} +(1- \psi^{l }) \psi^{u } \psi^{r }(1 - \psi^{d})  \Big]  ~ 
      \end{array}
\eeq
and 
\begin{eqnarray}
 \begin{array}{ll}
&
Z_{\langle l r \rangle}[\psi^l_i ,\psi^r_j] = \psi^{l}_i \psi^{r}_j + (1-\psi^{l}_i)(1 - \psi^{r}_j) \ ,
\\
&
Z_{\langle u d \rangle}[\psi^u_i ,\psi^d_j] = \psi^{u}_i \psi^{d}_j + (1-\psi^{u}_i)(1 - \psi^{d}_j) \ . 
\end{array}
\label{eq:Zlr}
\end{eqnarray}
The first term $Z_v$ represents the shift in the partition function brought in by the
introduction of a new vertex which is connected with four rooted trees. 
The other terms $Z_{\langle l r \rangle}$ and $Z_{\langle u d \rangle}$ 
represent the shift in the partition function induced by the connection of two rooted trees 
(respectively one left and one right or one up and one down) through a link.
In terms of these quantities one can compute the intensive free-energy 
(here and in the following we normalize $f$ by the number of vertices) 
which characterizes the bulk properties of the system in the thermodynamic limit:
\beq\label{free-energy-Cavity}
 \begin{array}{l}
\displaystyle \beta f[a,b,c,d,\boldsymbol{\psi}_1,\boldsymbol{\psi}_2] = - \frac12 \Big( \ln Z_{v}[\boldsymbol{\psi}_1] +  \ln Z_{v}[\boldsymbol{\psi}_2] ~+
 \\ \vspace{-0.2cm} \\
\hspace{0.8cm}\displaystyle - \ln Z_{\langle l r \rangle}[\psi^l_1 ,\psi^r_2]  -  \ln Z_{\langle l r \rangle}[\psi^l_2 ,\psi^r_1]
- \ln Z_{\langle u d \rangle}[\psi^u_1 ,\psi^d_2] - \ln Z_{\langle u d \rangle}[\psi^u_2 ,\psi^d_1] \Big) 
\ . 
       \end{array}
\eeq
{In the l.h.s. of the above equation and in the following expressions related to
the free energy, we introduce the inverse 
temperature $\beta$, which enters in the definition of the fugacities once
we parametrize them through the energies of the vertices $e_i$, {\it i.e.} $w_i = e^{- \beta e_i}$.} 
The free-energy (\ref{free-energy-Cavity}) 
evaluated at the fixed points (\ref{Fixed_points}) reads as follows:
\beq\label{free-energy-fixed-points-8vertex}
\begin{array}{lll}
& \qquad\qquad
\displaystyle \beta f_{\rm PM} = \beta  f[a,b,c,d,\boldsymbol{\psi}_{\rm PM}]  = - \ln\left(\frac{a+b+c+d}{2}\right) 
\ , 
 \\ \vspace{-0.2cm} \\
& \qquad\qquad
\beta f_{\text{$a$-FM}} = \beta f[a,b,c,d,\boldsymbol{\psi}_{\text{$a$-FM}}]  = - \ln a 
 \ , 
 \\ \vspace{-0.2cm} \\
& \qquad\qquad
\beta f_{\text{$b$-FM}} = \beta  f[a,b,c,d,\boldsymbol{\psi}_{\text{$b$-FM}}] = - \ln b 
\ , 
 \\ \vspace{-0.2cm} \\
& \qquad\qquad
\beta f_{\text{$c$-AF}} =  \beta f[a,b,c,d,\boldsymbol{\psi}_{\text{$c$-AF}}] = - \ln c 
\ , 
 \\ \vspace{-0.2cm} \\
& \qquad\qquad
\beta f_{\text{$d$-AF}} = \beta  f[a,b,c,d,\boldsymbol{\psi}_{\text{$d$-AF}}]  = - \ln d
\ . 
\end{array}
\eeq

By comparing these free-energy densities    
one readily determines the phase diagram along with the critical hyper-planes separating the different phases.
For instance, from the condition $f_{\rm PM} = f_{a-{\rm FM}}$ one finds $a_c= b+c+d$. 
Surprisingly enough, it turns out that the critical planes have exactly the same 
parameter dependence as in the six- and in the eight-vertex model on the square lattice. 
The BP approximation yields, therefore, the exact 
topology of the phase diagram.

\subsubsection{Order parameters}\label{Sec:8vertex-OrderParam}

According to the definitions in eq.~(\ref{eq:magn}),  
we characterize the phases by direct and staggered magnetizations.
In particular, we define the following order parameters, each one associated with a
particular ordered phase and all of them vanishing in the PM state:\footnote{This is a generalization 
of the definitions given in eq.~(\ref{eq:magn}), needed in order 
to make the difference between FM orders dominated by $a$ or $b$ vertices, as well as AF orders dominated by $c$ or $d$.
It corresponds to consider $m_{\text{$a$-FM}} = \frac12 (m_+^x + m_+^y)$, 
$m_{\text{$b$-FM}} = \frac12 (m_+^x - m_+^y)$, 
$m_{\text{$c$-FM}} = \frac12 (m_-^x - m_-^y)$ and $m_{\text{$d$-FM}} = \frac12 (m_-^x + m_-^y)$.}
\beq\label{Eq:Order_Parameters_sv}
  \begin{array}{l}
\displaystyle m_{\text{$a$-FM}} = \frac{a}{Z_{v}} [ \psi^{l } \psi^{u } \psi^{r } \psi^{d} -
(1-\psi^{u })(1- \psi^{l })(1 - \psi^{r })(1 - \psi^{d }) ]  \ ,

 \\ \vspace{-0.2cm} \\
 
 \displaystyle m_{\text{$b$-FM}} = \frac{b}{Z_{v}}  [ (1 - \psi^{l }) \psi^{u }(1-  \psi^{r }) \psi^{d} - 
  \psi^{l }(1 - \psi^{u }) \psi^{r }(1 - \psi^{d}) ] \ , 

\\ \vspace{-0.2cm} \\

\displaystyle m_{\text{$c$-AF}} =  \frac{c}{Z_{v}} [ \psi^{u } \psi^{l }(1 - \psi^{r })(1 - \psi^{d } )
-  (1-\psi^{u })(1- \psi^{l }) \psi^{r }\psi^{d }]  \ ,

\\ \vspace{-0.2cm} \\
 
\displaystyle m_{\text{$d$-AF}} = \frac{d}{Z_{v}} [ (1- \psi^{l }) \psi^{u } \psi^{r }(1 - \psi^{d}) -
 \psi^{l }(1 - \psi^{u })(1 - \psi^{r })\psi^{d} ]  \ , 

  \end{array}
\eeq 
where $Z_v$ is the contribution of one vertex to the partition function defined in eq.~(\ref{eq:Zv}).
The order parameters ~(\ref{Eq:Order_Parameters_sv}) are easily evaluated at the fixed
points, eq.~(\ref{Fixed_points}). The PM solution $\boldsymbol{\psi}_{PM}$ yields vanishing  
magnetizations for all the order parameters. Conversely, any ordered solution $\boldsymbol{\psi}_{{\rm fp}=\ast}$ gives   
a saturated value of the associated magnetization $m_{{\rm fp}=\ast}=1$ (and zero for the other 
components $m_{{\rm fp}\neq\ast}=0$), corresponding to a completely frozen ordered phase. 

\subsubsection{Stability analysis}\label{Sec:Stabiliy_sv}

The stability of the fixed points is determined by the eigenvalues of the Jacobian matrix, 
\begin{equation}\label{Stability_Matrix}
M\equiv\frac{d \hat{\Psi}^{\alpha}}{d \psi^{\beta}}\Big{|}_{\boldsymbol{\psi}_{\rm fp}}
\ , 
\end{equation}
that describes the derivative  
 of the vector function $\hat{\bf{\Psi}} = (\hat{\Psi}^{u},\hat{\Psi}^{r},\hat{\Psi}^{l},\hat{\Psi}^{d})$
defined in eq.~(\ref{psiU_8vertex}) with respect to the fields $\{\psi^{\alpha}\}$, 
evaluated at the fixed point, ${\boldsymbol{\psi}_{\rm fp}}$. The eigenvectors are parametrized as 
 $(\delta\psi^u,\delta\psi^r,\delta\psi^l,\delta\psi^d)$. The solution ${\boldsymbol{\psi}_{\rm fp}}$ 
 becomes 
 unstable as soon as one of the eigenvalues becomes one, $|E_{max}|=1$.

Indeed, one can easily prove  that the condition $|E_{max}|=1$ 
corresponds to the divergence of the magnetic susceptibility $\chi$, and allows us to
identify the different phase transitions. 
 {The susceptibility to an infinitesimal 
 homogeneous field $h$ that couples to the arrow-spin polarization reads:
\beq
  \begin{array}{l}
\displaystyle \chi = \left. \lim_{N\to\infty} \frac{1}{N}  
\frac{{\rm d} \sum_{\langle i j \rangle} \langle s_{\langle i j \rangle} \rangle}{{\rm d} h}  \right|_{h=0}
=  \lim_{N\to\infty} \frac{1}{N}  \sum_{\langle i j \rangle ,\langle k l \rangle} \langle s_{\langle i j \rangle}s_{\langle k l \rangle}\rangle_{c} 
\displaystyle \propto   \sum_{\alpha, \beta = \langle u d \rangle, \langle l r \rangle} 
\sum_{r=1}^{\infty} \sum_{{\mathcal P}(r)}   \langle s_0^{\alpha} s_r^{\beta} \rangle_c 
  \end{array}
\eeq
 where in the last equality we used the homogeneity of the solution in the thermodynamic limit
 $N\to\infty$, and here $N$ is the number of spins.}
  The symbol ${\mathcal P}(r)$ indicates that the sum runs over
 all the paths that connect a given spin on an edge of type $\alpha$, supposed to be the centre (site denoted by $0$)
 of the tree, to all the spins that live on edges of type $\beta$ and located at a distance $r$ from $0$
 (in terms of the number of edges that make the path).
 As the tree has no loops, such paths are uniquely defined.
 The above formula can be simplified by first using the fluctuation-dissipation relation
\beq
  \displaystyle \langle s_0^{\alpha} s_r^{\beta} \rangle_c  = \frac{{\rm d} \langle s_r^{\beta} \rangle}{{\rm d} h_0^{\alpha}} \ ,
 \eeq
 where $h_0^{\alpha}$ is a field conjugated to $s_0^{\alpha}$, 
 and next using the chain rule
\beq
\displaystyle \frac{{\rm d} \langle s_r^{\beta}\rangle}{{\rm d} h_0^{\alpha}} = \frac{{\rm d}  \hat{\Psi}^{\gamma_{1}}}{{\rm d} \psi^{\alpha}_{0}}
\prod_{i = 2}^{r-1} \frac{{\rm d}  \hat{\Psi}^{\gamma_{i}}}{{\rm d}  \psi^{\gamma_{i-1}}}
 \frac{{\rm d} \langle s_r^{\beta}\rangle}{{\rm d} \psi^{\gamma_{r-1}}}
 \; , 
\eeq
where the particular values taken by $\{\gamma_{i} \} \in \{u,d,l,r\}$ 
depend on the path followed. Each derivative is finally evaluated at
the fixed point. Then, 
defining the vectors $| v_{\alpha} \rangle$ such that $v_{\alpha}^{\beta} = \frac{{\rm d}  \hat{\Psi}^{\beta}}{{\rm d}\psi^{\alpha}}$,
and $| w_{\alpha} \rangle$ such that $w_{\alpha}^{\beta} =  \frac{{\rm d}  \hat{\Psi}^{\alpha}}{{\rm d}\psi^{\beta}}$,
one obtains
\beq
  \begin{array}{l}
\displaystyle \displaystyle \chi =  
 \sum_{\alpha, \beta = \langle u d \rangle, \langle l r \rangle} 
\sum_{r=1}^{\infty} \sum_{{\mathcal P}(r)} \;   \langle s_0^{\alpha} s_r^{\beta} \rangle_c 
 \\ \vspace{-0.2cm} \\
\hspace{0.3cm} \displaystyle = \sum_{\alpha, \beta = \langle u d \rangle, \langle l r \rangle} \;  \sum_{r=1}^{\infty} 
 \sum_{{\mathcal P}(r)} \; \ \frac{{\rm d}  \hat{\Psi}^{\gamma_{1}}}{{\rm d}\psi^{\alpha}_{0}}
\prod_{i = 2}^{r-1} \frac{{\rm d}  \hat{\Psi}^{\gamma_{i}}}{{\rm d}  \psi^{\gamma_{i-1}}}
 \frac{{\rm d} \langle s_r^{\beta}\rangle}{{\rm d} \psi^{\gamma_{r-1}}} 
 
 \\ \vspace{-0.2cm} \\
\hspace{0.3cm}  \displaystyle \propto \sum_{\alpha, \beta \; = \; \langle u d \rangle, \langle l r \rangle} \;  \sum_{r=1}^{\infty} 
 \;  \langle v_{\alpha} | M^{r-2} | w_{\beta} \rangle 


\simeq    \sum_{r=1}^{\infty} \mbox{Tr} M^r . 
  \end{array}
\eeq
As long as the absolute value of the eigenvalues remains smaller than one, {\it i.e.} $|E_{max}|<1$, 
the series converges yielding a finite  susceptibility. Otherwise, if $|E_{max}| \ge 1$ the series diverges
yielding an infinite susceptibility. 

Let us  focus on the stability of the PM
phase $\boldsymbol{\psi}_{\rm PM}$, {\it i.e.} $\psi^l = \psi^r = \psi^d = \psi^u = \frac12$. 
 The  matrix $M$  evaluated in the PM solution is
 \beq\label{matrix_Jac}
M_{\boldsymbol{\psi}_{\rm PM}}
= 
\left[
 \begin{matrix}
 \displaystyle \frac{a+b-c-d}{a+b+c+d}  &  \displaystyle \frac{a-b+c-d}{a+b+c+d}  & 
 \displaystyle  \frac{-a+b+c-d}{a+b+c+d}  & 0 \\ \\
  \displaystyle \frac{a-b+c-d}{a+b+c+d}  &  \displaystyle \frac{a+b-c-d}{a+b+c+d}  & 0 
& \displaystyle \frac{-a+b+c-d}{a+b+c+d}  \\ \\
\displaystyle \frac{-a+b+c-d}{a+b+c+d}  & 0 & \displaystyle \frac{a+b-c-d}{a+b+c+d}  &  \displaystyle \frac{ a-b+c-d}{a+b+c+d}  \\ \\
0 & \displaystyle \frac{-a+b+c-d}{a+b+c+d} & \displaystyle \frac{a-b+c-d}{a+b+c+d} & \displaystyle \frac{a+b-c-d}{a+b+c+d}
 \end{matrix}
\right] \\ \ .
\eeq
Its eigenvalues are
\beq\label{Eig_stab_sv}
  \begin{array}{l}
  \displaystyle     E^{\rm PM}_1 =\frac{3 a-b - c- d}{a+b+c+d} \ , 
                  \qquad\qquad
 \displaystyle    E^{\rm PM}_2 =\frac{- a + 3b - c - d}{a+b+c+d} \ , 
                          \\ \vspace{-0.2cm} \\
 \displaystyle E^{\rm PM}_3 =\frac{a+b- 3 c+ d}{a+b+c+d} \ , 
          \qquad\qquad
 \displaystyle  E^{\rm PM}_4 = \frac{a+b+c-3 d}{a+b+c+d} \ . 
  \end{array}
\eeq
In the order $(\delta\psi^u,\delta\psi^r,\delta\psi^l,\delta\psi^d)$  the corresponding eigenvectors can be written:
${\rm v}_1=(1,1,1,1)$, ${\rm v}_2=(1,-1,-1,1)$, ${\rm v}_3=(-1,1,-1,1)$ and ${\rm v}_4=(-1,-1,1,1)$.
In general, the eigenvalue $E^{\rm PM}_1$ regulates the instability towards the $a$-FM, $E^{\rm PM}_2$ towards the $b$-FM, 
$E^{\rm PM}_3$ towards the $c$-AF, and $E^{\rm PM}_4$  towards the $d$-AF. 

One reckons that the stability of PM solution can be stated in terms of the condition 
\beq
\left| \frac{ (1 + E^{\rm PM}_3)(1 + E^{\rm PM}_4) - 
(1 - E^{\rm PM}_1)(1 - E^{\rm PM}_2)}{(1 + E^{\rm PM}_3)(1 + E^{\rm PM}_4) + (1 - E^{\rm PM}_1)(1 - E^{\rm PM}_2) } \right|<1, \
\eeq
which in terms of $a,b,c,d$ reads 
\begin{equation}
\left| \frac{a^2+b^2-c^2-d^2}{2(ab+cd)}\right|=|\Delta_8|<1 \ .
\end{equation}
Therefore, the stability analysis of the PM phase also yields the 
same condition for the phase boundaries as in the exact solution 
of the $2D$ model on the square lattice [see eq.~(\ref{eq:Lambda8}) and Fig.~\ref{fig:phase-diagram-8vertex}]. 

A similar analysis can be carried out to evaluate the stability of the FM and AF phases.
One has to evaluate the matrix~(\ref{Stability_Matrix}) at the ordered solutions' fixed points.
Considering the fixed point $\boldsymbol{\psi}_{\text{$a$-FM}}$ as an example,
one obtains the four eigenvalues: 
\beq\label{Eig_stab_sv_FM}
  \begin{array}{l}
  \displaystyle     E^{\text{$a$-FM}}_1 = \frac{b+c+d}{a} \ , 
                  \qquad\qquad
 \displaystyle    E^{\text{$a$-FM}}_2 = \frac{b-c-d}{a} \ , 
                          \\ \vspace{-0.2cm} \\
 \displaystyle E^{\text{$a$-FM}}_3 = \frac{b-c+d}{a} \ , 
          \qquad\qquad
 \displaystyle  E^{\text{$a$-FM}}_4 = \frac{b+c-d}{a} \ , 
  \end{array}
\eeq
and the corresponding eigenvectors are the same (in the same order) 
that we have already discussed for the PM solution.
From eq.~(\ref{Eig_stab_sv_FM}), one sees that $E^{\text{$a$-FM}}_1 = 1$ as soon
as $a=b+c+d$, {\it i.e.} when $\Delta_8=1$, while for larger values of $a$
the $a$-FM fixed point is stable. Moreover, the solution may develop
multiple instabilities when some vertices are missing. Analogous results hold 
for the other ordered phases.

Finally, let us remark that the model also has discontinuous transition points
separating  ordered phases, for instance the $a$-FM and the $c$-AF at $a=c$ and $b=d=0$, 
the $b$-FM and the $c$-AF at $b=c$ and $a=d=0$,
the $c$-AF and the $d$-AF at $c=d$ and $a=b=0$, etc.

\subsubsection{The six-vertex model: phase diagram and discussion}

We found four different fixed points in the six-vertex model ($d=0$).
These characterize
the (i) $a$-FM phase, (ii) $b$-FM phase, (iii) $c$-AF phase and (iv) 
PM phase (we will distinguish between the 
PM phase found on the tree and the SL phase in $D=2$ in ways that we will describe below). 

The transition lines were found in two equivalent ways. 
On the one hand, we compared the free energies of the PM and ordered
phases as given in eqs.~(\ref{free-energy-Cavity}). On the other hand, 
we analyzed when one of the eigenvalues of the stability matrix becomes equal to $1$, signaling the instability of the 
considered phase. We found that, as in the exact solution of the model in $2D$, the phase transitions are  
controlled by the anisotropy parameter $\Delta_6$ defined in eq.~(\ref{delta6}), with the 
transition lines determined by $|\Delta_6|=1$.  Interestingly enough, the eigenvalue 
$E_4^{\rm PM}$ equals 1 when $d=0$, $\forall \ a, \ b, \ c$, meaning that the whole PM phase is critical in the sense that the
magnetic susceptibility diverges for $d=0$.   
This is reminiscent of the critical properties of the SL (or Coulomb) phase in $D=2$.
Analogously, if $a=0$ then $E_1^{\rm PM}=-1$ $\forall\  c, \ b, \ d$; if $b=0$ then 
$E_2^{\rm PM}=-1$ $\forall \ a, \ c, \ d$
and if $c=0$ then $E_3^{\rm PM}=1$ $\forall\  a, \ b, \ d$. These observations suggest that in fact the  PM phase 
becomes critical whenever one of the four kinds of vertex is absent.

The transitions between the PM and the ordered phases are all discontinuous (as it can be seen from the singular behavior
of the free-energy):  
the (possibly staggered) magnetization jumps from zero to one. The fact that the magnetization 
saturates at its maximum value in the whole ordered phases indicates that in the FM and AF phases the order is 
prefect and frozen.

Even though the transitions are discontinuous, they are characterized by the absence of metastability and hysteresis, and 
they are associated to a diverging susceptibility. In fact, approaching the transition from the two sides,
both the PM and the FM solutions become unstable. 
This kind of transition corresponds to a  multi-critical point
of infinite order and it is called KDP transition in the literature~\cite{Benguigui77}.
Note that if one focuses on the transition line, and plugs the critical value $a_c=b+c$ 
into eqs.~(\ref{psiU_8vertex}) and (\ref{eq-cav}), assuming the homogeneity of the solution, $\psi^{\alpha}=\psi$ 
$\forall \alpha$, the equations take the trivial form $\psi=\psi$, meaning that all
values of $\psi$ are solutions of the self-consistent equations and all values of magnetization are equally
probable. Actually, the free-energy for the same 
critical value of $a_c$ does not depend on $\psi$ and it is thus the same for
all values of the magnetization.
The PM-AF transition shares the same properties of the KDP 
transitions between the FM and the PM phases. 

{From Eq.~(\ref{free-energy-fixed-points-8vertex}) one sees that 
the entropy per vertex $S = \beta^2 \frac{\partial^2 f}{\partial \beta^2}$
in the spin-ice point, 
$a=b=c=1$ is $S^{sv} = \beta f_{\rm PM} =  \ln 3/2 \simeq 0.405$, {\it i.e.} 
the Pauling result for water ice~\cite{Pauling1935} 
(see Fig.~\ref{fig:free-energy-6vertex}). The superscript $sv$ indicates that such 
quantity is derived within the Bethe ``single vertex" approximation, to distinguish
it from the results that will be obtained in the following with the tree of plaquettes.}

\subsubsection{The eight-vertex model: phase diagram and discussion}

The addition of $d$ vertices  does not change the properties of the
$a$-FM, $b$-FM and $c$-AF phases. As vertices of kind $d$ are allowed, 
another AF phase, denoted $d$-AF, emerges, corresponding to a phase with staggered
order of four-in and four-out vertices. The transition from the PM to the $d$-AF phase is also discontinuous and 
the $d$-AF phase is completely frozen as well. Moreover, as for the six-vertex model both the FM and the PM solutions 
become unstable on the transition planes.
When all the fugacities are finite  the 
four eigenvalues of the stability matrix within the PM phase 
become all smaller than one (except, of course, on the critical transition planes). 
This means that the PM phase of the eight-vertex model is no longer critical. 
The transition lines are altogether characterized by the  condition
$|\Delta_8| = 1$, as for the exact solution of the $2D$ model, with $\Delta_8$ given in eq.~(\ref{eq:Lambda8}).

\subsubsection{Summary}

In short, the location of the transition lines of the $2D$ six- and eight-vertex models are reproduced 
exactly by the calculations on the tree of single vertices. 
However, the critical properties  on the tree  are different from the ones of the $2D$ model. The 
absence of loops `freezes' the ordered phases and makes all transitions discontinuous, with 
the (possibly staggered) magnetization equal to one in the whole ordered phases. While this is true for the PM-FM transition
of the six-vertex model in $2D$, this is no longer true neither for the PM-FM transition of the {eight-vertex model}, nor
for the PM-AF transitions of the six- and eight-vertex models (in particular, in the former case the PM-AF transition
is of KT type).

\subsection{The six- and eight-vertex models on a tree of plaquettes}
\label{Sec:Bethe_plaquette}

In this Subsection we study the six- and eight-vertex models  on a tree of plaquettes.
The calculations  proceed along the
same lines as for the single vertex tree. They become, though,
 more involved, since the number of configurations allowed
on a plaquette is quite large.

\subsubsection{Recursion equations}

\begin{figure}[ht]
\centering
 \includegraphics[scale=1.2]{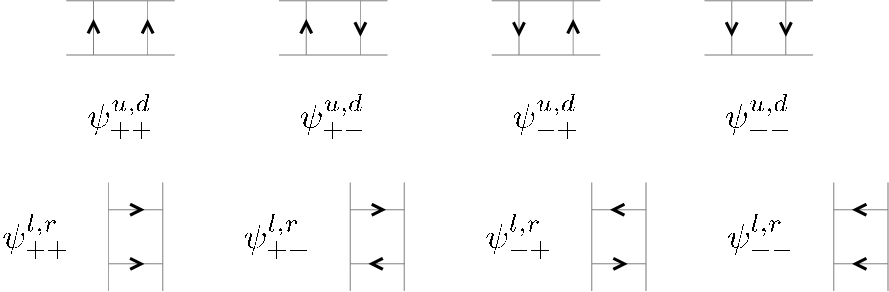}
\vspace{0.1cm}
\caption{\small Definitions of $\{\psi_{++}^{\alpha},\psi_{-+}^{\alpha},\psi_{+-}^{\alpha},\psi_{--}^{\alpha}\}_{\alpha = u,l,r,d}$ used 
in the recursion equations for the plaquette model. First line: $\psi^{u,d}$  where the first index $\pm$ denotes the value of the 
spin on the left and the second index denotes the spin on the right. Second line: $\psi^{l,d}$  where the first index $\pm$ denotes 
the value of the spin on the top and the second index denotes the spin on the bottom.
}    
    \label{Fig:psi_plaquette}
\end{figure}

Each rooted tree of plaquettes has two missing edges, which means that one has 
to write appropriate self-consistent equations for the joint probability 
of the two arrows lying on those edges. 
The analogue of $\psi^{\alpha}$,  which in the previous section
described the marginal probability of the arrow to point up (or right, depending
on whether we are considering a vertical or an horizontal edge), now becomes a probability vector with four
components. The marginal probability to find a pair of arrows with values $++$, $-+$, $+-$, $--$
will be denoted
by ${\boldsymbol \psi}^{\alpha} = \{\psi^{\alpha}_{++}, \psi^{\alpha}_{-+}, 
\psi^{\alpha}_{+-}, \psi^{\alpha}_{--} \}$.\footnote{In the case of the tree of simple vertices the
bold symbol referred to the vector associated to the four directions $\alpha$, while
here it is signaling that already for one direction one has to deal with multiple probabilities.} 
The superscript $\alpha = u, d, l, r$ denotes, just as before, whether the pair of arrows are on the 
missing edges of an up, down, left, right rooted tree.   
This is illustrated in Fig.~\ref{Fig:psi_plaquette}. We keep the convention on the choice of the signs
of the spins; namely, the spins take positive values if the arrows point up or right. 
Moreover, we assume that the first bottom index indicates the state of the arrow that is on
the left, for the vertical edges, and on the top, for the horizontal ones. Consequently, the
second bottom index refers to the value taken by the spin sitting on the right (respectively the bottom)  
edge for vertical (respectively horizontal) edges.

\begin{figure}[h]
\begin{center}
\includegraphics[scale=1]{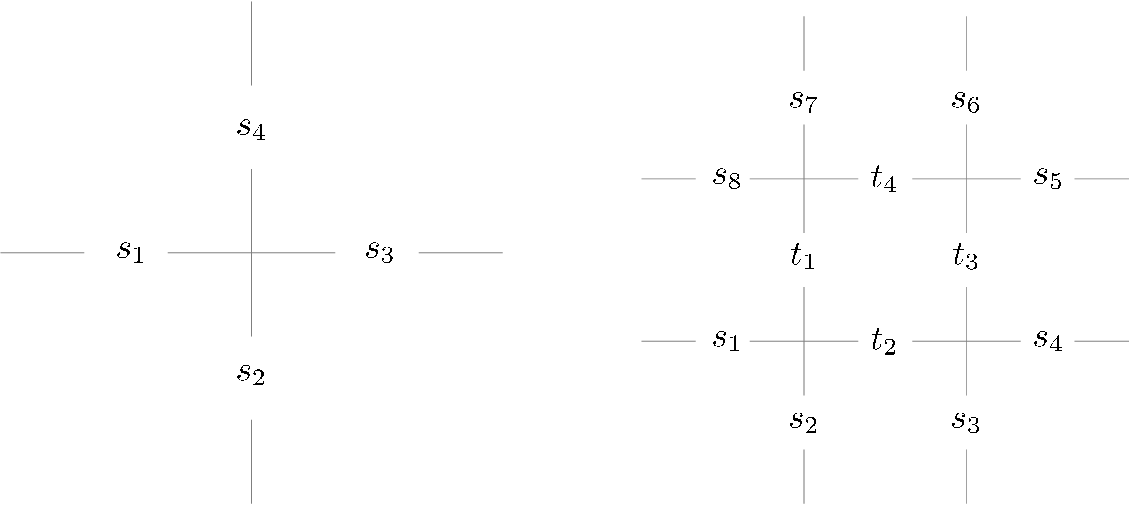}
\end{center}
\caption{\small 
Left panel: representation of the variables $\{s_i\}$ used in the definition
of the vertex weight in eq.~(\ref{vertex_weight_8vertex}). Right
panel: numbering assigned to the spin/arrow variables in eqs.~(\ref{Eq_psi_plaquette})
and~(\ref{Eq_free_energy_plaquette}). We denote by $\boldsymbol{S_P}=\{ s_1,s_2,s_3,s_4,s_5,s_6,s_7,s_8,t_1,t_2,t_4\}$ 
the set of spin variables on a plaquette.}
\label{Fig:vertex_weight}
\end{figure}

The weights of the vertex can be written as follows
\beq
w_{s_1,s_2,s_3,s_4}(a,b,c,d) = \frac{1}{4} \Big[ a' (1+s_1 s_2 s_3 s_4) +  b' (s_1 s_3 +  s_2 s_4)  + 
 c' (s_1 s_4 + s_2 s_3) + d' (s_1 s_2 + s_3 s_4)  \Big] 
 \\
 \label{vertex_weight_8vertex}
\eeq
where $s_1,\dots,s_4$ are taken as in Fig.~\ref{Fig:vertex_weight}, and
\beq\label{change_weights}
  \begin{array}{ll}
 \displaystyle  a' = \frac{1}{2} (a+b+c+d) \ , \hspace{1.5cm}
 \displaystyle  b' = \frac{1}{2} (a+b-c-d) \ , 
\\ \vspace{-0.2cm} \\
 \displaystyle  c' = \frac{1}{2} (a-b+c-d)  \ , \hspace{1.5cm}
 \displaystyle  d' = \frac{1}{2} (a-b-c+d) \ .
  \end{array}
\eeq

Similarly, we introduce a new parameterization of the probability vector,  
\beq
  \begin{array}{ll}
\displaystyle \psi^{\alpha}_{s_i s_j} & = 
\displaystyle \frac12  \Big[ (1 + s_i s_j) \frac{\psi_{++}^{\alpha}+ \psi_{--}^{\alpha}}{2} + (s_i + s_j)  \frac{\psi_{++}^{\alpha}-\psi_{--}^{\alpha}}{2} +

\\ \vspace{-0.2cm} \\

 & \;\;\; \displaystyle \qquad + (1 - s_i s_j) \frac{\psi_{+-}^{\alpha}+\psi_{-+}^{\alpha}}{2} + (s_i - s_j)  \frac{\psi_{+-}^{\alpha}-\psi_{-+}^{\alpha}}{2} \Big]

\\ \vspace{-0.2cm} \\

& = \displaystyle \frac14  \Big[ 1 + s_i s_j  \, s^{\alpha} + (s_i + s_j) \, p^{\alpha} + (s_i - s_j)  \, q^{\alpha} \Big]
\ , 
  \end{array}
\eeq
where we introduced a new set of variables defined as 
\begin{equation}
{\boldsymbol \phi}^{\alpha} = ( p^{\alpha}, s^{\alpha}, q^{\alpha}) 
\equiv ( \psi_{++}^{\alpha}- \psi_{--}^{\alpha} ,\psi_{++}^{\alpha}+\psi_{--}^{\alpha}- \psi_{-+}^{\alpha}-\psi_{+-}^{\alpha}, \psi_{+-}^{\alpha}-\psi_{-+}^{\alpha})\ .
\end{equation} 
We exploited the fact that, 
due to the normalization conditions, only three variables are truly independent for each edge direction. 
In the PM phase there is no symmetry breaking and $p^{\alpha}= q^{\alpha}=0$.
The ordered phases and the corresponding phase transitions are  
characterized by the spontaneous symmetry breaking associated to FM, $p^{\alpha}\neq0$, or
AF, $q^{\alpha}\neq 0$, order respectively.

For the sake of completeness we also report the inverse mapping:
\beq\label{change_var}
  \begin{array}{ll}
\displaystyle \psi_{++}^{\alpha}=\frac14 (1 + s^{\alpha} +  2 p^{\alpha})  \ , \qquad\qquad
\displaystyle \psi_{+-}^{\alpha} = \frac14  (1 - s^{\alpha} + 2 q^{\alpha})  \ , \\ \vspace{-0.2cm} \\
\displaystyle  \psi_{-+}^{\alpha} = \frac14  (1 -  s^{\alpha} - 2 q^{\alpha})  \ , 
\qquad\qquad
\displaystyle \psi_{--}^\alpha =\frac14  (1 + s^{\alpha} - 2 p^{\alpha}) \ . 
 \end{array}
\eeq
{Moreover we define $\boldsymbol{S_P}$ the set of 
spins lying on the $2\times2$ plaquette of four vertices, as shown in Fig.~\ref{Fig:vertex_weight}.}
The self-consistent equations for the probability vector describing up, left, right and down rooted trees
now read
\beq\label{Eq_psi_plaquette}
  \begin{array}{ll}
 \displaystyle \psi^{u}_{s_2 s_3} & = \hat{\Psi}_{s_2 s_3}^{u}[ a,b,c,d, {\boldsymbol \psi}^{u},{\boldsymbol \psi}^{l},{\boldsymbol \psi}^{r},{\boldsymbol \psi}^{d}]  
 
 \\  \vspace{0.2cm} \\
 
&  \displaystyle  = {\frac{1}{z^u}} \sum_{\boldsymbol{S_P}\backslash \{s_2,s_3\} } 
 w_{s_1,s_2,t_2,t_1} w_{t_2,s_3,s_4,t_3} w_{t_4,t_3,s_5,s_6} w_{s_8,t_1,t_4,s_7} \psi^{l}_{s_8 s_1} \psi^{u}_{s_7 s_6} \psi^{r}_{s_5 s_4}
 \ , 
 
 \\  \vspace{0.2cm} \\
 
 \displaystyle \psi^{l}_{s_5 s_4} & = \hat{\Psi}_{s_5 s_4}^{l}[ a,b,c,d, {\boldsymbol \psi}^{u},{\boldsymbol \psi}^{l},{\boldsymbol \psi}^{r},{\boldsymbol \psi}^{d}]  
 
 \\ \vspace{-0.2cm} \\
 
 &  \displaystyle = {\frac{1}{z^l}} \sum_{\boldsymbol{S_P}\backslash \{s_5,s_4\} } 
  w_{s_1,s_2,t_2,t_1} w_{t_2,s_3,s_4,t_3} w_{t_4,t_3,s_5,s_6} w_{s_8,t_1,t_4,s_7} \psi^{d}_{s_2 s_3} \psi^{l}_{s_8 s_1} \psi^{u}_{s_7 s_6}
 \ , 
 \\  \vspace{0.2cm} \\
 
 \displaystyle \psi^{d}_{s_7 s_6} & = \hat{\Psi}_{s_7 s_6}^{d}[ a,b,c,d, {\boldsymbol \psi}^{u},{\boldsymbol \psi}^{l},{\boldsymbol \psi}^{r},{\boldsymbol \psi}^{d}]  
 
 \\ \vspace{-0.2cm} \\
 
 &  \displaystyle = {\frac{1}{z^d}}  \sum_{\boldsymbol{S_P}\backslash \{s_6,s_7\} } 
  w_{s_1,s_2,t_2,t_1} w_{t_2,s_3,s_4,t_3} w_{t_4,t_3,s_5,s_6} w_{s_8,t_1,t_4,s_7} \psi^{l}_{s_8 s_1} \psi^{d}_{s_2 s_3} \psi^{r}_{s_5 s_4}
\ , 
\\  \vspace{-0.2cm} \\
 
 \displaystyle \psi^{r}_{s_8 s_1} & = \hat{\Psi}_{s_8 s_1}^{r}[ a,b,c,d, {\boldsymbol \psi}^{u},{\boldsymbol \psi}^{l},{\boldsymbol \psi}^{r},{\boldsymbol \psi}^{d}]  
 
 \\ \vspace{-0.2cm} \\
 
 &  \displaystyle = {\frac{1}{z^r}} \sum_{\boldsymbol{S_P}\backslash \{s_8,s_1\} }  
  w_{s_1,s_2,t_2,t_1} w_{t_2,s_3,s_4,t_3} w_{t_4,t_3,s_5,s_6} w_{s_8,t_1,t_4,s_7} \psi^{u}_{s_7 s_6} \psi^{r}_{s_5 s_4}
 \psi^{d}_{s_2 s_3}
 \ ,
  \end{array}
\eeq 
and the normalization constants are given by 
$z^{\alpha} = \sum_{s_i, s_j} \psi_{s_i s_j}^{\alpha}$.

It is more convenient to work with the variables 
${\boldsymbol \phi}^{\alpha} = (\phi_1^{\alpha}, \phi_2^{\alpha}, \phi_3^{\alpha}) = (p^{\alpha}, s^{\alpha}, q^{\alpha})$ 
for which one can readily derive the set of self-consistent equations from eqs.~(\ref{Eq_psi_plaquette}),
\begin{eqnarray}
  \begin{array}{ll}
p^{\alpha} = \hat{\Phi}^{\alpha}_1 [ a,b,c,d, {\boldsymbol \phi}^{u},{\boldsymbol \phi}^{l},{\boldsymbol \phi}^{r},{\boldsymbol \phi}^{d}]  
\equiv 
\hat{\Psi}_{++}^{\alpha} -  \hat{\Psi}_{--}^{\alpha} 
\ , 
\label{Eq_psq_plaquette-p}
\\ \vspace{-0.2cm} \\
s^{\alpha} = \hat{\Phi}^{\alpha}_2 [ a,b,c,d, {\boldsymbol \phi}^{u},{\boldsymbol \phi}^{l},{\boldsymbol \phi}^{r},{\boldsymbol \phi}^{d}]  
\equiv 
\hat{\Psi}_{++}^{\alpha} +  \hat{\Psi}_{--}^{\alpha} - \hat{\Psi}_{-+}^{\alpha} -  \hat{\Psi}_{+-}^{\alpha}
\ , 
\label{Eq_psq_plaquette-s}
\\ \vspace{-0.2cm} \\
q^{\alpha} = \hat{\Phi}^{\alpha}_3 [ a,b,c,d, {\boldsymbol \phi}^{u},{\boldsymbol \phi}^{l},{\boldsymbol \phi}^{r},{\boldsymbol \phi}^{d}]  
\equiv 
\hat{\Psi}_{-+}^{\alpha} -  \hat{\Psi}_{+-}^{\alpha} \ ,
\label{Eq_psq_plaquette-q}
  \end{array}
\end{eqnarray}
with $\alpha=u,l,r,d$. The argument of the functions $\hat{\Psi}_{\pm \pm}^{\alpha}$ of the right hand side 
can be expressed in terms of the variables $p^{\alpha}$, $s^{\alpha}$, and $q^{\alpha}$ 
using the transformations~(\ref{change_var}). 

\subsubsection{Free-energy}

The free-energy per vertex can be generically written as:
\beq\label{Eq_free_energy_plaquette}
  \begin{array}{l}
 \displaystyle \beta f[a,b,c,d,{\boldsymbol \psi}^{u},{\boldsymbol \psi}^{l},{\boldsymbol \psi}^{r},{\boldsymbol \psi}^{d}] =
  \frac14 \Big( \ln   \sum_{\substack{s_1,s_2}} \psi^{d}_{s_1 s_2} \psi^{u}_{s_1 s_2}  +
  \ln  \sum_{\substack{s_1,s_2}} \psi^{l}_{s_1 s_2} \psi^{r}_{s_1 s_2} - \ln Z_{pl}  \Big)
  \end{array}
\eeq
with the plaquette partition function given by
\beq\label{Eq_Zpl}
Z_{pl} = \sum_{\boldsymbol{S_P}} 
 w_{s_1,s_2,t_2,t_1} w_{t_2,s_3,s_4,t_3} w_{t_4,t_3,s_5,s_6} w_{s_8,t_1,t_4,s_7} \psi^{l}_{s_8 s_1} \psi^{u}_{s_7 s_6} \psi^{r}_{s_5 s_4}  \psi^{d}_{s_2 s_3} \ ,
\eeq
where one can rewrite  ${\boldsymbol \psi}^\alpha$ in terms in 
${\boldsymbol \phi}^\alpha$ via eq.~(\ref{change_var}).

\subsubsection{Order parameters}

The definitions of the order parameters (\ref{Eq:Order_Parameters_sv}) given in Sec.~\ref{Sec:8vertex-OrderParam}
have to be generalized to take into account the $\boldsymbol{S_P}$ variables. 
The order parameter associated to the $a$-FM transition reads:
\beq\label{Eq_Magn_energy_plaquette}
  \begin{array}{l}
 \displaystyle m_{\text{$a$-FM}}[a,b,c,d,{\boldsymbol \psi}^{u},{\boldsymbol \psi}^{l},{\boldsymbol \psi}^{r},{\boldsymbol \psi}^{d}] =

 \\\vspace{-0.2cm} \\
 
  \qquad\quad    \displaystyle   \frac{1}{Z_{pl}} \Big(   \sum_{\boldsymbol{S_P}} 
\mathcal{O}_{\text{$a$-FM}}[\boldsymbol{S_P}] w_{s_1,s_2,t_2,t_1} w_{t_2,s_3,s_4,t_3} w_{t_4,t_3,s_5,s_6} w_{s_8,t_1,t_4,s_7} \psi^{l}_{s_8 s_1} \psi^{u}_{s_7 s_6} \psi^{r}_{s_5 s_4}  \psi^{d}_{s_2 s_3}   \Big)
 
  \end{array}
\eeq
with $\mathcal{O}_{\text{$a$-FM}}[\boldsymbol{S_P}] = \frac18 \Big( \sum_{i = 1,\dots,8} \frac{s_i}{2}+ \sum_{i = 1,\dots,4} t_i \Big)$ and
$Z_{pl}$ defined in eq.~(\ref{Eq_Zpl}).
The other order parameters are obtained from eq.~(\ref{Eq_Magn_energy_plaquette})  by substituting 
the quantity $\mathcal{O}_{\text{$a$-FM}}$ with other staggered magnetizations suited to 
capture the different ordered phases.
More explicitly, they are
\beq\label{Eq_Observable_plaquette}
  \begin{array}{l}
\displaystyle \mathcal{O}_{\text{$b$-FM}}[\boldsymbol{S_P}]  = 
\frac18 \, \Big[ \, \frac{1}{2}\, ( - s_1 + s_2 + s_3 - s_4 - s_5 + s_6 + s_7 - s_8 ) + t_1 - t_2 + t_3 - t_4 \Big]
\; , 
 \\\vspace{-0.2cm} \\
\displaystyle  \mathcal{O}_{\text{$c$-FM}}[\boldsymbol{S_P}]  = 
\frac18 \, \Big[ \, \frac{1}{2} \, ( - s_1 + s_2 - s_3 - s_4 + s_5 -  s_6 + s_7 + s_8 ) - t_1 + t_2 + t_3 - t_4 \Big]
\; , 
 \\\vspace{-0.2cm} \\
\displaystyle  \mathcal{O}_{\text{$d$-FM}}[\boldsymbol{S_P}]  = 
\frac18 \, \Big[ \,\frac{1}{2} \, ( s_1 + s_2 - s_3 + s_4 - s_5 -  s_6 + s_7 - s_8 ) - t_1 - t_2 + t_3 + t_4 \Big] \ .
  \end{array}
\eeq

\subsubsection{Stability analysis} 

Similarly to what done for the model on the tree of vertices, we investigate the stability of the solutions by studying the stability matrix
\beq\label{Matrix_stab_plaq}
M^{\alpha,\beta}_{i,j} = 
\left.
\frac{{\rm d} \hat{\Phi}^{\alpha}_i}{{\rm d} \phi^{\beta}_j} 
\right|_{\hat{\Phi}_{\bf fp}} 
\ , \qquad
\hspace{0.5cm} \alpha,\beta=u,l,d,r \ , \qquad \hspace{0.5cm} i,j = 1, 2, 3 \ .
\eeq
In general, this is a  12$\times$12 matrix but, depending on which
solution one is studying, it might break into disjoint blocks. 
Moreover, the entries of such matrix have certain symmetry properties
that simplify the calculation.
From the stability analysis it turns out that the values of the fugacities where the PM solution becomes unstable 
also coincide with the transition planes between the different phases, corresponding to the points where the free-energies of 
the different phases cross.
We adopt the following parametrization of the 12 components of the eigenvectors
\begin{equation}
\delta \varphi = 
(\delta p^u,\delta s^u,\delta q^u;\delta p^d,\delta s^d,\delta q^d;\delta p^l,\delta s^l,\delta q^l;\delta p^r,\delta s^r,\delta q^r) \ .
\end{equation}

\subsubsection{Fixed points}

Let us discuss each fixed-point solution -- phase -- separately.

\vspace{0.25cm}
\noindent
{\bf The paramagnetic phase.}
The PM  fixed point is of the form ${\boldsymbol \phi}_{\rm PM} \equiv {\boldsymbol \phi}^u_{\rm PM} = {\boldsymbol \phi}^l_{\rm PM}
={\boldsymbol \phi}^r_{\rm PM} = {\boldsymbol \phi}^d_{\rm PM} = (p_{\rm PM}=0,s_{\rm PM},q_{\rm PM}=0)$ with $s_{\rm PM}$ determined by 
eq.~(\ref{Eq_psq_plaquette-s}). For convenience we introduce the quantities
\beq\label{change_xyzt}
\displaystyle x = \frac{a-b}{a+b+c+d} \ , \hspace{0.75cm} \displaystyle y = \frac{c-d}{a+b+c+d} \ , \hspace{0.75cm}
\displaystyle z = \frac{a+b-c-d}{a+b+c+d}  \ .
  \eeq
These relations can be easily inverted as
\beq\label{normalized_weights}
  \begin{array}{l}
\displaystyle \frac{a}{a+b+c+d} = \frac14 (1 + z + 2x)\ , \hspace{0.8cm}
\displaystyle \frac{b}{a+b+c+d} = \frac14 (1 + z - 2x)\ , \\  \vspace{-0.2cm} \\
\displaystyle \frac{c}{a+b+c+d} = \frac14 (1 - z + 2y)\ , \hspace{0.8cm}
\displaystyle \frac{d}{a+b+c+d} = \frac14 (1 - z - 2y)  \ .
  \end{array}
\eeq
Thus the fugacities of the vertices are linear functions of $x$, $y$ and $z$ up to a normalization factor $(a+b+c+d)^{-1}$. 
The self-consistent equation for $s_{\rm PM}$ now takes the form
\beq\label{eq_para8_plaquette}
(a + b + c + d)^4(x^2 - y^2) (1 + z^2)   \Big[ 1 + 2\,  \frac{\Upsilon+1}{\Upsilon-1} \, s_{\rm PM} 
- 2 \, \frac{\Upsilon+1}{\Upsilon-1}  \, s_{\rm PM}^3 - s_{\rm PM}^4 \Big] = 0 \ ,
\eeq
with 
\beq\label{Eq_def_Upsilon}
\Upsilon(a,b,c,d) = \frac{4 x^2 + z^2 - 1}{4 y^2 + z^2 - 1} = \frac{(a+b)(c+d) - (a-b)^2}{ (a+b)(c+d) - (c-d)^2} \ .
\eeq 
Apart from the trivial solutions $s_{\rm PM}=1,-1$, the PM solution is given by
\beq\label{sol_para_8vertex}
s_{\rm PM} = \frac{1-\sqrt{\Upsilon}}{1+\sqrt{\Upsilon}} \ .
\eeq
Equation~(\ref{sol_para_8vertex}) implies that the PM solution depends upon
a single parameter $\Upsilon$, and that it remains unchanged if the vertex weights $a,b,c,d$ are modified 
without changing the value of $\Upsilon$.
The limit of infinite temperature of the eight-vertex model, {\it i.e.} $a=b=c=d$,
corresponds to $s_{\rm PM}=0$ which 
implies $\psi_{++}^{\alpha}= \psi_{--}^{\alpha}=\psi_{+-}^{\alpha}= \psi_{-+}^{\alpha}=1/4$,
$\forall \alpha \in \{u,l,r,d\}$. We anticipate that this result will also be found
in the sixteen-vertex model. {This is in sharp contrast with the limit of infinite 
temperature of the six-vertex model $a=b=c$ and $d=0$ (more generally, whenever one of the four kinds of vertices 
is absent) which corresponds to a non-trivial solution $s_{\rm PM}\neq 0$ implying 
$\psi_{++}^{\alpha}= \psi_{--}^{\alpha}\neq \psi_{+-}^{\alpha} = \psi_{-+}^{\alpha}$}. 

By inserting the PM solution in the free-energy density  (\ref{Eq_free_energy_plaquette}) one obtains
\begin{eqnarray}
\beta \, f_{\rm PM}(a,b,c,d) & \!\!\! = \!\!\! &
- \ln\left(\frac{a + b + c + d}{2}\right)
\nonumber\\
&& - \frac14 \ln\Big[1 +  \frac{(x^2 - y^2)^2 (-3 + 2 x^2 + 2 y^2 - z^2)}{(-1 + 2 x^2 + 
     2 y^2 + z^2)} \Big] \ ,
\end{eqnarray}
where $x$, $y$ and $z$ are given in terms of the vertex weights in eq.~(\ref{change_xyzt}). 
This function is clearly invariant under the exchange of $a$ with $b$, $c$ with $d$, and 
the simultaneous exchange of the weights of the FM vertices with the
AF ones.

\vspace{0.25cm}
\noindent
{\bf Ferromagnetic phases.}
In order to study the ordered phases we introduce the following functions of four variables:
\begin{eqnarray}
&& 
\nu(\{ w_m \}) = \frac{w_1^2 - w_2^2 - (w_3+w_4)^2}{w_1^2 - w_2^2 - (w_3 - w_4)^2}
\ , 
\label{nu_plaq}
\\
&&
\Sigma(\{ w_m \}) =  \frac14 \ln \Big[ \frac{w_1^2 - w_2^2 - w_3^2 - w_4^2 }{
2 w_3^2 w_4^2 ( w_1^2 + w_2^2 ) + 
     (w_1^4 + w_3^2 w_4^2) (w_1^2 - w_2^2 - w_3^2 -  w_4^2) } \Big]
     \ , 
     \label{sigma_plaq}
\end{eqnarray}
with $\{ w_m \} = w_1, \ w_2, \ w_3, \ w_4$,
and
\beq\label{mu_plaq}
  \begin{array}{l}
\displaystyle \mu(\{ w_m \}) = \frac{(w_1^2 - w_2^2 - (w_3 - w_4)^2) (w_1^2 - w_2^2 - (w_3 + w_4)^2) }{(w_1^2 - w_2^2 - w_3^2 - 
   w_4^2) \sqrt{(w_1^2 - w_2^2 - w_3^2 - w_4^2)^2 - 
  4 w_3^2 w_4^2}} 
   \\ \vspace{-0.2cm} \\
\displaystyle 
\qquad \times \frac{(2  w_3^2 w_4^2 + 
   w_1^2 (w_1^2 - w_2^2 - w_3^2 - w_4^2)) w_1^2}{ (w_1^2 (w_1^2 (w_1^2 - w_2^2 - w_3^2 - w_4^2) + 2 w_3^2 w_4^2) - 
   w_3^2 w_4^2 (-w_1^2 - w_2^2 + w_3^2 + w_4^2))} \ .
     \end{array}
\eeq
Note that $\nu$, $\Sigma$ and $\mu$ are symmetric under the exchange of $w_3$ and $w_4$.

The $a$-FM phase is homogeneous along all the
directions 
and it is given by
\beq\label{sol_F1_plaq}
{\boldsymbol \phi}_{\text{$a$-FM}}^{\alpha} = {\boldsymbol \phi}_{\text{$a$-FM}} = \Big(p_{\text{$a$-FM}}= \sqrt{\nu(a,b;c,d)},s_{\text{$a$-FM}}
=
1,q_{\text{$a$-FM}}=0 \Big)
 \ , 
\eeq
$\forall \alpha=u,l,r,d$. 
This implies that $\psi_{+-} = \psi_{-+} = 0$. Spin reversal symmetry is spontaneously broken
since $\psi_{++}\neq\psi_{--}$.
The associated free-energy and magnetization read
    \beq\label{f_m_solF1_plaq}
  \begin{array}{l}
 \displaystyle \beta \, f_{\text{$a$-FM}} (a,b,c,d)= \Sigma(a,b ; c,d)
 \ , 
   \\ \vspace{-0.2cm} \\
 \displaystyle m_{\text{$a$-FM}}(a,b,c,d)= \mu(a,b ; c,d) \ .
    \end{array}
\eeq

The extension of these results to the other FM phase is straightforward.
The $b$-FM is characterized by the same functions, with the exchange of $b$ and $a$.
One finds ${\boldsymbol \phi}_{\text{$b$-FM}}^{l,r} = \Big(p_{\text{$b$-FM}}= - \sqrt{\nu(b,a;c,d)},s_{\text{$b$-FM}}=1,q_{\text{$b$-FM}}=0 \Big)$
for the horizontal edges while the positive sign remains on the vertical ones. 

\vspace{0.25cm}
\noindent
{\bf Antiferromagnetic phases.}
AF order can also be characterized by the functions $\nu$, $\Sigma$ and $\mu$ defined in 
eqs.~(\ref{nu_plaq}),  (\ref{sigma_plaq})  and (\ref{mu_plaq}). In particular, 
for the $c$-AF phase, 
\beq
{\boldsymbol \phi}_{\text{$c$-AF}}^{\alpha} = {\boldsymbol \phi}_{\text{$c$-AF}} = 
\Big(p_{\text{$c$-AF}}= 0,s_{\text{$c$-AF}}=-1,q_{\text{$c$-AF}}=\sqrt{\nu(c,d;a,b)} \Big) 
\ . \label{eq:solAF}
\eeq
This  implies $\psi_{++}=\psi_{--}=0$ and a staggered order
with $\psi_{+-}\neq \psi_{-+}$. Spin reversal symmetry {\it and} translational invariance
are simultaneously broken. The free-energy and staggered magnetization read
 \beq
  \begin{array}{l}
 \displaystyle  \beta \, f_{\text{$c$-AF}} (a,b,c,d)= \Sigma(c,d ; a,b) 
 \ , 
   \\ \vspace{-0.2cm} \\
 \displaystyle m_{\text{$c$-AF}}(a,b,c,d)= \mu(c,d ; a,b)
 \label{eq:staggered-magn-8V-AF-plaquette}
 \ . 
    \end{array}
\eeq
The AF solution dominated by $d$ vertices is of the same form as 
the one above, with the exchange of $c$ and $d$, and 
$q_{\text{$d$-AF}}^{l,r}=-q_{\text{$c$-AF}}^{l,r}$
meaning that the $c$-AF and $d$-AF solutions only differ by a sign along the horizontal edges. 
This is due to the fact that a $d$ vertex can be obtained from a $c$ vertex by 
reversing its horizontal arrows. 

\subsubsection{The six-vertex model: phase diagram and discussion}

The PM solution,
eq.~(\ref{sol_para_8vertex}), evaluated at $d=0$ describes the PM phase of the six-vertex model.
The fixed point is characterized by a value of $s_{\rm PM}$ in the interval $-1 \leq s_{\rm PM} \leq 0$. 
As for the model on the tree of vertices,  
as soon as one of the fugacities is set to zero, the stability 
matrix~(\ref{Matrix_stab_plaq}) has an eigenvalue whose absolute value is equal to one in the whole phase. 
This calculation 
allows to show explicitly how the introduction of a hard constraint can  
drastically change the collective behavior of the system: 
in our case, the addition of a hard constraint turns the 
conventional PM phase into one with a diverging susceptibility and a soft mode.  The normalized eigenvector 
associated to this mode is of the form 
\begin{equation}
\delta \varphi_{\rm PM} = (1/2,0,0;  - 1/2,0,0;- 1/2 ,0,0; 1/2,0,0)\ .
\end{equation}

The PM-FM transitions are of the same type as the ones found in the single vertex
problem. Eqs.~(\ref{sol_F1_plaq})  and (\ref{f_m_solF1_plaq}) evaluated at  $d=0$ 
yield a discontinuous transition towards a frozen
phase with $p_{\text{$a$-FM}}=1$, {\it i.e.} $\psi_{++}=1$, 
at $a_c=b+c$ (or $b_c=a+c$) where $s_{\rm PM}=0$ $\forall\, b, \ c$, accordingly to eq.~(\ref{sol_para_8vertex}).
The free-energy density in the frozen phase, $f_{\text{$a$-FM}} = - \ln a$, 
and the magnetization, $m_{\text{$a$-FM}} =  1$, are identical to the exact results on 
the square lattice. At the transition, $\beta f_{\rm PM}(a_c,b,c) = \beta \, f_{\text{$a$-FM}}(a_c,b,c) = - \ln a_c$.

\begin{figure}[h]
\begin{center}
\includegraphics[scale=0.28]{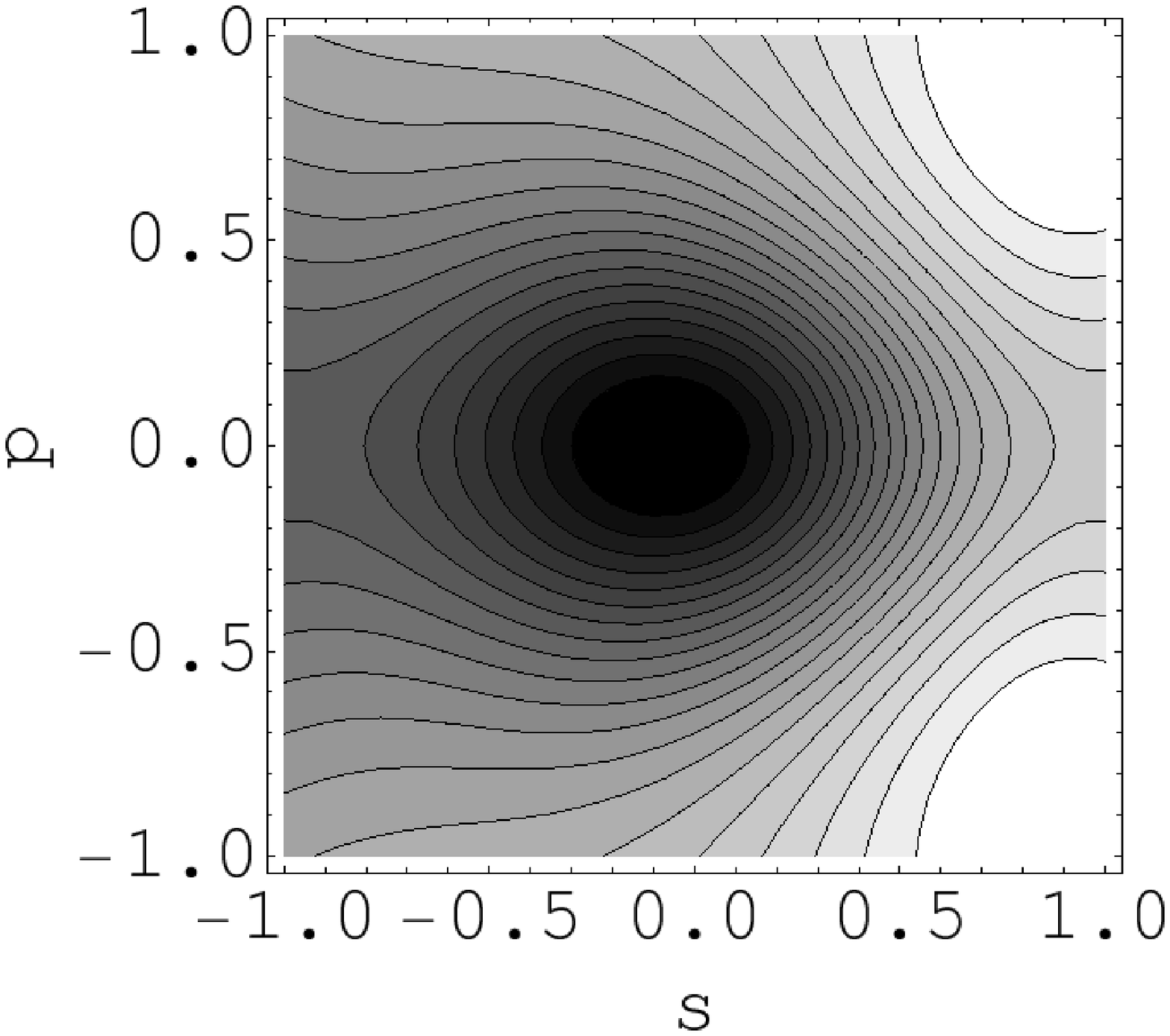}
\includegraphics[scale=0.28]{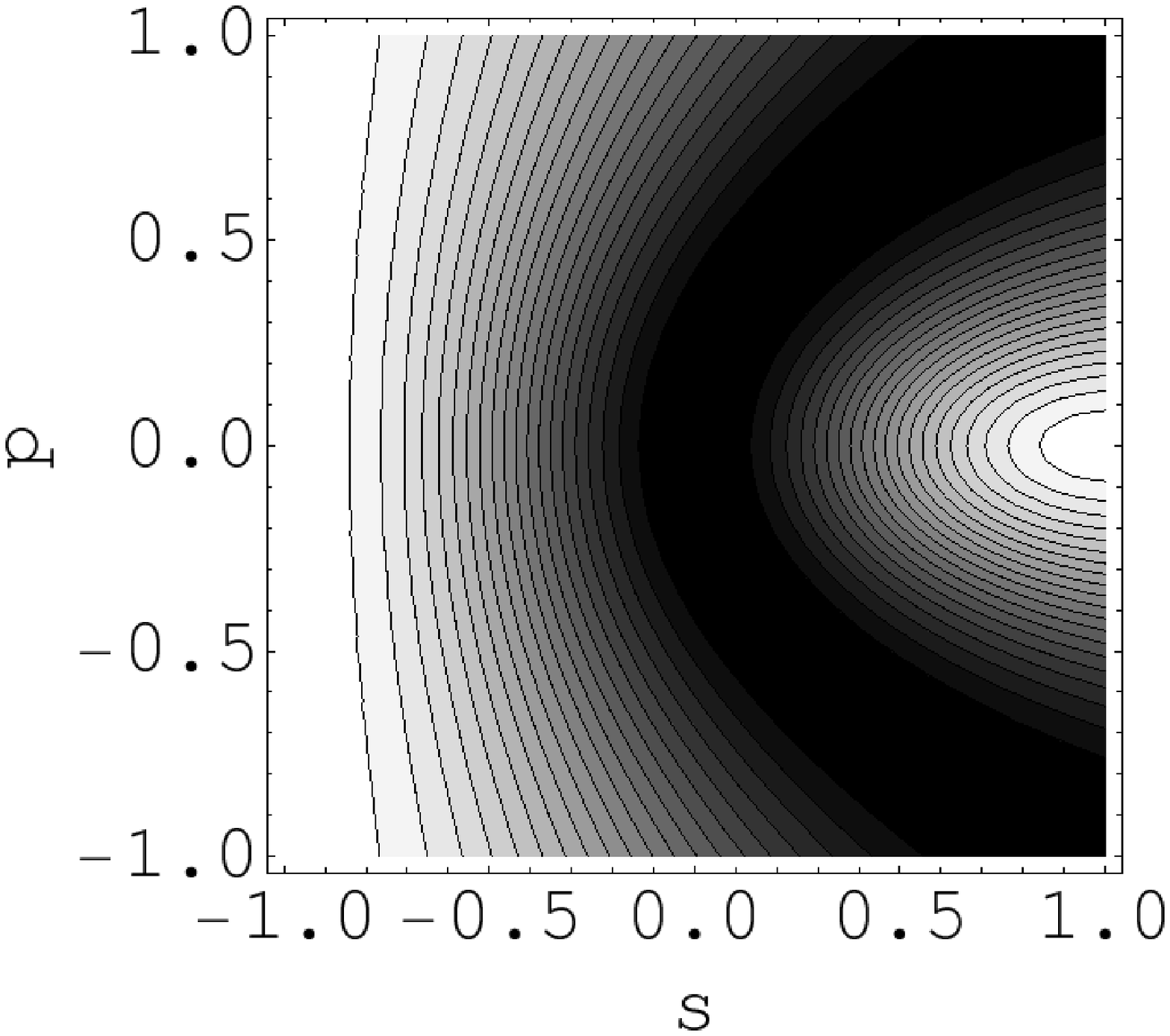}
\includegraphics[scale=0.28]{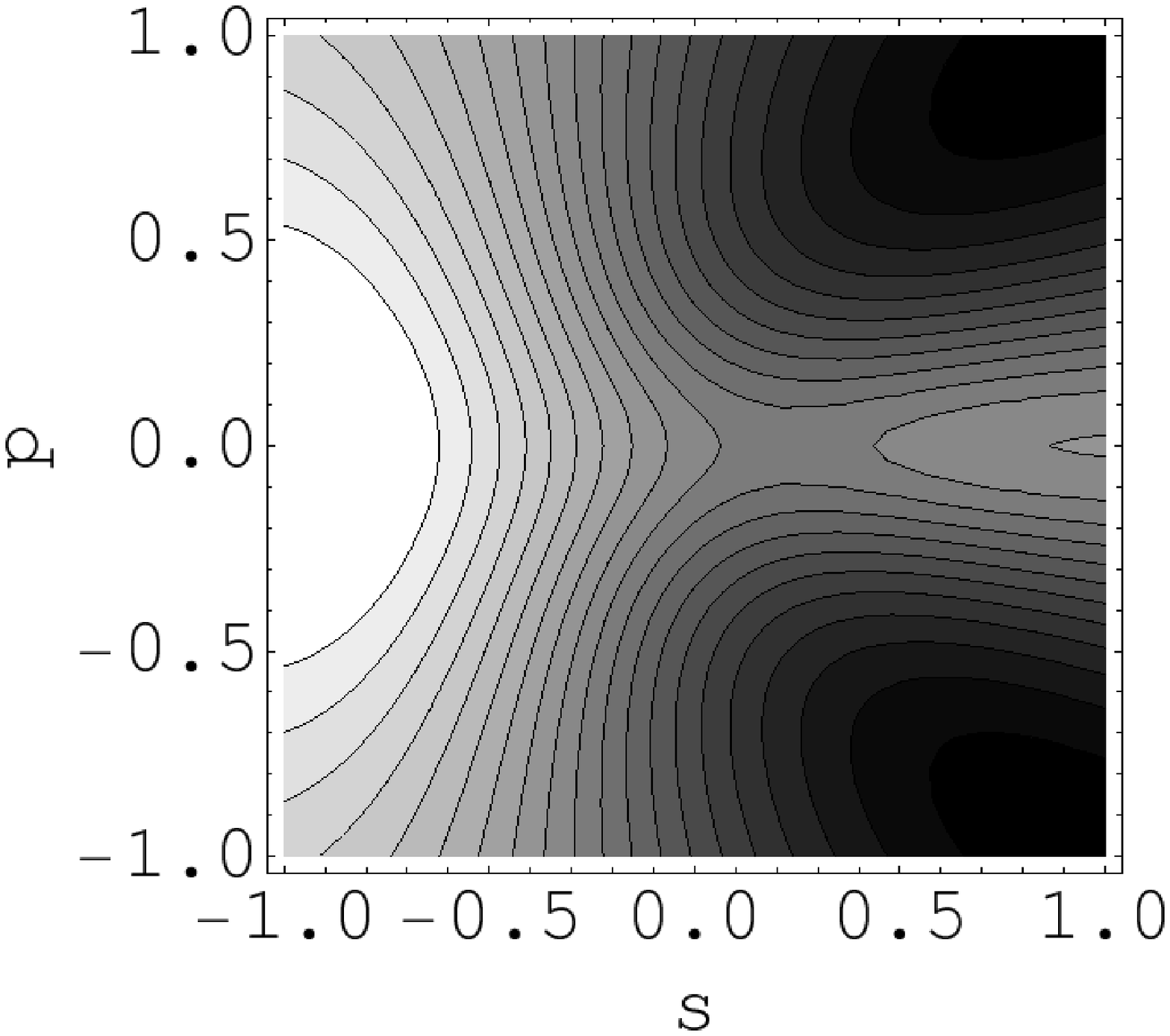}
\caption{\small (Color online.) Projection of the free energy in the plane $(s,p,q=0)$, 
where $p^{\alpha}=p$, $s^{\alpha}=s$ and $q^{\alpha}=0$ $\forall \alpha$,
for some value of the fugacities for the six-vertex model.
The free energy decreases from light to dark colors and the minima are
in the black regions. The three panels show from left to right the PM 
phase  $a<b+c$, the critical point $a=b+c$ and the FM phase
 $a>b+c$. The figure shows that at the transition point the free energy has an entire line
 of degenerate minima.}
\label{fig:free-energy-6vertex-levelcurves}
\end{center}
\end{figure}

On the transition lines both the PM and FM solutions
become unstable. One can check that fixing $a=b+c$ (or equivalently $b=a+c$)
and looking for a solution of the type $\phi_{a-{\rm FM}}^c = (p_{a-{\rm FM}}^c, s_{a-{\rm FM}}^c,q_{a-{\rm FM}}^c=0)$ 
the two equations for $p^c_{a-{\rm FM}}$ and $s^c_{a-{\rm FM}}$ become linearly dependent and
simultaneously satisfied by the condition $(p_{a-{\rm FM}}^c)^2 - s_{a-{\rm FM}}^c = 0$,  
that describes an entire line of fixed points joining the FM solution ($p_{a-{\rm FM}}=1,s_{a-{\rm FM}}=1$)
to the PM one  ($p_{\rm PM}=0,s_{\rm PM}=0$). In Fig.~\ref{fig:free-energy-6vertex-levelcurves}
we show the free energy in the plane of the solutions $(s,p)$, with $s^{\alpha}=s$, $p^{\alpha}=p$ 
and $q^{\alpha}=0$ $\forall \alpha$, and the free energy diminishes from light to dark colors.
The three panels correspond to different values of the fugacities and 
represent from left to right a paramagnetic minimum $|\Delta_6| < 1$, 
the $a$-FM critical point $\Delta_6 = 1$ associated to a degenerate line of minima, 
and the $a$-FM phase with $\Delta_6 > 1$.

\begin{figure}[h]
\begin{center}
\includegraphics[scale=0.58]{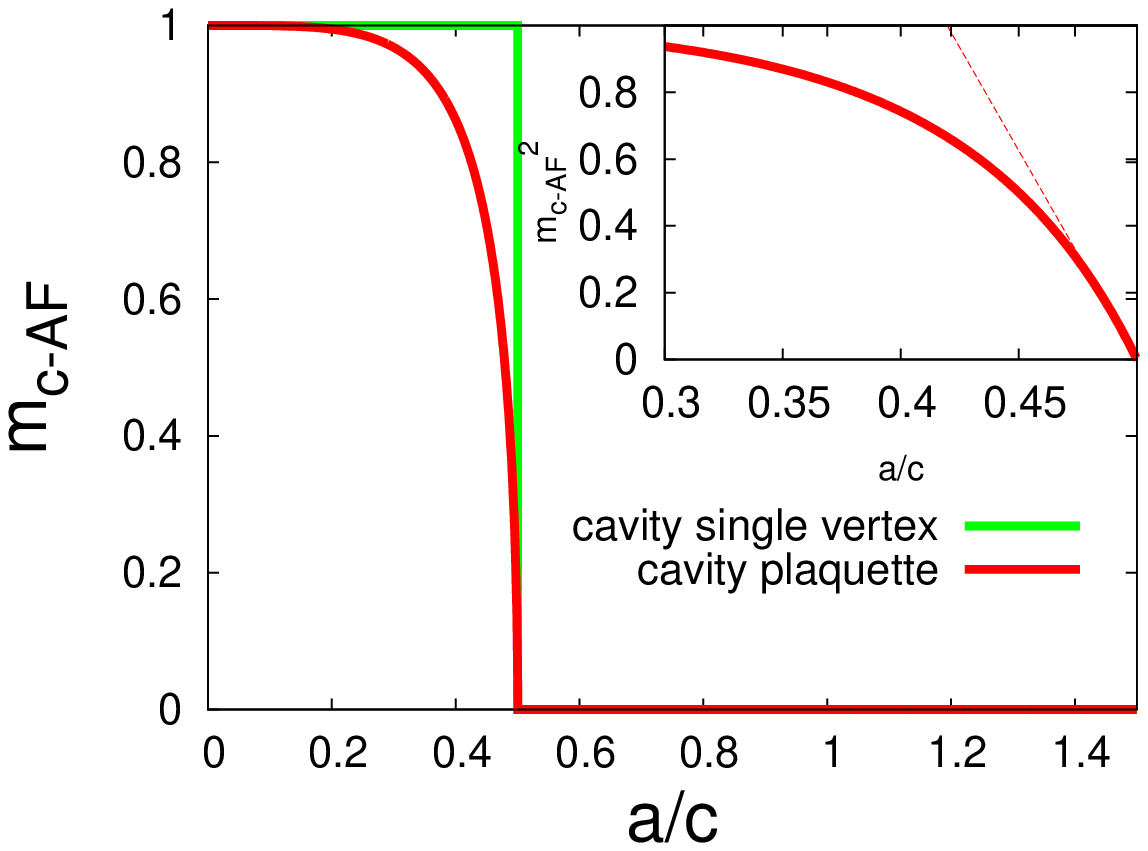}
\includegraphics[scale=0.58]{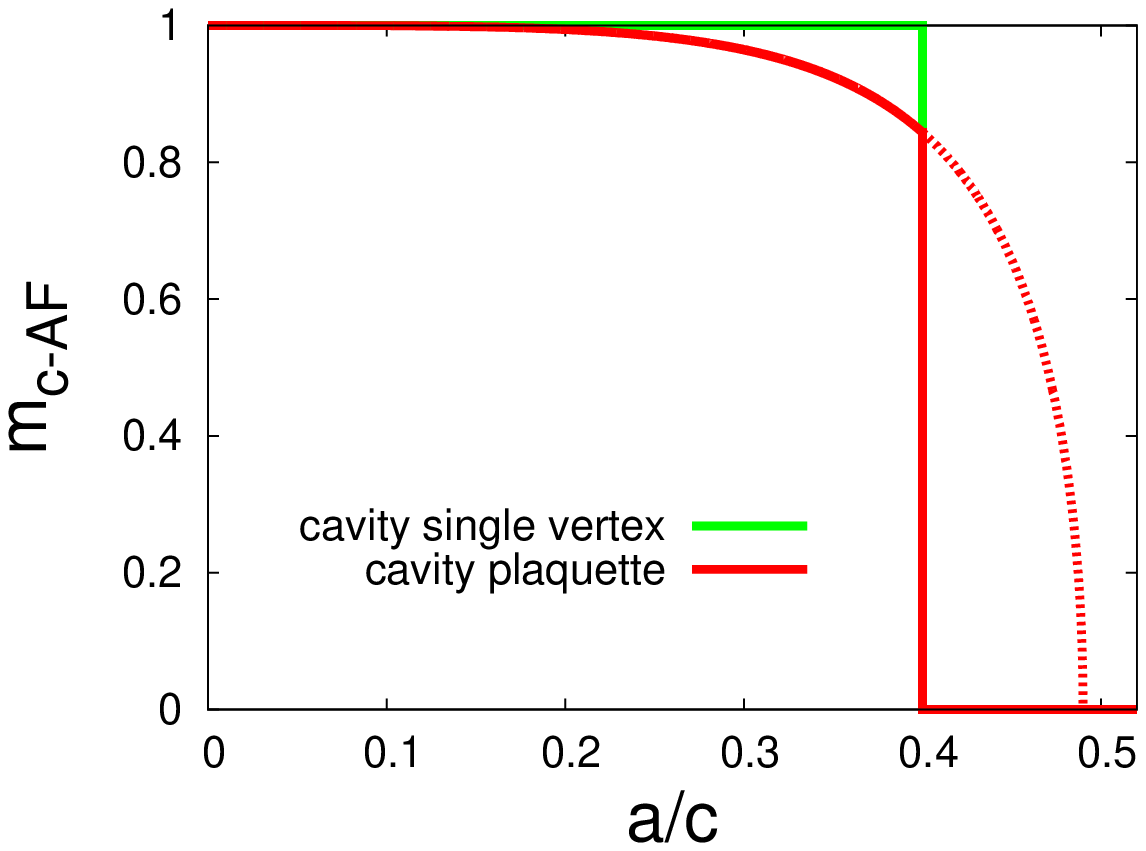}
\caption{\small (Color online.) The staggered magnetization, $m_{\text{$c$-AF}}$,  on a
tree of single vertices (green curves) and of plaquettes (red curves). 
Left panel: the  six-vertex model  with $a=b$. The  green step shows the transition towards a 
completely AF ordered phase (single vertices) while the red curve shows a continuous transition 
(plaquettes), see eq.~(\ref{eq:staggered-magn-6V-AF-plaquette}). 
The inset shows the mean-field exponent $\beta = 1/2$, $m_{\text{$a$-FM}}^2 \simeq [(a-a_c)/c]$.
Right panel: the eight-vertex 
model with $a=b$ and $d/c=0.2$.  PM-to-frozen AF transition on the single-vertex tree (green) and 
discontinuous  behavior on the plaquette tree (red), see eq.~(\ref{eq:staggered-magn-8V-AF-plaquette}), 
showing a jump towards the AF phase (with a non-frozen staggered order due to thermal fluctuations). The dashed red line
shows the continuation of the FM solution~(\ref{f_m_solF1_plaq}) beyond the transition point where the 
PM phase is favored. 
}
\label{fig:magnetization-6vertex}
\end{center}
\end{figure}

The PM-AF transition for the tree of plaquettes is 
still placed at $c_c=a+b$ (as for the tree of single {vertex} and for the exact result) 
but it is {qualitatively} different  from  the one found with the  
single vertex. The small loops of four spins allow for fluctuations in the AF  phase and
the transition becomes a continuous one with a singularity 
of the second derivative of the free-energy.
One can indeed check that $f_{\rm PM}(a,b,c_c,0)$ and  $f_{\text{$c$-AF}}(a,b,c_c,0)$
have the same first derivatives with respect to the fugacities
at the critical point.
At the AF-PM transition the PM solution reaches the critical value $s_{\rm PM}=-1$ 
and for larger values of $c$ the PM solution becomes imaginary. 
The magnetization, $m_{\text{$c$-AF}}$, 
is given by
\beq
\displaystyle m_{\text{$c$-AF}}(c ; a,b) = \frac{\sqrt{2} (a + b)^3 [(a + b)^2 + a b] }{\sqrt{ a b } 
[(a + b)^4 + a b ((a + b)^2 + a b)]}  \sqrt{\frac{c - c_c}{c_c}}
+ {\mathcal O}\left[ \big( \frac{c - c_c}{c_c}\big)^{3/2}\right] 
\label{eq:staggered-magn-6V-AF-plaquette}
\eeq
close to the transition line, 
which gives the mean-field classical exponent $\beta=1/2$ (see Fig.~\ref{fig:magnetization-6vertex}).

\begin{figure}[h]
\begin{center}
\includegraphics[scale=0.21]{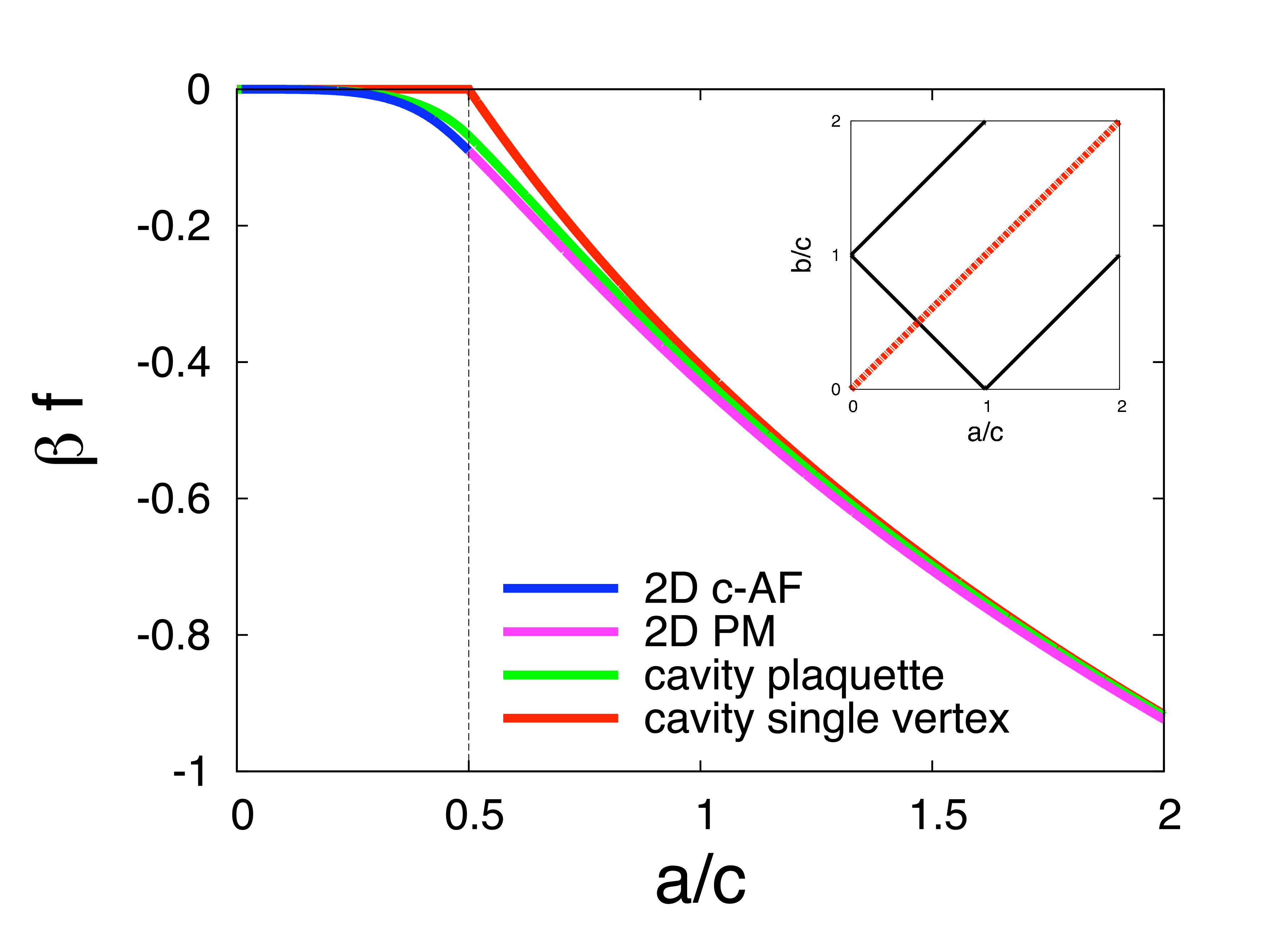}
\includegraphics[scale=0.21]{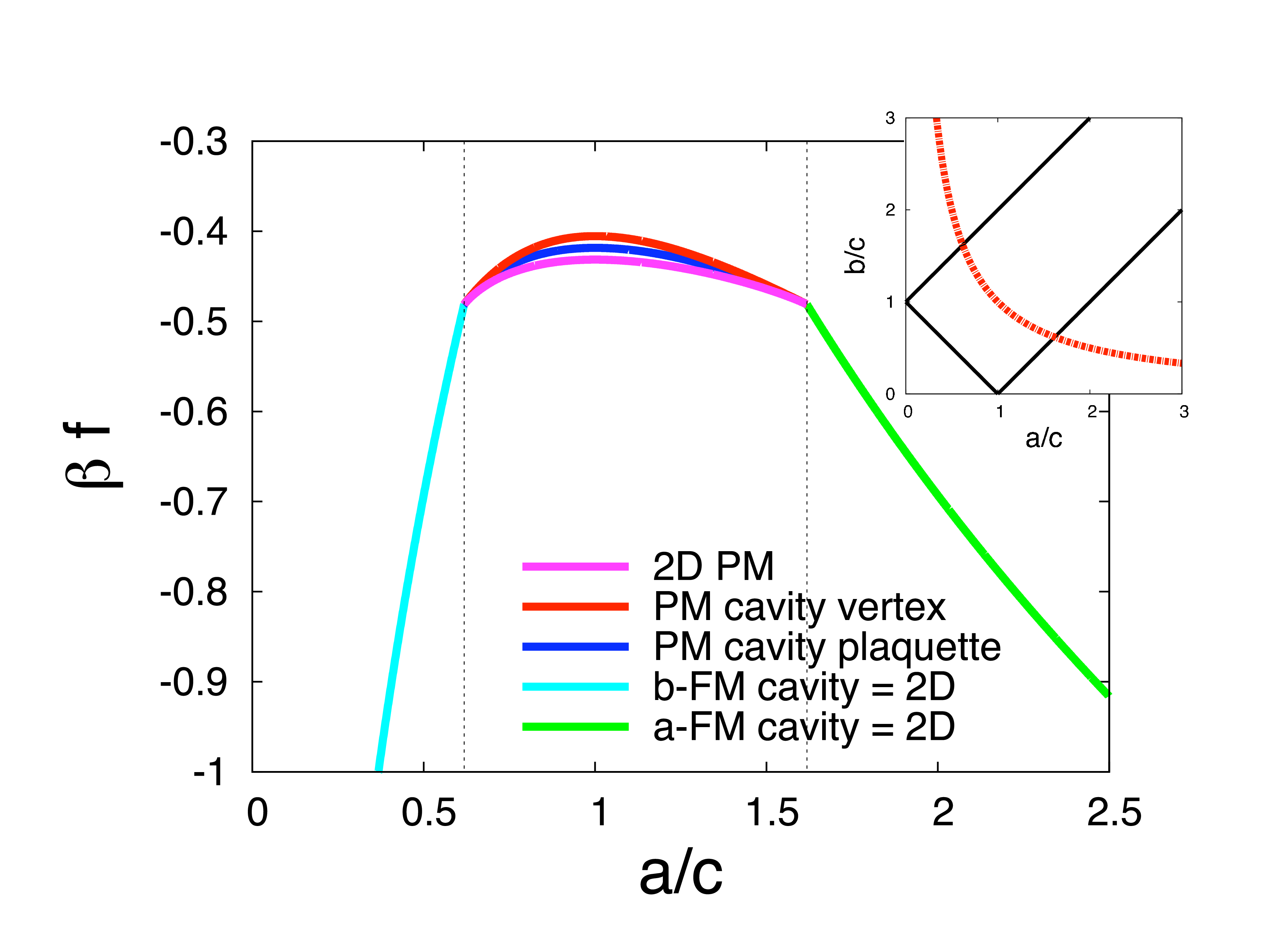}
\caption{\small (Color online.) Free-energy density of the six-vertex model
along the two paths sketched in the insets (red dotted lines), $a/c=b/c$ (left) and $a\, b=c^2$ (right).
Left panel: free-energy in the PM phase ($a/c>1/2$) and
AF phase ($a/c<1/2$) on the Bethe lattice of single vertices (red curve) and plaquettes (green curve),
 and of the exact results on the $2D$ model
(pink and blue lines).  
Right panel: free-energy of the PM phase  for $0.6 \lesssim a/c \lesssim 1.6$
on the Bethe lattice of single vertices (red curve), of plaquettes (blue curve) and in the $2D$ model
(pink line). In the FM phases, both single vertex and plaquette trees lead to the
same (exact) free-energy (green and light blue lines). 
}
\label{fig:free-energy-6vertex}
\end{center}
\end{figure}

In Fig.~\ref{fig:free-energy-6vertex} we compare the free-energy
 of the model on the tree of single vertices ($f_{sv}$), on the tree of plaquettes ($f_{pl}$), and 
 the exact free energy of the model on the $2D$ square lattice 
($f_{2D}$)~\cite{BaxterBook}.
The left panel of Fig.~\ref{fig:free-energy-6vertex} shows the free-energy in  the PM and
 AF phases as a function of $a/c$, moving along the line $a=b$.
 The figure clearly shows that the discontinuity of $f_{sv}$ at the transition is smoothed out  
by the inclusion of small loop fluctuations.
In the right panel we show the free-energy in the PM and
FM phases, as a function of $a/c$ moving along the line $a b =c^2$. 

{In the spin-ice point $a=b=c=1$ the entropy per vertex $S^{pl}$ of 
the six-vertex model on a tree of plaquettes is
\begin{equation}
S^{pl} = - \frac{1}{4}\ln \frac{3}{16}\simeq 0.418 \ .
\end{equation} 
This result is closer to the exact value obtained by Lieb for the $2D$ 
model $S^{2D} = (3/2) \ln (4/3) \simeq 0.4315$ (see Fig.~\ref{fig:free-energy-6vertex}), 
than the value obtained for the single vertex tree $S^{sv}$.}

As shown in Fig.~\ref{fig:free-energy-6vertex}  the free-energy of the frozen FM phases is the same 
for the square lattice model, the tree of single vertices and the tree of plaquettes. Instead, in the PM and AF phases the
mean-field approach does not give the exact result. The plaquette geometry yields a better quantitative
estimation of the thermodynamics properties of the model compared to the single vertex geometry 
($f_{2D} \le f_{pl} \le f_{sv}$).\footnote{The mean-field approximation can be interpreted as a variational
principle on the free energy of the system. As a result, a more accurate mean-field approximation yields
a smaller free energy, closer to the true one of the $2D$ system.}

The properties discussed so far are a general consequence of the form of 
the functions defined in eqs.~(\ref{nu_plaq}), (\ref{sigma_plaq}) and (\ref{mu_plaq}),  
independently of the precise specification of their arguments. As a consequence, similar conclusions are drawn when any
of the four types of vertices is missing.
Note  that for $w_1>w_2,w_3$  and $w_4=0$ one recovers:
\begin{eqnarray}
&& 
\nu(w_1,w_2 ; w_3,0) = \nu(w_1,w_2;0,w_4) = 1
\ , 
\nonumber\\
&& 
\Sigma(w_1,w_2 ; w_3,0) = \Sigma(w_1,w_2; 0,w_4) = - \ln w_1
\ , 
\nonumber\\
&& 
\mu(w_1,w_2 ; w_3,  0)  = 1 
\nonumber
\ , 
\end{eqnarray}
while for $w_2 = 0$ and $w_1 > w_3,w_4 \neq 0$ they all take a non trivial value. 
This means that whenever one of the fugacities of the 
AF vertices ($c$ or $d$) vanishes, the FM  phases are frozen (since $\mu=1$ and $\Sigma\equiv \mbox{cte}$)
while the AF transition is continuous. 
The same holds for the AF order, which becomes frozen, in absence of one of the FM vertices [see eq.~(\ref{f_m_solF1_plaq})]. 

\subsubsection{The eight-vertex model: phase diagram and discussion}

For the eight-vertex model the plaquette geometry gives quite different results with respect
to the tree of single vertices.
Indeed, on the tree of single vertices the addition of vertices of type $d$
does not change the nature of the transitions
(which are all discontinuous and of critical-to-frozen type as explained in 
Sec.~\ref{Sec_6_and_8_vertex_single_vertex}), 
whereas on the tree of plaquettes it does.
However, even though using the plaquette geometry the nature of the phase transitions are changed with respect
to the tree of single vertices, 
their location is not modified (the location of the critical planes will still coincide with
the exact solution).
On the other hand, as far as the PM phase is concerned, both within the single vertex mean-field approximation 
and the plaquette one, the criticality and the ``soft modes'' disappear as soon as $a,b,c,d\neq 0$.

When $a,b,c,d\neq0$
the free-energy at the transition planes shows a singularity in its first derivatives corresponding to 
a first-order phase transition.  Indeed, one can check that 
 \begin{equation}
 f_{\rm PM}(w_2+w_3+w_4,w_2 ; w_3, w_4) = \Sigma(w_2+w_3+w_4,w_2 ; w_3,w_4)  
 \ ,
 \end{equation} 
 where $w_3$ and $w_4$ are the statistical weights of  FM (resp., AF) vertices
 if the transition under consideration is a PM-FM (resp., a PM-AF) one. 
Consequently, the magnetization at the transition shows a finite jump towards a 
non-frozen ordered phase. The $a/c$ dependence of the staggered magnetization 
$m_{\text{$c$-AF}}$ is displayed in the right panel of Fig.~\ref{fig:magnetization-6vertex} for the tree of 
single vertices (green curve) and plaquettes (red curve).

Let us now analyze the $a$-FM-PM transition
and focus on the transition plane, $a_c=b+c+d$. 
If we plug a solution of the type 
$\phi^c_{\text{$a$-FM}} = (p^c_{\text{$a$-FM}}, s^c_{\text{$a$-FM}},q^c_{\text{$a$-FM}}=0)$,
into the self-consistent mean-field equations,  
we find once again the fixed point is undetermined. 
The equations for $p^c_{\text{$a$-FM}}$ and $s^c_{\text{$a$-FM}}$ 
 become dependent.
The relation between $p^c_{\text{$a$-FM}}$ and $s^c_{\text{$a$-FM}}$ 
defines a line of fixed points joining $\phi_{\rm PM}[b+c+d,b,c,d]$
and $\phi_{\text{$a$-FM}}[b+c+d,b,c,d]$. The relation between $p^c_{\text{$a$-FM}}$ and 
$s^c_{\text{$a$-FM}}$ is given by the condition
\beq\label{Eq_critical_ferromagnet}
  \begin{array}{l }
\displaystyle  4 \left[\frac{ (c+d) b}{2 c d} + 1\right] (p^c_{\text{$a$-FM}})^2 - 4 \frac{(c+d) b}{2 c d}  
s^c_{\text{$a$-FM}} + (1 - s^c_{\text{$a$-FM}})^2 = 0 \ .


   \end{array}
\eeq

Similar considerations hold for the other transitions occurring in the model. 
For instance, at the transition
towards the $c$-AF phase, $c_c = a+b+d$, one can identify 
a line of fixed points of the form $\phi_{\text{$c$-AF}}^c = (p_{\text{$c$-AF}}^c =0, s_{\text{$c$-AF}}^c,q_{\text{$c$-AF}}^c)$,
where $s_{\text{$c$-AF}}^c$ and $q_{\text{$c$-AF}}^c$ are constrained to 
fulfill a given condition similarly to~(\ref{Eq_critical_ferromagnet}) for the $a$-FM transition.

For the disordered points lying on the surfaces
$a+d = c+b$ or $a+c = d+b$ in the PM phase we have $s_{\rm PM}=0$. 
The free-energy at these points computed with the 
 single vertex tree, eq. (\ref{free-energy-fixed-points-8vertex}), 
is the same as the free-energy obtained using the plaquette model and coincides with
the exact result of the $2D$ model~\cite{BaxterBook}. This can be understood
in terms of the transformation of eq.~(\ref{change_weights}), which maps these particular points in the
PM phase of the phase diagram
into the frozen FM phase of the six-vertex model (see next section).
 Since the particular {structure} of the graph is irrelevant for the FM (frozen) phase of the six-vertex
model, the free-energy on the Bethe lattice coincides with the exact result on the square lattice in $2D$.
 
 \subsection{Duality in the eight-vertex model}\label{secion_duality}

Both the tree of single vertices and the tree of plaquettes yield the same critical
planes, given by the condition $|\Delta_8| = 1$, which also coincides with the
exact result in $2D$. This is surprising, since in general the
location of the critical points depends on the characteristics of the lattice.
This result can be interpreted in terms of a duality transformation connecting the disordered phase
with $a \simeq b \simeq c \simeq d $ 
and the ordered phase $a \gg b, c, d$. Such transformation is {independent of} the particular
structure of the lattice and only depends upon 
the connectivity. For the $2D$ model on the square lattice the duality is discussed in Sec.~10.2 
of~\cite{BaxterBook},
and it can be easily generalized to any graph of connectivity four. It connects the 
partition function of the model with fugacities $a, b, c, d$ 
to the one with fugacities $a', b', c'$ and $d'$, under the mapping~(\ref{change_weights}).
The transformation~(\ref{change_weights}) is an involution and one may express $a$, $b$, $c$, $d$ in terms 
of $a^{\prime}$, $b^{\prime}$, $c^{\prime}$, $d^{\prime}$ exactly in 
the same way.

At the level of our computation with the plaquette one can 
note that $\Upsilon[a',b',c',d']=\nu[a,b,c,d]$ or, similarly, $\Upsilon[a,b,c,d]=\nu[a',b',c',d']$. 
Moreover,  the duality holds for the free-energies 
$f_{\rm PM}[a',b',c',d'] =$$ f_{\text{$a$-FM}}[a,b,c,d]$.
Then, one can map one solution into the other,
\beq
   \begin{array}{rcl}
   \label{symmetry_fields}
\displaystyle s_{\rm PM}[a,b,c,d] &=& 
\displaystyle
\frac{\psi_{--}^{\rm PM}[a,b,c,d]-\psi_{+-}^{\rm PM}[a,b,c,d]}{\psi_{++}^{\rm PM}[a,b,c,d]+
\psi_{+-}^{\rm PM}[a,b,c,d]} 
\\ \vspace{-0.2cm} \\
\displaystyle \hspace{0.6cm} &=& 
\displaystyle
\frac{s_{\text{$a$-FM}} - p_{\text{$a$-FM}}[a',b',c',d']}{1 + p_{\text{$a$-{\rm FM}}}[a',b',c',d']}
\\ \vspace{-0.2cm} \\
\displaystyle \hspace{2.8 cm} &=& 
\displaystyle
\frac{\psi_{--}^{\text{$a$-FM}}[a',b',c',d'] - \psi_{+-}^{\text{$a$-FM}}[a',b',c',d'] }{\psi_{++}^{\text{$a$-FM}}[a',b',c',d'] + \psi_{+-}^{\text{$a$-FM}}[a',b',c',d'] } 
\ , 
   \end{array}
   \eeq
and {\it vice versa}, as well as the thermodynamic quantities.
The transition point can be recognized as the fixed point of the transformation (\ref{change_weights})
which is consistent with the exact result in $2D$, $a_c = b+c+d$.
Thanks to the mapping of eq.~(\ref{symmetry_fields}) the infinite temperature solution $\phi_{\rm PM}[a,a,a,a]$
can be mapped into the completely ordered state $\phi_{\text{$a$-FM}}[a',0,0,0]$.
However, at the transition point $\phi_{\rm PM}[b+c+d,b,c,d] \neq \phi_{\text{$a$-FM}}[b+c+d,b,c,d]$ and the
line described by eq.~(\ref{Eq_critical_ferromagnet}) connects the two solutions.
The same duality holds for the solution of the single vertex Bethe lattice where 
the free-energy of the PM phase can be mapped into the free-energy of the
 completely frozen FM solution under the mapping  given by eq.~(\ref{change_weights}).

The fixed point of~(\ref{change_weights}) gives the transition point for the $a$-FM transition. 
This is enough to recover the transition planes relative to the other ordered phases 
using the symmetries that connect the different types of ordered phases, as discussed in~\cite{BaxterBook}.

\subsection{The sixteen-vertex model on a tree of vertices}\label{16vertex_single_vertex}

In the previous sections we compared our mean-field approaches with the exact results
of the six- and eight-vertex models in $2D$, showing that the BP computation allows one to describe all the 
expected phases of the models and yields remarkably accurate results compared to the $2D$ case. 
We now focus on the mean-field analysis of the complete sixteen-vertex which includes
also charge $\pm 2$ defects, see Fig.~\ref{fig:sixteen-vertex}.

\subsubsection{Recursion relations and fixed points}

We call $\psi^{i^\alpha\to j^\beta}_{\zeta}$ the probability for the root vertex $i$ on a rooted  tree with missing edge 
$\langle i^\alpha j^\beta \rangle$ of type $\zeta$ with 
$\zeta \in \chi_v^{16} \{v_1,v_2,...,v_{16}\}$. Similarly to 
the analysis of the six- and eight-vertex models, we introduce 
the cavity probabilities $\psi^{\beta}_{i}\equiv \psi_\zeta^{i^\alpha\to j^\beta}(+1)$ to find
a positive spin on the missing edge of the rooted tree; these probabilities satisfy
\begin{align}
&\psi^{u}_i = \psi^{i^d \to  j^u}_{v_1} + \psi^{i^d \to  j^u}_{v_3} + \psi^{i^d \to  j^u}_{v_6} + \psi^{i^d \to  j^u}_{v_8}
+   \psi^{i^d \to  j^u}_{v_{9}} + \psi^{i^d \to  j^u}_{v_{11}} + \psi^{i^d \to  j^u}_{v_{13}} + \psi^{i^d \to  j^u}_{v_{16}}
\ , 
\nonumber \\ 
&\psi^{d}_i = \psi^{i^u \to  j^d}_{v_1} + \psi^{i^u \to  j^d}_{v_3} + \psi^{i^u \to  j^d}_{v_5}+ \psi^{i^u \to  j^d}_{v_7} 
+   \psi^{i^u \to  j^d}_{v_{10}} + \psi^{i^u \to  j^d}_{v_{11}} + \psi^{i^u \to  j^d}_{v_{14}} + \psi^{i^u \to  j^d}_{v_{16}}
\ , 
\label{def_prob16}
\\ 
&\psi^{l}_i = \psi^{i^r \to  j^l}_{v_1} + \psi^{i^r \to  j^l}_{v_4} + \psi^{i^r \to  j^l}_{v_6}+ \psi^{i^r \to  j^l}_{v_7} 
+   \psi^{i^r \to  j^l}_{v_{9}} + \psi^{i^r \to  j^l}_{v_{12}} + \psi^{i^r \to  j^l}_{v_{14}} + \psi^{i^r \to  j^l}_{v_{16}}
\ , 
\nonumber \\ 
&\psi^{r}_i = \psi^{i^l \to  j^r}_{v_1} + \psi^{i^l \to  j^r}_{v_4} + \psi^{i^l \to  j^r}_{v_5} + \psi^{i^l \to  j^r}_{v_8}
+   \psi^{i^l \to  j^r}_{v_{9}} + \psi^{i^l \to  j^r}_{v_{11}} + \psi^{i^l \to  j^r}_{v_{14}} + \psi^{i^l \to  j^r}_{v_{15}} 
\ . 
\nonumber 
\end{align}
Along the same line of reasoning outlined in Sec.~\ref{Sec_6_and_8_vertex_single_vertex}, we derive  
the set of self-consistent equations for the probabilities defined above:
\beq
 \begin{array}{ll}
&
\displaystyle  \psi^{u } 
\displaystyle    =  \hat{\Psi}^u[a, b, c, d, e, \psi^u , \psi^d , \psi^l , \psi^r ]  
\\ \vspace{-0.2cm} \\
& =   \displaystyle \frac{1}{z^{u }} \Big[ a~  \psi^{l } \psi^{u } \psi^{r } +
b ~(1 - \psi^{l }) \psi^{u }(1-  \psi^{r }) 
 +  c~ (1-\psi^{u })(1- \psi^{l }) \psi^{r }
 + d ~ \psi^{l }(1 - \psi^{u })(1 - \psi^{r })   
\\ \vspace{-0.2cm} \\

& \;\;\;\; + e ~\Big( \psi^{l } (1 - \psi^{u }) \psi^{r }  + \psi^{l } \psi^{u }(1 - \psi^{r }) + 
 ( 1 - \psi^{l })  \psi^{u } \psi^{r } + (1 - \psi^{l })(1 - \psi^{u })(1 - \psi^{r })
 \Big) \Big]

\\ \vspace{-0.2cm} \\
&
\displaystyle  \psi^{l } 
\displaystyle    =  \hat{\Psi}^l[a, b, c, d, e, \psi^u , \psi^d , \psi^l , \psi^r ] 
\\ \vspace{-0.2cm} \\
& 
      =       \displaystyle \frac{1}{z^{l }} \Big[ a~ \psi^{d } \psi^{l }  \psi^{u } +
b~(1-\psi^{d }) \psi^{l }(1 - \psi^{u } )
+  c  ~\psi^{d }(1 - \psi^{l }) (1 - \psi^{u })  + d~ (1 - \psi^{d })(1 - \psi^{l })  \psi^{u } 
\\ \vspace{-0.2cm} \\
& \;\;\;\;
+ 
e ~\Big( \psi^{d }  \psi^{l } (1 - \psi^{u })  + \psi^{d }(1 -  \psi^{l }) \psi^{u }   
 +  
  (1 - \psi^{d } ) \psi^{l }  \psi^{u }  + (1 - \psi^{d })(1 - \psi^{l })(1 - \psi^{u })
 \Big) \Big]

\\ \vspace{-0.2cm} \\
&
\displaystyle  \psi^{d } \displaystyle    =  \hat{\Psi}^d[a, b, c, d, e, \psi^u , \psi^d , \psi^l , \psi^r ] 
\\ \vspace{-0.2cm} \\
& 
     =     \displaystyle \frac{1}{z^{d }} \Big[ a ~\psi^{r } \psi^{d } \psi^{l }   +
b~ (1 - \psi^{r }) \psi^{d }(1 - \psi^{l })
  + c ~(1 - \psi^{r })(1 - \psi^{d }) \psi^{l }
  + d ~ \psi^{r } (1 - \psi^{d })(1 - \psi^{l }) 
\\ \vspace{-0.2cm} \\
& \;\;\;\; + 
e~ \Big( (1 - \psi^{r })  \psi^{d }  \psi^{l }  +  \psi^{r }(1 - \psi^{d } ) \psi^{l }   + 
  \psi^{r }  \psi^{d }(1 -  \psi^{l }) + (1 - \psi^{r }) (1 - \psi^{d })(1 - \psi^{l } )
 \Big) 
 \Big]

\\ \vspace{-0.2cm} \\
&
\displaystyle  \psi^{r } \displaystyle    =  \hat{\Psi}^u[a, b, c, d, e, \psi^u , \psi^d , \psi^l , \psi^r ] 
\\ \vspace{-0.2cm} \\
& 
    =     \displaystyle \frac{1}{z^{r }} \Big[ a \psi^{u }  \psi^{r } \psi^{d }   +
b(1 - \psi^{u })  \psi^{r } (1 - \psi^{d })
+  c  \psi^{u } (1 - \psi^{r } )(1 - \psi^{d } )  +  d (1 - \psi^{u })(1 - \psi^{r }) \psi^{d }
\\ \vspace{-0.2cm} \\
& \;\;\;\; + e \Big( \psi^{u } (1 - \psi^{r })  \psi^{d }    +   \psi^{u } (1 - \psi^{r })  \psi^{d }    
 +  
  \psi^{u } (1 - \psi^{r })  \psi^{d }  + (1 - \psi^{u } ) (1 - \psi^{r }) (1 - \psi^{d })
 \Big) 
 \Big]
\\
&
\vspace{-0.2cm}
    \end{array}
    \\
    \label{psiU_16vertex}
\eeq
where $z^{\alpha}$ are normalization constants.\footnote{The $z^{\alpha}$ differ from 
the ones defined in eq.~(\ref{psiU_8vertex}), due to the contributions
from $e \neq 0$} In order to describe AF phases, these equations must be studied on 
bipartite graphs as in eq.~(\ref{eq-cav}).

For simplicity, here and in the following we consider the case in which the vertices $v_9,\dots,v_{16}$
have the same weight $e$; arrow reversal symmetry holds for the statistical weights
of all other vertices as well. However
these assumptions could be easily released.
The fixed points of eqs.~(\ref{psiU_16vertex}) on one of the two sub-lattices, say $A_1$, are
\begin{eqnarray}\label{eq:Fixed-points-16-sv}
&&
\psi^u_{PM} = \psi^l_{PM} = \psi^r_{PM} = \psi^d_{PM} = \frac12 \ ,
\nonumber \\
&&
\psi^u_{\text{$a$-FM}} = \psi^l_{\text{$a$-FM}} = \psi^r_{\text{$a$-FM}} =  \psi^d_{\text{$a$-FM}} = H_{+}(a,b,c,d,e) 
\ , 
\nonumber \\
&&
\psi^u_{\text{$b$-FM}} = \psi^d_{\text{$b$-FM}} =
H_{+}(b,a,c,d,e)  \ , \hspace{0.9cm} \psi^l_{\text{$b$-FM}} =  \psi^r_{\text{$b$-FM}}=H_{-}(b,a,c,d,e)
\ , 
\\
&&
\psi^u_{\text{$c$-AF}} = \psi^l_{\text{$c$-AF}} = 
H_{+}(c,b,a,d,e)  \ , \hspace{0.9cm}  \psi^r_{\text{$c$-AF}} =  \psi^d_{\text{$c$-AF}} = H_{-}(c,b,a,d,e) 
\ ,
\nonumber \\
&&
\psi^u_{\text{$d$-AF}} = \psi^r_{\text{$d$-AF}} = 
H_{+}(d,b,c,a,e) \ ,  \hspace{0.9cm}  \psi^d_{\text{$d$-AF}} =  \psi^l_{\text{$d$-AF}} = H_{-}(d,b,c,a,e) 
\nonumber
\ , 
\end{eqnarray}
with 
$$\displaystyle H_{\pm}(w_1,w_2,w_3,w_4,w_5) 
= \frac12 \pm \frac{w_1 - w_2 - w_3 - w_4 - 2 w_5}{2\sqrt{(w_1 - w_2 - w_3 - w_4)^2 - 4 w_5^2}} \ .
$$

 \subsubsection{Free-energy, stability and order parameters}

In presence of vertices of kind $e$, the vertex partition function reads 
\beq\label{eq:Zv16}
 \begin{array}{l}
Z_{v}[\psi^l, \psi^r,\psi^u,\psi^d]    =  a~ \Big[ \psi^{l } \psi^{u } \psi^{r } \psi^{d} +
(1-\psi^{u })(1- \psi^{l })(1 - \psi^{r })(1 - \psi^{d }) \Big]  ~
\\ \vspace{-0.2cm} \\
 \hspace{3.5cm} + ~ b~ \Big[ (1 - \psi^{l }) \psi^{u }(1-  \psi^{r }) \psi^{d} +  \psi^{l }(1 - \psi^{u }) \psi^{r }(1 - \psi^{d}) \Big] ~
 \\ \vspace{-0.2cm} \\
   \hspace{3.5cm} + ~   c ~\Big[ (1-\psi^{u })(1- \psi^{l }) \psi^{r }\psi^{d } + \psi^{u } \psi^{l }(1 - \psi^{r })(1 - \psi^{d } )\Big]
 \\ \vspace{-0.2cm} \\
   \hspace{3.5cm} + ~ d ~\big[ \psi^{l }(1 - \psi^{u })(1 - \psi^{r })\psi^{d} +(1- \psi^{l }) \psi^{u } \psi^{r }(1 - \psi^{d})  \Big]  ~ 
 \\ \vspace{-0.2cm} \\
 \hspace{3.5cm}  + ~e~ \Big[ \psi^{l } (1 - \psi^{u }) \psi^{r }\psi^{d }  + (1-\psi^{l })  \psi^{u }(1-\psi^{r })(1-\psi^{d })  
 \\ \vspace{-0.2cm} \\
 \hspace{4.4cm}  +\, \psi^{l } \psi^{u }(1 - \psi^{r })\psi^{d }  +(1- \psi^{l })(1- \psi^{u }) \psi^{r }(1-\psi^{d })  
 \\ \vspace{-0.2cm} \\
 \hspace{4.4cm}  + \,  ( 1 - \psi^{l })  \psi^{u } \psi^{r }\psi^{d } +    \psi^{l }(1 - \psi^{u })(1 - \psi^{r })(1 - \psi^{d }) 
 \\ \vspace{-0.2cm} \\
 \hspace{4.4cm}    +\, (1 - \psi^{l })(1 - \psi^{u })(1 - \psi^{r })\psi^{d }  + \psi^{l } \psi^{u }\psi^{r }(1-\psi^{d })   \Big] \ ,
      \end{array}
\eeq
which in the limit $e\to 0$ gives back eq.~(\ref{eq:Zv}).
The free-energy reads
\beq\label{free-energy-Cavity16}
  \begin{array}{l}
\displaystyle \beta f[a,b,c,d,e,\boldsymbol{\psi}_1,\boldsymbol{\psi}_2] = - \frac12 \Big( \ln Z_{v}[\boldsymbol{\psi}_1] +  \ln Z_{v}[\boldsymbol{\psi}_2] ~+
 \\ \vspace{-0.2cm} \\
 \qquad
\displaystyle - \ln Z_{\langle l r \rangle}[\psi^l_1 ,\psi^r_2]  -  \ln Z_{\langle l r \rangle}[\psi^l_2 ,\psi^r_1]
- \ln Z_{\langle u d \rangle}[\psi^u_1 ,\psi^d_2] - \ln Z_{\langle u d \rangle}[\psi^u_2 ,\psi^d_1] \Big) \ ,
       \end{array}
\eeq
where $Z_{\langle l r\rangle}$ and  $Z_{\langle u d\rangle}$ are 
defined in eqs.~(\ref{eq:Zlr}).
The free-energies of
the different phases is obtained evaluating eq.~(\ref{free-energy-Cavity16}) at the fixed points
given by eq.~(\ref{eq:Fixed-points-16-sv}):
\beq\label{free-energy-fixed-points-16vertex}
\begin{array}{l}
\displaystyle \beta f_{\rm PM} = \beta  f[a,b,c,d,e,\boldsymbol{\psi}_{\rm PM}]  = - \ln\Big[\frac{a+b+c+d + 4 e}{2}\Big] 
\ ,
 \\ \vspace{-0.2cm} \\
\displaystyle \beta f_{\text{$a$-FM}} = \beta f[a,b,c,d,e,\boldsymbol{\psi}_{\text{$a$-FM}}]  = - \ln\Big[a - \frac{2 e^2}{- a + b + c + d}\Big] 
\ , 
 \\ \vspace{-0.2cm} \\
\displaystyle \beta f_{\text{$b$-FM}} = \beta  f[a,b,c,d,e,\boldsymbol{\psi}_{\text{$b$-FM}}] = - \ln\Big[b - \frac{2 e^2}{a - b + c + d}\Big] 
\ , 
 \\ \vspace{-0.2cm} \\
\displaystyle \beta f_{\text{$c$-AF}} =  \beta f[a,b,c,d,e,\boldsymbol{\psi}_{\text{$c$-AF}}] = - \ln\Big[c - \frac{2 e^2}{a - c + b + d}\Big] 
\ , 
 \\ \vspace{-0.2cm} \\
\displaystyle \beta f_{\text{$d$-AF}} = \beta  f[a,b,c,d,e,\boldsymbol{\psi}_{\text{$d$-AF}}]  = - \ln\Big[d - \frac{2 e^2}{a - d + b + c}\Big] 
\ . 
\end{array}
\eeq

The order parameters are defined as in Sec.~\ref{Sec:8vertex-OrderParam}
with the addition of the contributions from the new vertices. For instance, the
FM order parameter $m_{\text{$a$-FM}}$ reads:
\beq\label{Eq:Order_Parameters_16v_sv}
 \begin{array}{l}
\displaystyle m_{\text{$a$-FM}} = \frac{1}{Z_{v}} \Big[ a \, \Big( \psi^{l } \psi^{u } \psi^{r } \psi^{d} -
(1-\psi^{u })(1- \psi^{l })(1 - \psi^{r })(1 - \psi^{d }) \Big) 
 \\ \vspace{-0.2cm} \\
\; \; \displaystyle 
\qquad \qquad
+ \frac{e}{2} \Big(    
 - \psi_r - \psi_u - \psi_d  - \psi_l 
 
  + 2 \psi_r \psi_u 
  +2 \psi_l \psi_r
  +2 \psi_u \psi_l
  + 2 \psi_d \psi_l
     \\ \vspace{-0.2cm} \\
     \qquad\qquad\qquad
  + 2 \psi_d \psi_u
  + 2 \psi_d \psi_r
  - 2 \psi_l \psi_r \psi_u
  - 2 \psi_d \psi_r \psi_u
  - 2 \psi_d \psi_l \psi_r
  - 2 \psi_d \psi_l \psi_u \Big) \Big]
  \end{array}
\eeq
 where the normalization is now the one defined in eq.~(\ref{eq:Zv16}). 
 The other order parameters characterizing the other ordered phases have a similar expression.
The PM solution $\boldsymbol{\psi}_{\rm PM}$ is the same as for the eight-vertex model
and yields vanishing magnetization for all the order parameters.
Differently from the six- and the eight-vertex model, where the order parameters
jump discontinuously to one at the phase transition on the tree of single vertices, when $e \neq 0$
any ordered FM or AF solution is associated to a continuous transition. 
For instance, for the $a$-FM phase transition, plugging 
$\boldsymbol{\psi}_{\text{$a$-FM}}$ from eq.~(\ref{eq:Fixed-points-16-sv}) 
into eq.~(\ref{Eq:Order_Parameters_16v_sv}) 
one obtains:
\beq\label{m_a-FM_16vertex}
m_{\text{$a$-FM}} = \frac{\sqrt{(a - b - c - d)^2 - 4 e^2}}{a -  b - c - d } \ .
\eeq
The expansion close to the critical point $a_c=b+c+d+2 e$ 
yields the expected mean-field classical exponent $\beta=1/2$:
\beq\label{exp-m_a-FM_16vertex}
m_{\text{$a$-FM}} = \frac{\sqrt{a-a_c}}{\sqrt{e}}
+ {\mathcal O}\left[ \big(a - a_c\big)^{3/2}\right] \ .
\eeq
In the limit $e \to 0$ one recovers from eq.~(\ref{m_a-FM_16vertex}) 
the value $m_{\text{$a$-FM}}=1$,
characteristic of the frozen (on the tree of single vertices) ordered phase in the eight-vertex model, 
due to a divergence of the coefficient of the singular term $\sqrt{a-a_c}$.
The magnetizations $m_{\text{$b$-FM}}$,  $m_{\text{$c$-AF}}$ 
and $m_{\text{$d$-AF}}$ are zero in the $a$-FM phase. 
Analogous results hold for the other phase transitions and the corresponding order parameters;
these results can be obtained by simply 
exchanging the parameter $a$ and the opportune vertex weight 
in eqs.~(\ref{m_a-FM_16vertex}) and (\ref{exp-m_a-FM_16vertex}). 

We study the stability properties of the PM phase by analyzing the eigenvalues of the 
stability matrix introduced in eq.~(\ref{matrix_Jac}) for $e \geq 0$.
The eigenvalues associated to the PM solution $\boldsymbol{\psi}_{\rm PM}$ are
\beq\label{stability_16_single_vertex}
  \begin{array}{l}
  \displaystyle     E^{\rm PM}_1 =\frac{3 a-b - c- d}{a+b+c+d+4e}
  \ , 
 \qquad\qquad\qquad
 \displaystyle    E^{\rm PM}_2 =\frac{- a + 3b - c - d}{a+b+c+d+4e}
 \ , 
                          \\ \vspace{-0.2cm} \\
 \displaystyle E^{\rm PM}_3 =\frac{a+b- 3 c+ d}{a+b+c+d+4e}
  \ , \qquad\qquad\qquad
 \displaystyle  E^{\rm PM}_4 = \frac{a+b+c-3 d}{a+b+c+d+4e}
 \ , 
  \end{array}
\eeq
with the same eigenvectors as in Sec.~\ref{Sec:Stabiliy_sv}.
Overall the stability is controlled by the condition
\beq
\left| \frac{ (1 + E^{\rm PM}_3)(1 + E^{\rm PM}_4) - 
(1 - E^{\rm PM}_1)(1 - E^{\rm PM}_2)}{(1 + E^{\rm PM}_3)(1 + E^{\rm PM}_4) + (1 - E^{\rm PM}_1)(1 - E^{\rm PM}_2) } \right| 
= |\Delta_{16}^{sv}|<1
\eeq
with
\beq
\Delta_{16}^{sv} = \frac{a^2+b^2-c^2-d^2+2 (a + b - c - d)  e}{2 [c d + a b + e (a + b + c + d + 2 e)]} \ .
\eeq
This \emph{generalised anisotropy parameter} implies that, in the presence of a non-vanishing vertex 
weight $e$, all the transition lines
are shifted by  $2 e$ with respect to the value obtained for the eight-vertex model. This should be 
compared to the conjectured value for the $2D$ model, $\Delta_{16}$, given in eq.~(\ref{Delta16_2D_model}), 
and with the numerical results in $2D$ on the square lattice. 
The BP approach on the tree of vertices gives the correct 
qualitative behavior, meaning an increase of the PM phase with increasing $e$, but it does not
coincide with the numerical results.  
The conjectured $\Delta_{16}$ given in eq.~(\ref{Delta16_2D_model}) is closer to the numerics 
(this is not surprising, since its functional form was guessed
from the analysis of the numerical results). Unfortunately, there is no exact result in $2D$
for the sixteen-vertex model  
to compare with and we are not able to draw more precise conclusions on the anisotropy parameter.\footnote{Actually 
there is no guarantee of the existence of such a single 
$\Delta_{16}$ parameter for the sixteen-vertex model.} 

\subsection{The sixteen-vertex model on the tree of plaquettes}

The procedure described in the previous sections can be easily extended to 
analyze the sixteen-vertex model
on the tree of plaquettes. We use the definition 
 \beq\label{vertex_weight_16vertex}
   \begin{array}{ll}
\displaystyle w_{s_1,s_2,s_3,s_4}(a,b,c,d,e) 
\!\!\! & =\displaystyle  \frac{1}{4} \Big[ a' (1+s_1 s_2 s_3 s_4) +  b' (s_1 s_3 +  s_2 s_4)  + 
 c' (s_1 s_4 + s_2 s_3) 
 
  \\ \vspace{-0.2cm} \\
& \hspace{0.4cm} +~ d' (s_1 s_2 + s_3 s_4)  \Big] 
  \displaystyle + \frac{e}{2} (1 - s_1 s_2 s_3 s_4)
  \ , 
   \end{array}
\eeq
where $a'$, $b'$, $c'$ and $d'$ given in eq.~(\ref{vertex_weight_8vertex}).
Then, eqs.~(\ref{Eq_psi_plaquette}) and (\ref{Eq_psq_plaquette-p})
yield the fixed-points, eq.~(\ref{Eq_free_energy_plaquette}) gives the free-energies of the different 
phases and eqs.~(\ref{Eq_Magn_energy_plaquette}) determine
the order parameters of the sixteen-vertex model.
As  the expressions are rather long and involved we prefer not to write them down explicitly here. 

\subsubsection{Paramagnetic phase}

The PM solution is of the type 
${\boldsymbol \phi}_{\rm PM} \equiv {\boldsymbol \phi}^u_{\rm PM} = {\boldsymbol \phi}^l_{\rm PM}
={\boldsymbol \phi}^r_{\rm PM} = {\boldsymbol \phi}^d_{\rm PM} = (p_{\rm PM}=0,s_{\rm PM},q_{\rm PM}=0)$
and $s_{\rm PM}$ is the solution of the equation:
\beq\label{eq_para16_plaquette}
  \begin{array}{l}
\displaystyle 
(a + b + c + d)^4 (x^2 - y^2) \Big[
 \lambda_4 ~s_{\rm PM}^4 + \lambda_3~ s_{\rm PM}^3 + 
 \lambda_2 ~s_{\rm PM}^2 + \lambda_1 ~s_{\rm PM} +
 \lambda_0 \Big] = 0
     \end{array}
\eeq
with
\beq
  \begin{array}{l l}
& \displaystyle \lambda_0(x,y,z,u) =  \displaystyle (1+u)^2 + z^2 
\ , 

 \\ \vspace{-0.2cm} \\
 
& \displaystyle \displaystyle \lambda_1(x,y,z,u)  

 = \displaystyle - \, \frac{(1 + u)^4  + z^4 - 2 (1 + z^2 - u^2) ( x^2 + y^2) }{x^2 - y^2} 
 \  ,

 \\ \vspace{-0.2cm} \\
 
& \displaystyle \displaystyle \lambda_2(x,y,z,u)   = \displaystyle - 12 u
\ , 
      
       \\ \vspace{-0.2cm} \\
& \displaystyle \displaystyle \lambda_3 (x,y,z,u) 
 
   = \displaystyle  \, \frac{(1 - u)^4   + z^4 - 2 (1 + z^2 - u^2)(x^2 + y^2)}{x^2 - y^2} 
     = - \lambda_1(x,y,z,- u) 
     \ , 
   
   \\ \vspace{-0.2cm} \\
 
& \displaystyle \displaystyle \lambda_4(x,y,z,u)    =  \displaystyle - [ (1- u)^2 + z^2 ] = - \lambda_0 (x,y,z,-u)  \  .
     \end{array}
\eeq
The variables $x$, $y$ and $z$ were already defined in eq.~(\ref{change_xyzt}) and
$u = 4 e/(a+b+c+d)$. 
This equation reduces to that of the eight-vertex model, eq.~(\ref{eq_para8_plaquette}), when $u=0$.
The presence of a non-zero value of $u$ implies a much more complicated dependence
on the parameters as the roots of eq.~(\ref{eq_para16_plaquette}) are more involved.
We do not solve this equation explicitly; one can show that, as for the eight
vertex model, the limit of infinite temperature $a=b=c=d=e$ corresponds to the trivial solution $s_{\rm PM}=0$.
At linear order in $u$ one obtains
\beq
  \begin{array}{ll}
\displaystyle s_{\rm PM}  =  \frac{1 - \sqrt{\U}}{1 + \sqrt{\U}} + 
 \frac{8 (x^2 - y^2)}{ (1 + z^2) (-1 + 4 x^2 + z^2) (-1 + 4 y^2 + 
   z^2) } \, \,  \times 

   \\ \vspace{-0.2cm} \\
\qquad \qquad \displaystyle \frac{(1 + 4 (x^4 - 6 x^2 y^2 + y^4) - 4 (x^2 + y^2) z^2 - 
   z^4)}{(\sqrt{-1 + 4 y^2 + z^2} + \sqrt{-1 + 4 x^2 + z^2})^2} \, \,  u  + \mathcal{O}(u^2) \ ,
    \end{array}
\eeq
where $\U$ is the function defined in eq.~(\ref{Eq_def_Upsilon}) which characterizes the
PM phase when $u=0$.

\subsubsection{Ferromagnetic phase}

Solving numerically eqs.~(\ref{Eq_psq_plaquette-p}) for the sixteen-vertex model one finds that the 
$a$-FM fixed point takes the general form $p^u = p^l =  p^r =  p^d = p^{\text{$a$-FM}}$,
$s^u = s^l =  s^r =  s^d = s^{\text{$a$-FM}}$, $q^u = q^l = - q^r =  - q^d = q^{\text{$a$-FM}}$
with $q^{\text{$a$-FM}}$ slightly different from zero unless $c=d$ or $e=0$.
This implies that the corresponding eigenvector has a non-vanishing component also
along the ``direction" of $q$. However, $q^{\text{$a$-FM}}$ is very small and influences
very little the transition point, while the instability is mainly driven by the 
FM component which gives a non zero value of $p^{\text{$a$-FM}}$.
Therefore, while in the numerical study we used the exact expression of the eigenvector, in the 
analytic study of some asymptotic behaviors we disregarded the components contributing to the non-zero value 
of $q^{\text{$a$-FM}}$.
Accordingly, in order to study the stability we consider the simplified condition: 
\beq
\tilde{E}_{\text{$a$-FM}} = \sum_{\alpha = u,l,d,r} 
\frac{{\rm d} \, \hat{\phi}_1^{\beta}}{{\rm d} \, \phi_1^{\alpha}} \Big|_{{\boldsymbol \phi}_{\rm PM}} = 
\sum_{\alpha = u,l,d,r} 
\frac{{\rm d} \, \hat{\phi}_1^{\beta}}{{\rm d} \, p^{\alpha}} \Big|_{{\boldsymbol \phi}_{\rm PM}} = 1 \ ,
\eeq
where $\beta$ can be $u$, $d$, $l$ or $r$ (the result does not depend on $\beta$).
This condition translates into the equation:
\beq\label{eq_stab_16_plaquette}
  \begin{array}{l}
\displaystyle 
\tilde{\lambda}_3 ~ s_{c}^3 + 
\tilde{\lambda}_2 ~s_{c}^2 + \tilde{\lambda}_1 ~s_{c} +
\tilde{\lambda}_0  = 0
     \end{array}
\eeq
with 
\beq
  \begin{array}{l l}
 \displaystyle \tilde{\lambda}_0(x,y,z,u) = &    \displaystyle
 - (1 + u)^4 + x (1 + u)^3 + 2 x (x^2 + y^2 + 2 u y^2) + 
   \\ \vspace{-0.2cm} \\
 &  2 (x (1 + u)^2 + x^2 (1 + u + x) + (1 + u - x) y^2) z + (1 + u) (1 + 
    u + x) z^2
 \ , 
 \\ \vspace{-0.2cm} \\
\displaystyle \displaystyle \tilde{\lambda}_1(x,y,z,u) 
 =  &  \displaystyle 4 x^3 (1 + z) + (y^2 - x^2) (-1 + 4 u + (1 + u)^2 + (z - 1)^2 -   2 z) + 
  \\ \vspace{-0.2cm} \\
 &  \displaystyle 2 x (2 y^2 (-1 + z) + z (-u^2 + (1 + z)^2))
 \  ,
 \\ \vspace{-0.2cm} \\
  \displaystyle \tilde{\lambda}_2 (x,y,z,u)   =  & 
((u - 1)^2 - z^2) z^2 + 
 2 (x^2 + y^2) (-(1 - z)^2 + u^2 - z (1 + u )) + x (1 - u)^3  
   \\ \vspace{-0.2cm} \\
 & 
  \displaystyle
+  x \,( - 1 + 2 x^2 (1 + z) - 2 y^2 (-1 + z + 2 u) + (1 + z)^2  -
     u  z (4 + z - 2 u ))   
       \\ \vspace{-0.2cm} \\
 \displaystyle \displaystyle \tilde{\lambda}_3 (x,y,z,u) 

=  &  \displaystyle  \,  (-x^2 + y^2) ((-1 + u)^2 + z^2)  \  .
     \end{array}
\eeq
Altogether, eqs.~(\ref{eq_para16_plaquette}) and 
(\ref{eq_stab_16_plaquette}) give the value of $s_{c}$ and the critical point $a_c^{pl}$,
when the other four parameters are fixed.
The $b$-FM phase has the same properties as the $a$-FM one and 
the solution is of the
form: $p^u = -  p^l =  - p^r =  p^d = p^{\text{$b$-FM}}$, 
$s^u = s^l =  s^r =  s^d = s^{\text{$b$-FM}}$, $ q^l = q^d = - q^u =  - q^r = q^{\text{$b$-FM}}$
with $q^{\text{$b$-FM}}$ very small but different from zero unless $c=d$ or $e=0$.
If one exchanges the weights $b$ with $a$ while keeping fixed the other fugacities
the quantities above are precisely the same as for the $a$-FM phase.

\subsubsection{Ferromagnetic transition in the limit $a, b \gg c, d, e$}

\begin{figure}[h]
\centering
\includegraphics[scale=0.6]{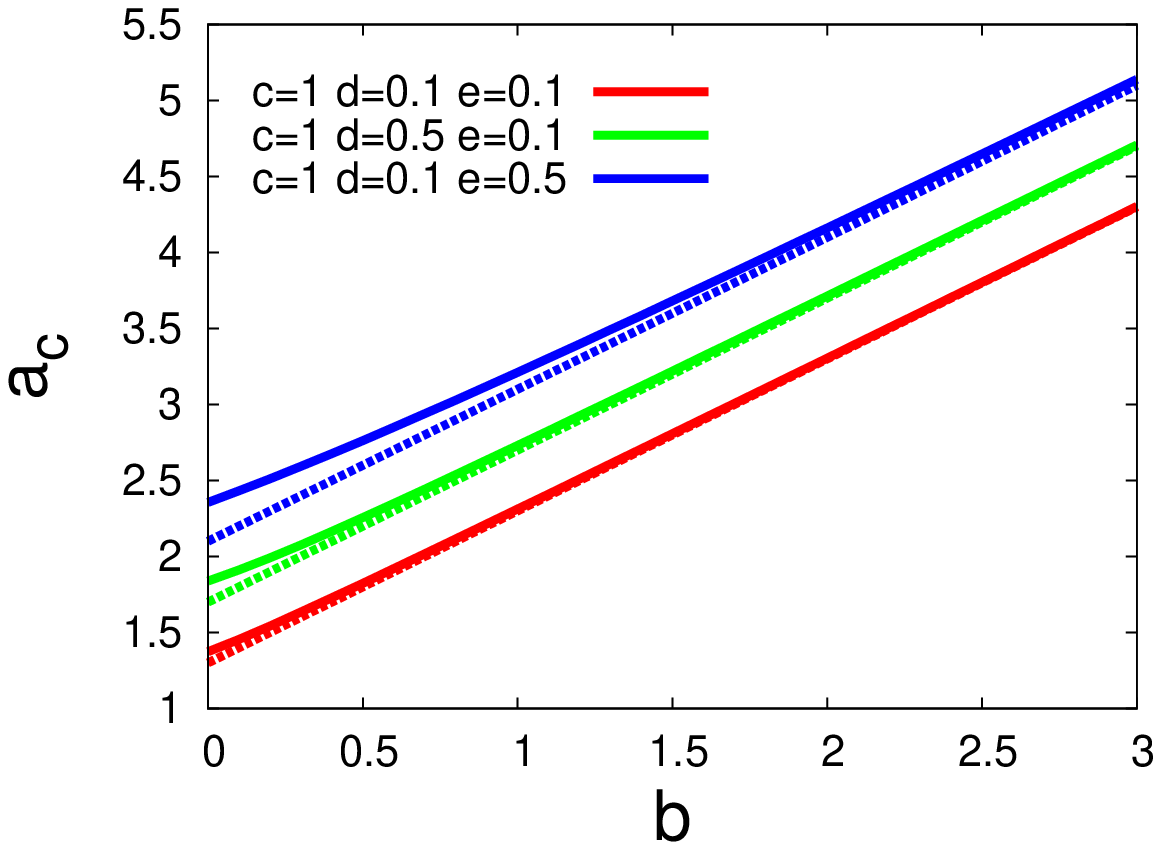}
\includegraphics[scale=0.6]{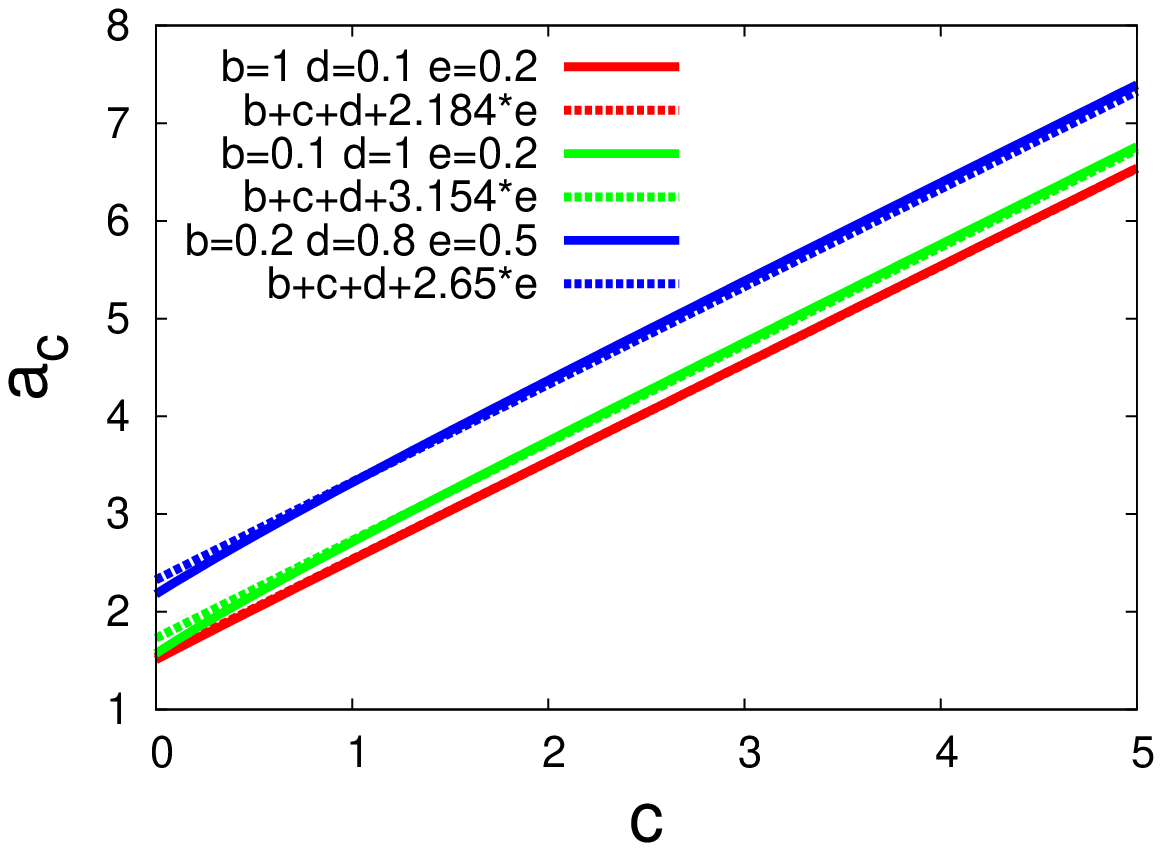}
\caption{\small Left panel: critical value $a_c^{pl}$ versus $b$ on the
tree of plaquettes for different values of the other parameters $c$, $d$
and $e$ (solid lines) compared to $a_c^{sv} = b + c + d + 2 e$ 
found on the single vertex tree (dashed lines). The plot shows that in the large $b$ limit 
the FM transition is well described by $a_c^{sv}$.
Right panel: critical value $a^{pl}_c$ versus $c$ on the tree of plaquette for different $b$, $d$, and $e$  (solid lines) 
compared to the asymptotic behavior following from eqs.~(\ref{eq_largec_1}) 
and (\ref{eq_largec_2}) (dashed lines). The variation of the prefactor multiplying $e$
shows that a linear fit in all fugacities (as for the single vertex approximation) does not work in this regime.
Note that such non linear effects can also be seen in the left panel, in the small $b$ regime,
where $c$ is the dominating weight.
The agreement with the asymptotic limit improves for increasing $c$.
This proves that the behavior $a_c^{pl} = c + \tau_{ac}(b,d,e)$ is justified.
}
\label{Fig_asymp_16vertex_plaq}
\end{figure}

Let us focus on the FM transition at large values of $a$
for the model on the tree of plaquettes.
In order to extract some analytic behavior we will focus now on the limit in 
which one of the four other fugacities is large
and, in particular, in the regime $a, b \gg c, d, e$ or $a, c \gg c, d, e$. 
This situation corresponds to the cases where two types of vertices have an energy 
much smaller than the others. 
In order to study the large $b$ behavior, on the
basis of the numerical results we take $a$
to be of the form $a^{pl}_c = b + \tau_{ab}$, where $\tau_{ab}(c,d,e)$ is some (small)
function of the remaining parameters. 
In this limit one has $x \simeq \frac{\tau_{ab}}{2 b}$, $y \simeq \frac{c - d}{2 b}$,
$z \simeq 1 - \frac {c + d}{2 b}$ and $u \simeq \frac{2 e}{b}$.
In the infinite $b$ limit eq.~(\ref{eq_para16_plaquette}) admits as 
the only acceptable solution $s_{\rm PM} = s_c = 0$, which does not depend on the parameters.
This is also confirmed numerically, where one sees that in the limit 
$a, b \gg c, d, e$ the value of $s_c$ at
the transition point goes to zero. The same result is obtained in 
the eight-vertex model in the same limit.
We then linearize eq.~(\ref{eq_para16_plaquette}) in $s_c$ and we 
expand the solution at leading order in $1/b$.
In conclusion, we find
\beq\label{Eq_asym_scr}
 \begin{array}{ll}
 \displaystyle s_c 
 \simeq  \frac{ \tau_{ab}^2 - (c - d)^2}{8 (c + d + 2 e) b} \ .
      \end{array}
\eeq
Plugging this result into eq.~(\ref{eq_stab_16_plaquette}) and taking the leading order
in $1/b$ one obtains $\tau_{ab}(c,d,e) = c + d + 2 e$, which implies
\beq
a_c^{pl} = a_c^{sv} = b + c + d + 2 e \qquad \qquad \qquad \text{for } a, b \gg c, d, e ,
\eeq
as for the single vertex tree. In the left panel of Fig.~\ref{Fig_asymp_16vertex_plaq} we show 
the comparison between the exact $a$-FM transition point for the plaquette (solid lines) and 
the one predicted by the single vertex (dashed lines), for different values
of the vertex weights $c, d$ and $e$. In the limit $a, b \gg c, d, e$ the two results turn out
to be remarkably close.

\subsubsection{Ferromagnetic transition in the limit $a, c \gg b, d, e$}

This limit is more involved and shows a different behavior from the one found above for both $a$ and $b$ large.
Supported by the numerical results we assume that the
critical $a_c$ scales as $a_c^{pl} = c + \tau_{ac}$, with $\tau_{ac}(b,d,e)$ some function of the other 
parameters. In this limit   $x \simeq \frac{1}{2} - \frac{3 b + d - \tau_{ac}}{4 c}$, 
$y \simeq \frac{1}{2} - \frac{b + 3 d + \tau_{ac}}{4 c}$,
$z \simeq \frac{b - d + \tau_{ac}}{2 c}$ and $u \simeq \frac{2 e}{c}$.
Plugging these results in eq.~(\ref{eq_para16_plaquette}) 
one sees that in limit of large $c$, differently from above, $s_c$
does not go to zero but it converges towards a finite value which has to satisfy the relation 
\beq\label{Eq_largec_eqpara}
b - d - \tau_{ac} + 4 (b + d + 4 e) s_c - 4 (b + d - 4 e) s_c^3 + (-b + d + \tau_{ac}) s_c^4 = 0 \ .
\eeq
Even though $u$ goes to zero with $c$, the solution to this equation is not 
the same as the one of the eight-vertex model in the limit of large $c$ and $a^{pl}_c = c + \tau_{ac}$, 
since an explicit dependence on $e$ remains. 
The large $c$ limit in eq.~(\ref{eq_stab_16_plaquette}) yields
\beq\label{Eq_largec_eqstab}
(-3 b - 5 d - 8 e + 3 \tau_{ac}) + (b - d + 3 \tau_{ac}) s_c + (b - d - 8 e + 3 \tau_{ac}) s_c^2 + (b - d - \tau_{ac}) s_c^3 = 0 \ .
\eeq
We can now use eq.~(\ref{Eq_largec_eqpara}) to determine $\tau_{ac}$, obtaining
\beq\label{eq_largec_1}
\tau_{ac} = b - d + \frac{16 \, e \, s_c}{1 - s_c^2} + \frac{4 \, (b + d) \, s_c}{1 + s_c^2}
\eeq
which substituted in~(\ref{Eq_largec_eqstab}) gives an equation for $s_c$:
\beq\label{eq_largec_2}
2 b + d - 5 e - \frac{8 e}{-1 + s_c} + (2 b + d - 6 e) s_c + (-d + e) s_c^2 - 
 \frac{4 e}{1 + s_c} - \frac{2 (b + d)}{1 + s_c^2} = 0 \ .
\eeq
Both equations~(\ref{eq_largec_1}) and (\ref{eq_largec_2}) are linear in the
parameters $b$, $d$ and $e$; instead,  the function $\tau_{ac}$ will generically show
a non linear behavior, brought about by the non trivial value of $s_c(b,d,e)$, which is the
asymptotic result in the large $c$ limit. In the right panel of Fig.~\ref{Fig_asymp_16vertex_plaq}
we show the critical lines for the $a$-FM transition obtained for the plaquette
(solid lines) and from the asymptotic expressions extracted from Eqs.~(\ref{eq_largec_1}) and (\ref{eq_largec_2})
for different values of $b, d$ and $e$ (dashed lines).

Let us finally mention that another result can be obtained  
when $b=c=d=0$ and $e>0$. In this limit we take $a_c^{pl} = \tau_{ae} e$ and 
solving eqs.~(\ref{eq_para16_plaquette}) and~(\ref{eq_stab_16_plaquette})   
one finds $\tau_{ae} \simeq 2.34$ and $s_c \simeq 0.144$.
Apart from this ``extreme" case  we did not explore the large $e$ regime.

\subsubsection{Antiferromagnetic phases and transitions}

The AF phases are characterized by similar solutions. In particular 
the phase dominated by vertices of type $c$ is described by the fixed point
$p^u = p^l = - p^r =  - p^d = p^{\text{$c$-AF}}$,
$s^u = s^l =  s^r =  s^d = s^{\text{$c$-AF}}$, and  
$q^u = q^l = q^r = q^d = q^{\text{$c$-AF}}$,
with $p^{\text{$c$-AF}}$ very small but different from zero, unless $a=b$ or $e=0$.
As we discussed for the FM transition, in order to 
identify the point of instability one can disregard
the component along $p$ as it is negligible for the AF transitions 
and study the quantity
\beq\label{Stab_16_pl_AF}
\tilde{E}_{\text{$c$-FM}} = \sum_{\alpha = u,l,d,r} 
\frac{{\rm d} \, \hat{\phi}_3^{\beta}}{{\rm d} \, \phi_3^{\alpha}} \Big|_{{\boldsymbol \phi}_{\rm PM}} = 
\sum_{\alpha = u,l,d,r} 
\frac{{\rm d} \, \hat{\phi}_3^{\beta}}{{\rm d} \, q^{\alpha}} \Big|_{{\boldsymbol \phi}_{\rm PM}} = 1 \ ,
\eeq
where $\beta$ can be $u$, $d$, $l$ or $r$. Equation~(\ref{Stab_16_pl_AF}) can be obtained
from~eq.~(\ref{eq_stab_16_plaquette}) by exchanging $a$ with $c$ and $b$ with $d$
and sending $s$ to $- s$.

The same results discussed for the critical FM surfaces 
hold when one considers an AF transition (for instance the 
one at large $c$). In this case the critical value $c_c^{pl}$ obtained
with the plaquette coincides with the
one of single vertex in the limit in which $c, d \gg a, b ,e$ while the case $c, a \gg b, d, e$ shows
more subtle non-linearities. This is in fact the case for the AF transition 
lines reported in Fig.~\ref{fig:phase_diagram-16vertex}.

\subsection{The sixteen-vertex model: phase diagram and discussion}

\begin{figure}[h!]
\centering
\includegraphics[scale=0.5,angle=-90]{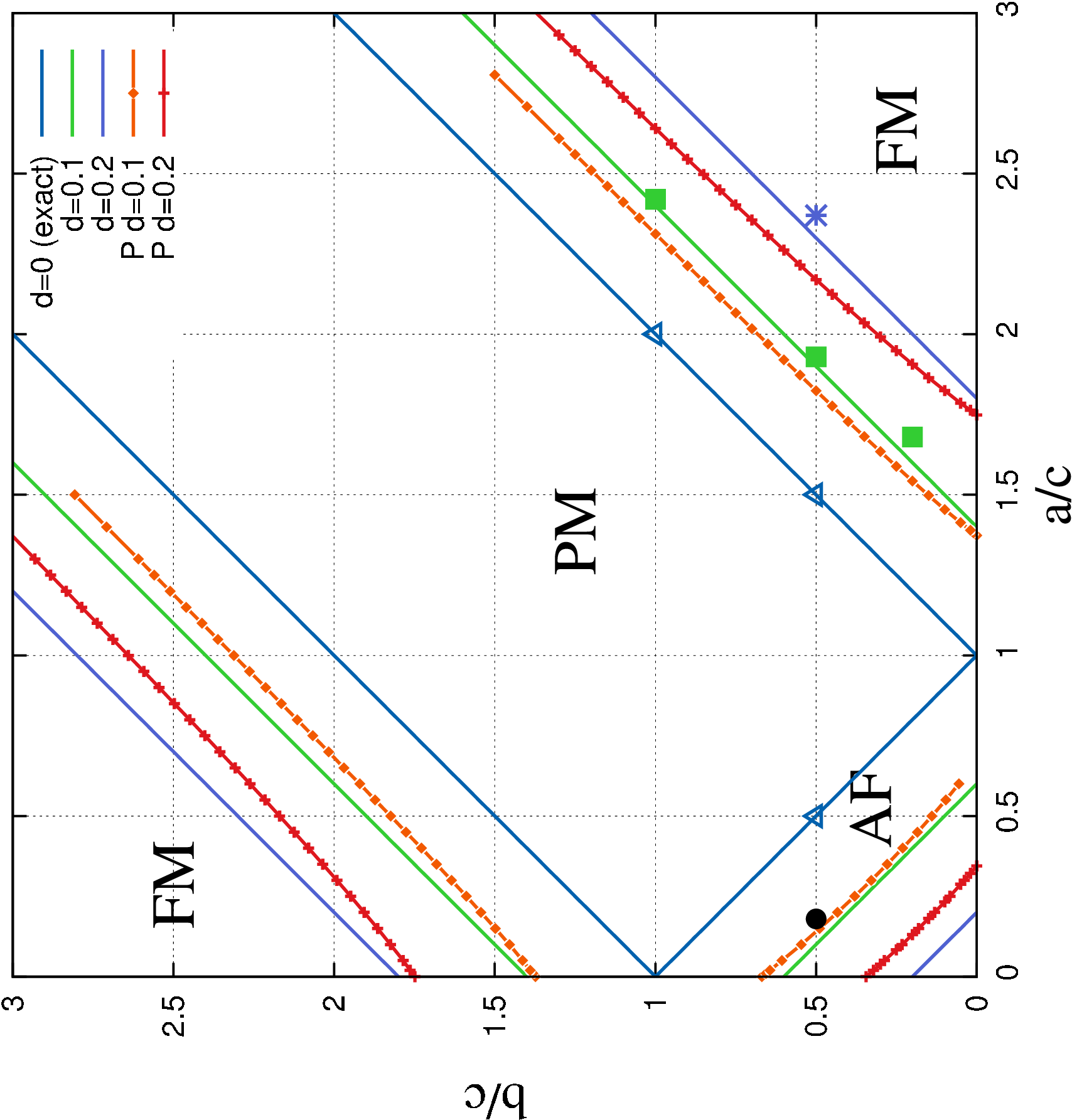}
\caption{\small Phase diagram of the sixteen-vertex model. The figure shows the projection of the transition surfaces 
on the plane of parameters
$(a/c,b/c)$, for two fixed values of $d/c=e/c=0.1$ and $d/c=e/c=0.2$. 
Orange and red dotted lines represent the results 
obtained with the cavity method for the tree of
plaquettes.  The critical lines obtained with the tree of single vertices 
are not shown but they are in very good agreement
with the results obtained with the plaquette for the $a$-FM and $b$-FM transitions in the regime $a,b \gg c, d, e$.
Green and violet  plain lines shows the proposed behavior of the transition lines
predicted by eq.~(\ref{Delta16_2D_model}) for the $2D$ model. Blue solid lines
correspond to the exact phase diagram in the limiting case $d=e=0$ (six-vertex model).
The same code of colors is used to indicate the transition points obtained with MC simulations for the
model defined on  the square lattice. In particular green squares are for $d/c=e/c=0.1$, 
the violet star for $d/c=e/c=0.2$ and blue triangles for $d=e=0$. 
We also indicate with a black dot a $c$-AF transition point
obtained with $d/c=e/c=0.05$.
}
\label{fig:phase_diagram-16vertex}
\end{figure}

In Fig.~\ref{fig:phase_diagram-16vertex} 
we plot the projection of the transition lines and the resulting phase diagram of the sixteen-vertex model 
on the plane $(a/c,b/c)$, for two fixed values
of $d/c=e/c$, showing the $a$-FM, the $b$-FM, the $c$-AF and the PM phases. 
The transition lines within the plaquette approximation can be either 
extracted from the analysis of the stability matrix~(\ref{Matrix_stab_plaq})
or one can directly detect the point where the self-consistent equations develop a symmetry breaking  
solution.
The results obtained with  the tree of plaquettes  are shown with 
 orange ($d/c=e/c=0.1$) and red ($d/c=e/c=0.2$) dotted lines; the transition lines
obtained according to the guessed value of the {anisotropy} parameter given in eq.~(\ref{Delta16_2D_model}) 
are drawn with green ($d/c=e/c=0.1$) and violet ($d/c=e/c=0.2$) straight lines while
the exact phase diagram of the six-vertex model for $d=e=0$ is shown with blue solid lines. 
The same code of color -- green, violet and blue -- is used to represent with squares, stars and triangles 
the results obtained by MC simulations
on the $2D$ square lattice. In addition we also indicate with a black dot an extra point  obtained by MC
simulations for the $c$-AF transition with $d/c=e/c=0.05$.
The results obtained with the single vertex are not shown in Fig.~\ref{fig:phase_diagram-16vertex}
however the $a$-FM and $b$-FM transitions lines found with the plaquette are in very good agreement
with those of the single vertex in the regime $a, b \gg c, d, e$.

The phase diagram of the model defined on the tree made of single vertices and of plaquettes 
is in qualitative agreement with the MC simulations on the $2D$ square lattice.
All the transitions are  
continuous when all possible vertices are present. 

The results on the tree of vertices predict a uniform
shift  of the critical lines, with respect to the eight-vertex model, 
by $2 e$.\footnote{This is obtained by comparing eq.~(\ref{stability_16_single_vertex})
with the criticality condition $|E_{\alpha}|= 1$.}  This implies, for instance, 
that the critical value of the $a$-FM transition occurs at $a_c^{sv} = b+c+d+2 e$.
This does not reproduce exactly the numerical results for the $2D$ model on the square lattice.
While for small values of $d$ and $e$, one numerically finds $a_c^{2D} \simeq b + c + d + 3 e$ 
(the results reported in~\cite{Levis2012}), 
for large values of these fugacities we see a deviation from the linear behavior. 
We do not have an analytic expression for the transition surfaces for the
tree of plaquettes but they can be found from the numerical solution of the self-consistent equations
with arbitrary precisions and their asymptotic expressions can be obtained in some particular limits. 
In particular, when $a, b \gg c, d, e$ for the FM transitions and when $c, d \gg a, b, e$
for the AF ones, one recovers the results of the single vertex. It would be interesting to see
whether asymptotically in the same regime a similar result holds for the $2D$ lattice. 
Other cases are generically accompanied by (small) non-linearities
in the fugacities. Comparing our analytical approaches with the MC data, we conclude that for the FM transition one finds  
$a_c^{sv} < a_c^{pl} < a_c^{2D}$ (where the superscript $sv$, $pl$ and $2D$ 
stand for single vertex, plaquette and $2D$ model respectively). Similar results hold for the other transitions.

\subsection{Fluctuations in the symmetry-broken phases}\label{Section:fluctuations}

In this subsection we discuss the nature of thermal fluctuations in the symmetry broken phases
of the sixteen-vertex model and we compare them to the ones found in the PM phase.
For simplicity we focus on the $a$-FM phase only, since the observations reported in the following
can be extended to the other ordered phases, as we show in the figures.
As we already mentioned, the solution ${\boldsymbol \phi}_{a-\rm FM}^{\alpha}$  
within the plaquette approximation is characterized 
by FM order signaled by a finite value of $p_{a-\rm FM}$ and by a small but non-vanishing
value of the AF field $q_{a-\rm FM}$. This field has not the same sign along all the
 directions, and overall it does not lead to a staggered magnetization. Moreover, it decreases upon
 increasing the FM order. Interestingly enough, such small AF field plays a crucial 
 role for thermal fluctuations in the symmetry broken phase.
 
\begin{figure}[h]
\centering
\includegraphics[scale=0.4]{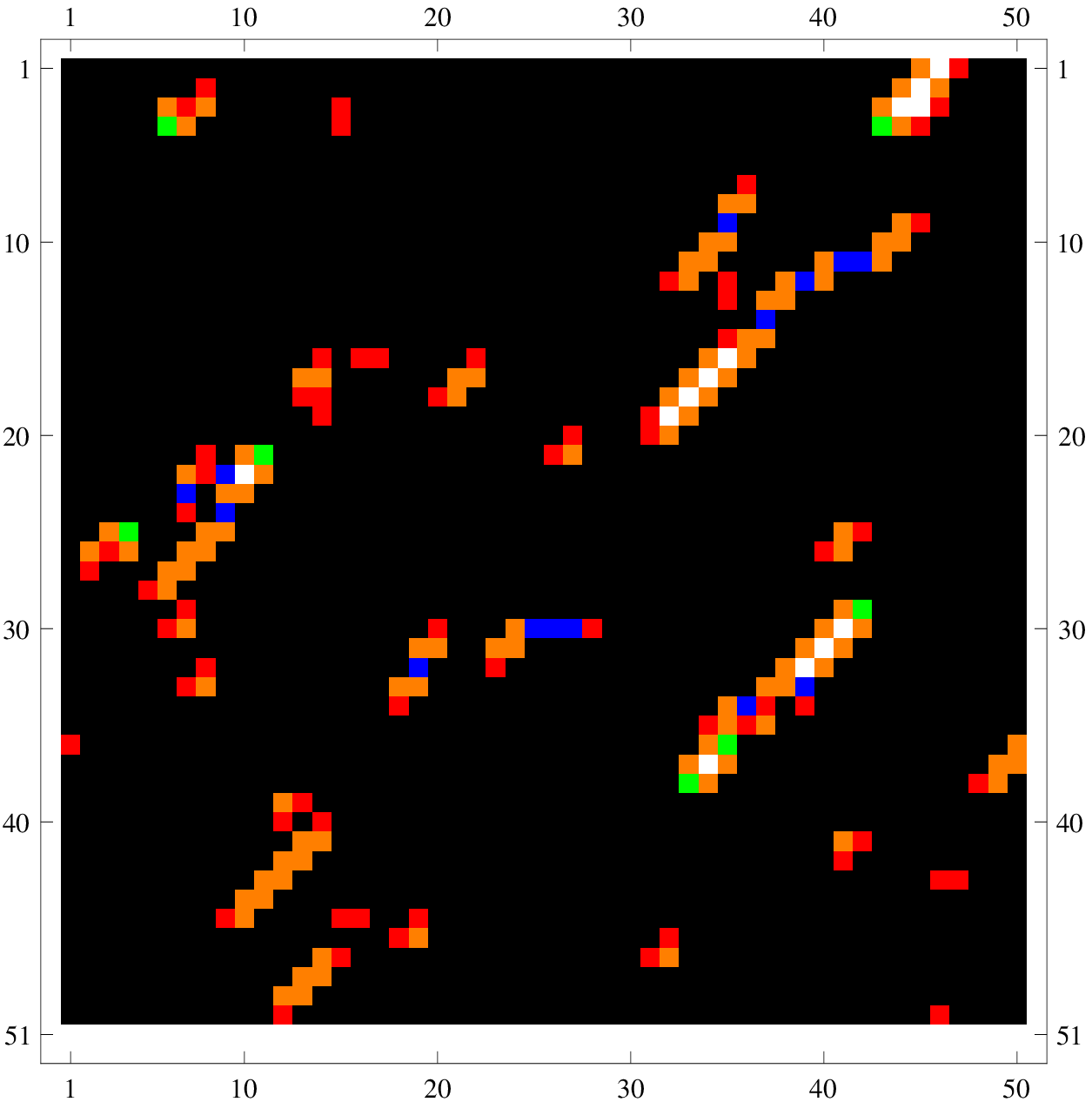}
\includegraphics[scale=0.4]{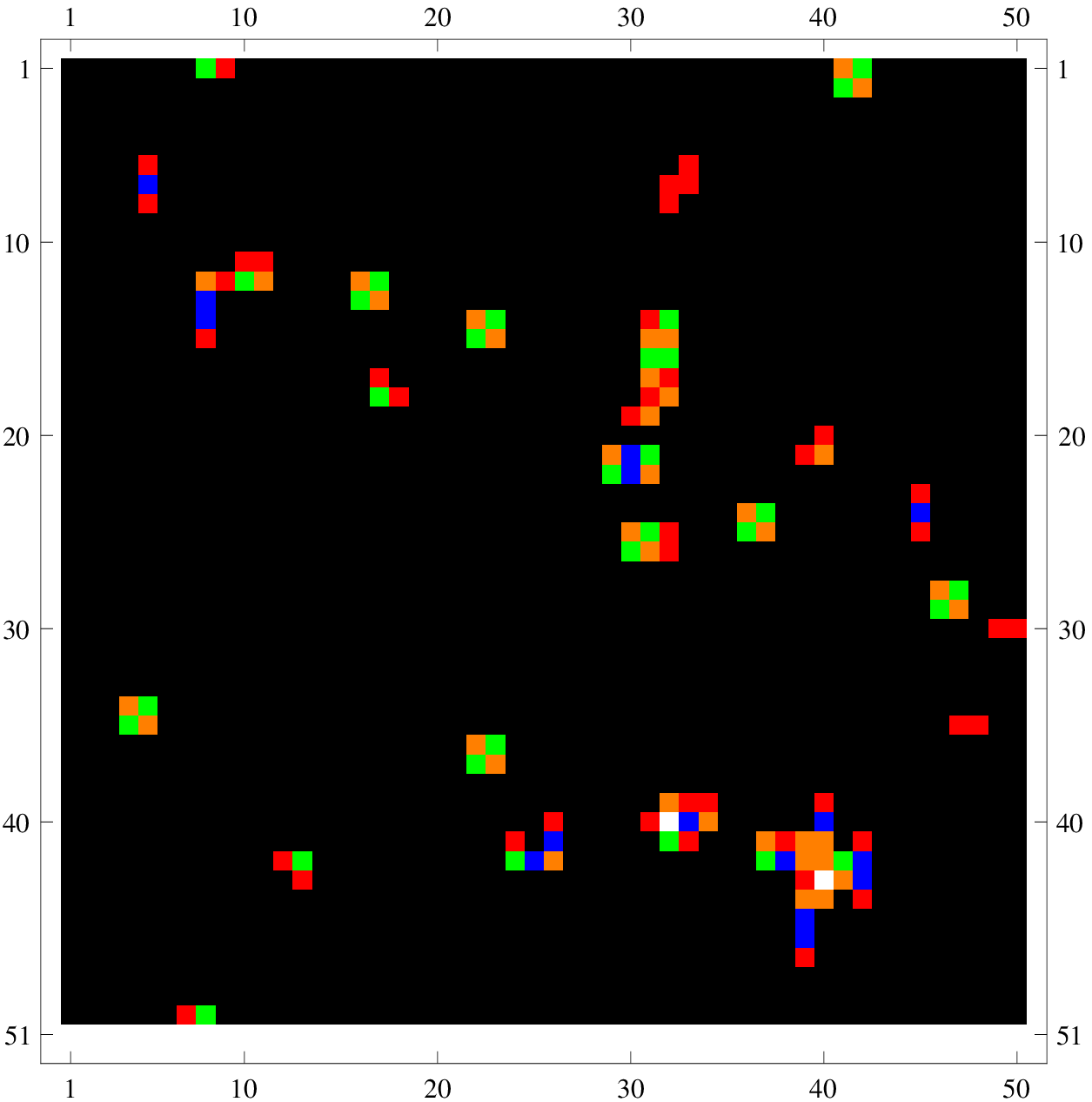}
\includegraphics[scale=0.4]{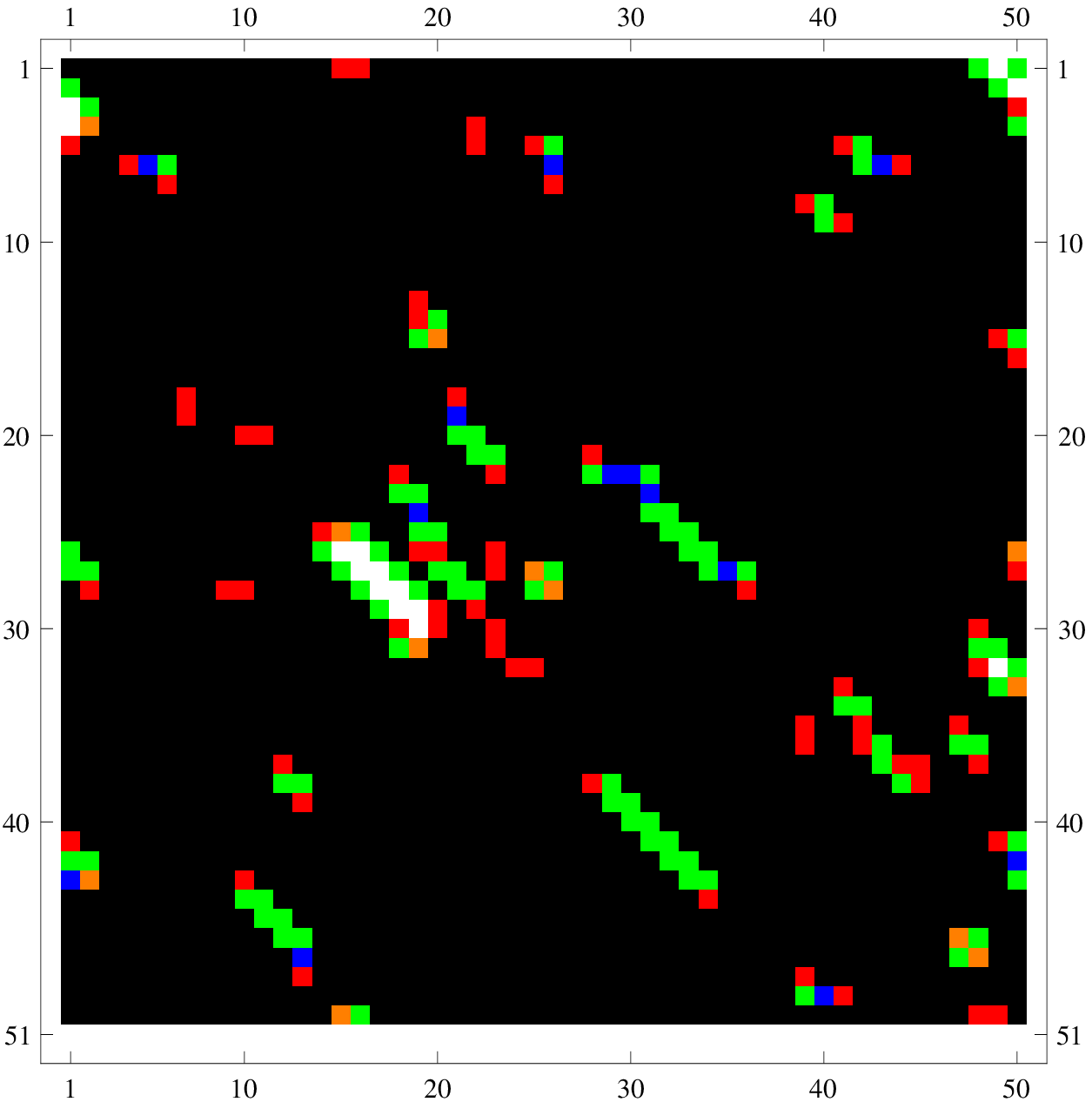}
\caption{\small MC $a$-FM equilibrium configurations of the $2D$ sixteen-vertex model for
different values of the ratio $c/d$, showing that fluctuations are anisotropic
for $c/d \neq 1$. The color code is the one used in Figs.~\ref{fig:eight-vertex} 
and~\ref{fig:sixteen-vertex}, blue is for vertices $b$, orange for $c$, green for $d$
and red for $e$. FM vertices of type $a$ which are 
colored in black ($v_1$) and white ($v_2$). 
 The size of the system is $L^2 = 50^2$.
Left panel:  $a=2.5$, $b=0.3$, $c=1.5>d=e=0.1$. Linear fluctuations made of $c$ vertices
extend along the down-left $\leftrightarrows$ up-right  diagonal direction.
Central panel: $a=2.5$, $b  = c = d =0.5$, $e=0.1$. Fluctuations are homogeneous
with no preferential direction.
Right panel: $a=2.5$, $b=0.3$, $d=1.5>c=e=0.1$. Linear fluctuations extend along the 
down-right $\leftrightarrows$ up-left diagonal and are  made of $d$ vertices.}
\label{fig:fluct-16vertex}
\end{figure}

 In  Fig.~\ref{fig:fluct-16vertex} we show some equilibrium
MC  configurations in the $a$-FM phase on the 2D lattice, 
 for different values of the ratio $c/d$.
The comparison between the three snapshots 
clearly shows that vertices $c$ and $d$ compete to form fluctuations
along opposite diagonal directions (in the left panel $c>d$ while in the right panel $c<d$) 
and that isotropy is recovered when their statistical weights are the same (central panel).
One can also find fluctuations of $e$ vertices alone, which are generically unstructured,
or of $b$ vertices, that consist in straight vertical and horizontal lines of defects 
with the same probability. In the following we concentrate on diagonal defects which are the origin
of anisotropic fluctuations, and which are related to the small AF field found in the mean-field solution
of the model on the tree of 
plaquettes.

\begin{figure}[h!]
\centering
\includegraphics[scale=1.25]{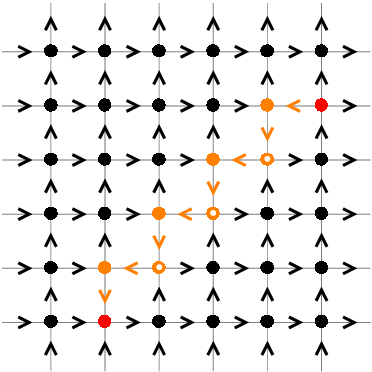}
\hspace{0.1cm}
\includegraphics[scale=1.25]{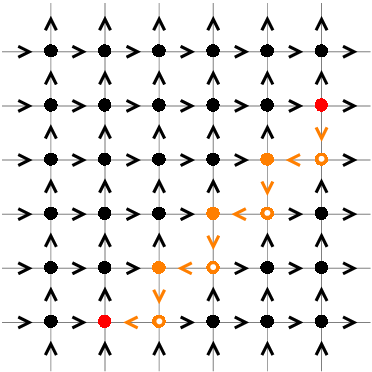}
\hspace{0.1cm}
\includegraphics[scale=1.25]{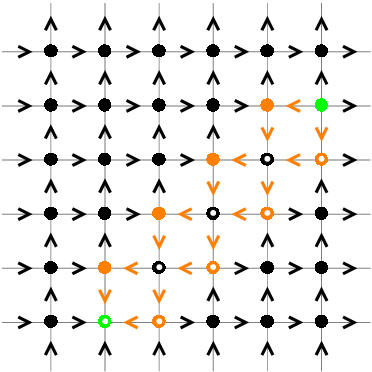}
\caption{\small (Color online.) Anisotropic fluctuations in the $a$-FM phase of the $2D$ sixteen-vertex model 
with $c>d$. We color the vertices according to the code used in Figs.~\ref{fig:eight-vertex}
and \ref{fig:sixteen-vertex}. We distinguish the
two vertices connected by arrow reversal symmetry by drawing a white dot in the middle. 
Arrows carry the color of the vertices in the background if they are compatible with the 
ordered phase, or the color of the neighboring vertices if they have an opposite orientation.
The first and second panel show two staircase lines of defects 
made by an alternating sequence of $c$ vertices (orange dots) 
between two vertices of type $e$ (red dots). In terms 
of arrows such defects can be viewed as a one dimensional domain.
The staircases can also be seen as double diagonal string of vertices 
along the down-left $\leftrightarrows$ up-right direction with $v_5$ above $v_6$ 
in both cases.
The third panel shows a thicker defect string. The AF vertices 
of type $c$ with opposite orientation
are now split apart and constitute the two boundaries of a domain containing FM
vertices with opposite orientation with respect to the background. 
The two extremities of the domain are here sealed 
by two opposite $d$-AF vertices (green dots). In the sixteen-vertex model
more complicated structures can also be found.
Note that if the symmetry were broken in favor of vertices $v_2$ one would find the same
type of fluctuations, with $v_5$ and $v_6$ exchanged.}
\label{fig:line_defect-FM}
\end{figure}
In the $a$-FM phase, when $c>d$, fluctuations constituted of diagonal strings of defects of $c$ vertices 
grow along the diagonal running from the up-right corner to
the down-left one. They can be mainly of two types: either
one-dimensional ``stairs" of arrows going in the opposite direction with respect 
to the ordered background and  extending between two vertices of type $e$; or 
thicker defects containing  domains of FM vertices with the opposite
magnetization with $c$ vertices on the boundary (see Figs.~\ref{fig:fluct-16vertex} and \ref{fig:line_defect-FM}).
As can be seen from Fig.~\ref{fig:line_defect-FM}, in both cases vertices of type $v_5$ are above vertices of type $v_6$
with respect to the diagonal axis that joins the two extremities
of the elongated defect. Reversing all the spins of the background (this corresponds to the situation
in which the opposite FM
order is the dominating minimum) amounts to invert the role of $v_5$ and $v_6$.
This implies that the breaking of reversal symmetry, accompanied by the selection of a given order, is also accompanied by
its own characteristic fluctuations, and by the selection of one particular pattern of AF vertices
among the two equivalent ones by arrow reversal.
This does not imply an overall staggered magnetization since the defects strings correspond to patches of
the two opposite AF configurations with equal probability (as in the first two upper panels in Fig.~\ref{fig:line_defect-FM}). 
When $d>c$ one has the very same scenario, with fluctuations organized in the
perpendicular direction and made of sequences of $v_7$ and $v_8$ vertices.
A similar phenomenon leading to striped domains takes place 
also out-of-equilibrium for the coarsening dynamics in the $a$-FM phase~\cite{Levis2012}.

\begin{figure}[h!]
\vspace{0.25cm}
\centering
\includegraphics[scale=1.25]{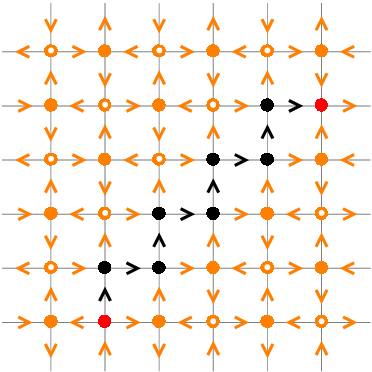}
\hspace{0.1cm}
\includegraphics[scale=1.25]{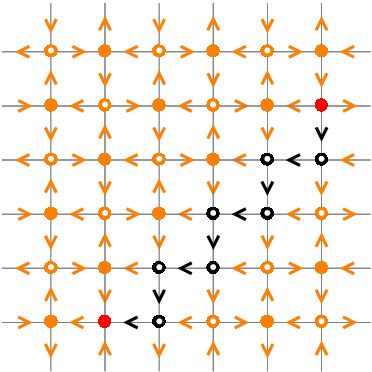}
\hspace{0.1cm}
\includegraphics[scale=1.25]{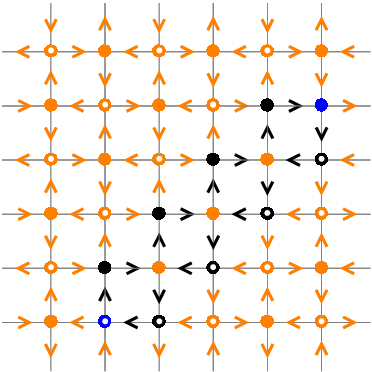}
\caption{\small Anisotropic fluctuations in the $c$-AF phase of the $2D$ sixteen-vertex model. 
The diagonal fluctuations are now driven by $a$ vertices (for $a>b$).
The first and the second panel show staircases made of $v_1$ or $v_2$ vertices and ending on $e$ vertices.  
On the diagonal joining the two vertices $e$ there can be both of them, 
while away from it, $v_1$ and $v_2$ occupy well defined positions
which depend on the ordered phase of the background 
(equivalently, in the AF case the positions of $v_1$ or $v_2$ vertices depend on which of 
the two sub-lattices the diagonal strings belong).
The third panel shows an example of a more spread domain,
analogue to the one presented in the third panel in Fig.~\ref{fig:line_defect-FM},  
with FM vertices on the boundary and AF ones in the interior.
}
\label{fig:line_defect-AF}
\end{figure}
Such anisotropic fluctuations cause the loss of the homogeneity of the solution
${\boldsymbol \phi}_{a-\rm FM}^{\alpha}$  
in the two directions. Isotropy is recovered in the case $c=d$, when there is
no preferred axis, as shown by the MC snapshots (correspondingly, 
the field $q_{a-\rm FM}$ vanishes in this case). 

\begin{figure}[h]
\centering
\includegraphics[scale=1]{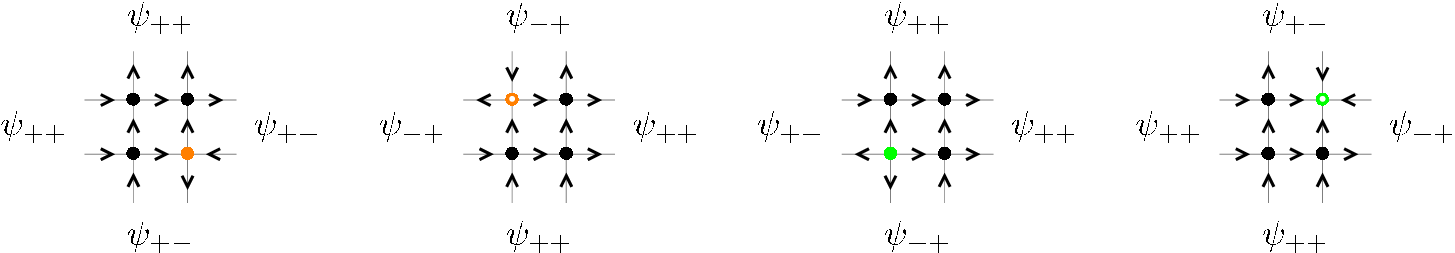}
\caption{\small (Color online.) Proposal for the anisotropic configurations of defects on the plaquette that contribute
the most to create an effective (small) AF field $q_{a-\rm FM}$ in the
$a$-FM solution. Black dots represent FM vertices associated to the 
symmetry broken phase, orange and green dots are AF vertices,
of type $c$ and $d$, that correspond to fluctuations.}
\label{fig:fluct-16vertex-plaquette}
\end{figure}

The AF field observed in the mean-field solution of the tree of plaquettes in the $a$-FM phase, 
is the signal of such fluctuations that
are induced by the anisotropy and the symmetry breaking characteristic of the
ordered phase. 
The configurations that contribute the most are those in which one vertex of type $c$ is sitting
in the down-right corner of the plaquette or the opposite vertex $c$ is in the upper-left corner, 
together with other 
three vertices of type $a$ (see Fig.~\ref{fig:fluct-16vertex-plaquette}).  

\begin{figure}[h]
\centering
\includegraphics[scale=0.4]{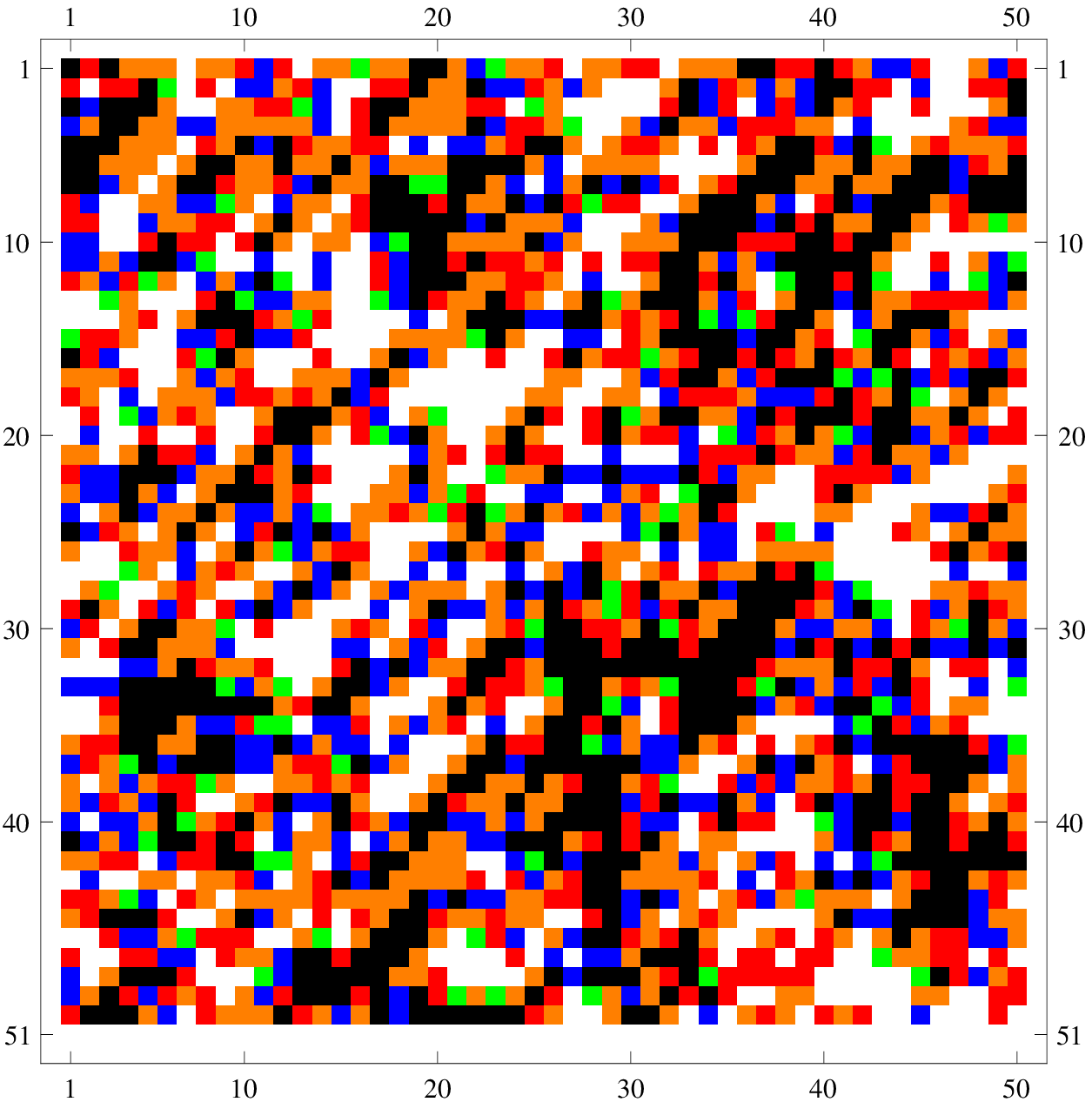}
\includegraphics[scale=0.41]{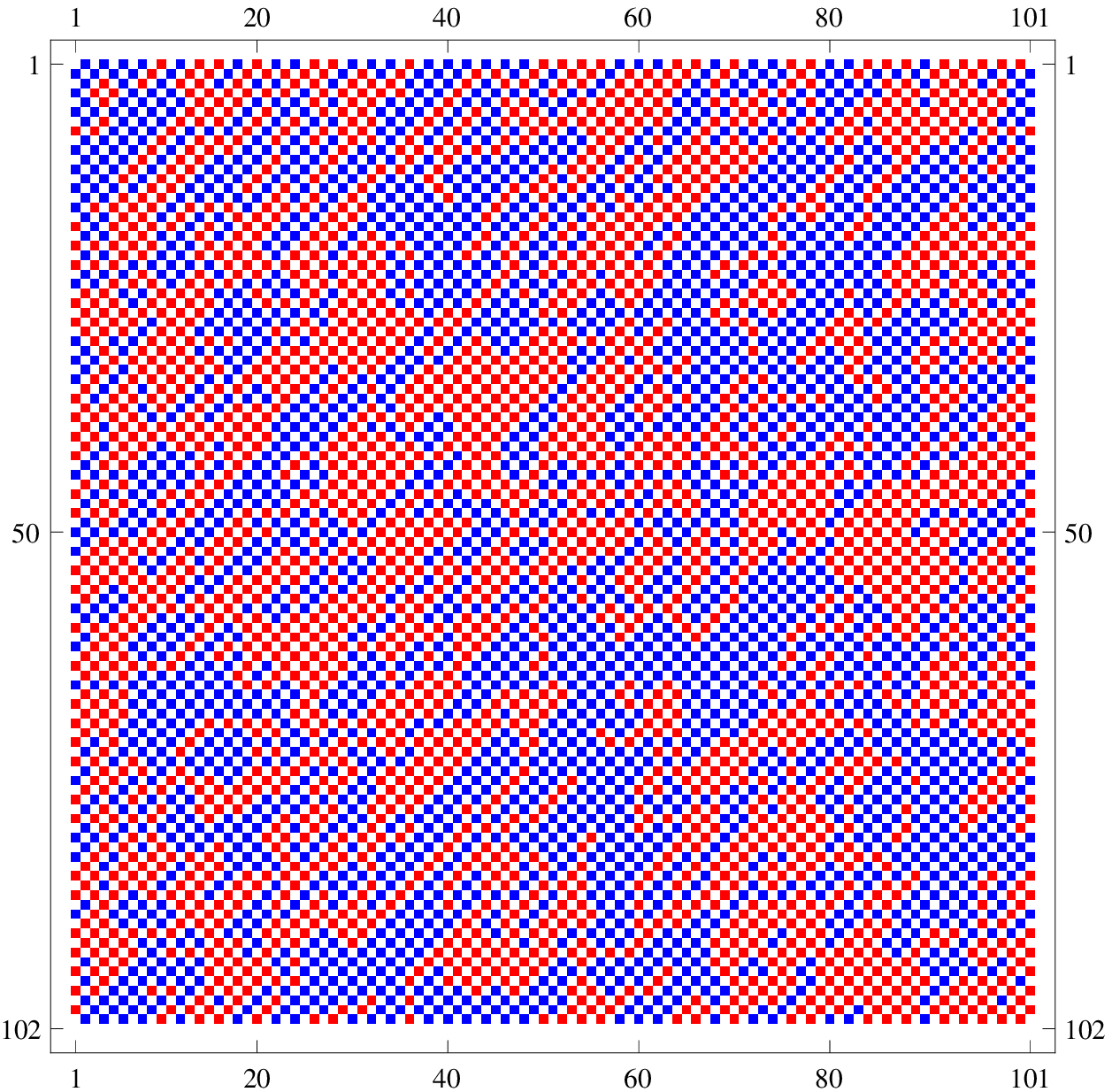}
\caption{\small (Color online.) MC equilibrium PM configurations in the $2D$ sixteen-vertex model with  
$a=2$, $b=0.7$, $c=1.2$, $d = e = 0.2$ and $L=50$. The left panel shows the lattice colored
according to the vertex configurations and following the convention used in Figs.~\ref{fig:eight-vertex}, \ref{fig:sixteen-vertex},
and~\ref{fig:fluct-16vertex}, where we colored $v_1$ vertices in black and $v_2$ in white (note that
these vertices are the ones 
with the largest weight). The right panel shows the same configuration where we colored
the arrow sites, in blue if $s_{ij}=+1$ and in red otherwise. The two figures demonstrate that  
the PM phase is anisotropic as well. For $a>c>d$ it is more likely to have boundaries between the domains of 
positive and negative arrows along the diagonal that joins the upper-right corner with the down-left one. Such 
boundaries are formed by $c$ vertices (in orange in the left panel). }
\label{fig:16vertex-conf-Para}
\end{figure}

At the level of the tree of single vertices such anisotropy can be detected by looking 
at the derivatives ${\rm d} \psi^{\alpha} / {\rm d} \psi^{\beta} |_{\boldsymbol{\psi}_{\text{$a$-FM}}}$, 
which take different values depending on $\alpha$ and $\beta$ and which 
ultimately determine the susceptibility along certain paths. 
In particular, the ``diagonal" derivatives ${\rm d} \psi^u / {\rm d} \psi^r$ and ${\rm d} \psi^u / {\rm d} \psi^l$
become equal only when $c=d$, while for $c>d$ one has 
${\rm d} \psi^u  / {\rm d} \psi^r > {\rm d} \psi^u / {\rm d} \psi^l$.
Such derivatives are generically different along different directions also
when evaluated in the PM solution signaling 
that also the disordered phase is anisotropic (apart from some particular choices of the
parameters), as shown in Fig.~\ref{fig:16vertex-conf-Para}.

 In this case,  if $a>b,c,d,e$ it will be more likely to have vertices of type $a$,
 albeit there is no spin reversal symmetry and a finite global magnetization since vertices $v_1$ and $v_2$ appear
 with the same frequency.
Therefore, the diagonal fluctuations at the boundary  and within these domains 
can be either the ones typical of domains of $v_1$ vertices or those found in domains of $v_2$ vertices.  
All AF patterns (and the AF boundaries in
Fig.~\ref{fig:fluct-16vertex-plaquette}) are equally probable and    
the position of $v_5$ or $v_6$ can be interchanged without any energy cost.

\section{Summary and conclusions}
 
Let us start by briefly reviewing the main results found in this paper.

Table~\ref{table:six-eight} summarizes the phase transitions found in the 
six- and eight-vertex models on the tree of single vertices and plaquettes. The entries
in blue correspond to the transitions which are of the
same order on the Bethe lattice and on the $2D$ square lattice, 
while red entries correspond to the transitions that are of different type.

\begin{table}[ht]
\centering
\begin{tabular}{|c|c|c|c|c|}
\hline
 & 6V single vertices & 8V single vertices & 6V plaquettes & 8V plaquettes \tabularnewline
\hline 
\hline 
PM-FMs  & \textcolor{blue}{Frozen} - to - \textcolor{blue}{sPM} & \textcolor{red}{Frozen} - to - \textcolor{blue}{PM}  &  \textcolor{blue}{Frozen} - to - \textcolor{blue}{sPM}  & 
\textcolor{red}{Discont} - to - \textcolor{blue}{PM} \tabularnewline
\hline 
PM-AFs & \textcolor{red}{Frozen}-to-\textcolor{blue}{sPM} & \textcolor{red}{Frozen}-to-\textcolor{blue}{PM} &  \textcolor{red}{Cont.}-to-\textcolor{blue}{sPM} &\textcolor{red}{Discont.}-to-\textcolor{blue}{PM}
\tabularnewline
\hline 
\end{tabular}
\vspace{0.4cm}
\caption{\small Transitions in the six- and eight-vertex models on tree-like graphs.  
We call sPM the (critical) PM phases with a maximal eigenvalue of the stability matrix identical to one, which are
reminiscent of the critical (Coulomb) PM phases in $2D$.
The comparison with the behavior of $2D$ models on the square lattice 
is encoded by the following color rule: the entries of the table are colored in blue whenever the transitions  
of the $2D$ model and on the Bethe lattice are of the same kind;
contrarily, we use
red when  the transition of the models on the tree are not of the same type as the $2D$ one. 
We recall that in $2D$  the PM-AFs transition in the six-vertex model is of Kosterlitz-Thouless type while 
it is second order  in the eight-vertex model.
We stress that the plaquette model provides an important improvement compared to the single vertex 
approach as, for instance, it is able to capture the anisotropic 
fluctuations in the ordered phases. }
\label{table:six-eight}
\end{table}

In the $2D$ six-vertex model the
disordered phase is critical (a SL), the FMs are frozen and the AF phase has fluctuations. 
In the tree of single vertices
the PM phase (that we call sPM in the table) is similar to a critical phase in the sense that one of the eigenvalues
of the stability matrix is identical to one, yielding a diverging susceptibility.\footnote{Models defined 
on a tree cannot have a power-law decay of the correlation functions as in the critical phases of 
finite dimensional systems, due to the fact that the number of neighbors of a given site at a given distance grows exponentially
with the distance itself.} 
The FMs and AF phases are frozen. 
In the mean-field treatment with the plaquettes 
the pseudo-critical nature of the PM phase is maintained, the FM phases remain frozen but the 
AF phase is not, as in the $2D$ model. 

In the $2D$ eight-vertex model the disordered phase is no longer critical and the FM and AF phases 
are no longer frozen. Moreover the transitions are all continuous. In the trees of single vertices and plaquettes the PM phase 
is also a conventional one, meaning that the stability analysis does not lead to a critical eigenmode. 
As for the FM and AF phases, they are frozen on the tree of single vertices but 
they are not frozen on the tree of plaquettes. With the latter improved treatment we find discontinuous transitions that are also 
associated to instability, as in other KDP-like transitions. However, differently from 
the cases encountered before, the transitions are now towards non-frozen ordered phases.

In conclusion, in these two (integrable) cases the plaquette version 
of the mean-field approach allows  us to 
introduce fluctuations in the ordered phases, making the transitions softer and
closer to the ones in the $2D$ case. 

The location of the transition lines for the six- and the eight-vertex models is independent of the 
particular structure of the lattice (provided that it has coordination equal to four) and it
is identified by the condition $|\Delta_8|=1$. This shows that the critical planes are 
insensitive to the local
 geometry of the lattice and to the length of the loops, since they only depend on the local connectivity of the 
 graph. This result can be understood in terms of the duality transformation discussed 
 in~\cite{BaxterBook} which can be extended to the model on the tree.
 On the contrary, for the sixteen-vertex model we found that the location of 
 critical points depends on the graph and a duality transformation as the one that holds for the 
 eight-vertex model is not easily generalizable to the ``complete" model for all values of the parameters. 
  While in $2D$ the criticality condition $|\Delta_8|=1$ is defined by the singularity 
 of the free energy, in our mean-field approach we recover it as a special combination of the eigenvalues
 of the stability matrix that captures the instability towards the four possible ordered phases.
 We used the same procedure  to extract a mean-field $\Delta_{16}^{sv}$
 that characterizes the critical surfaces of the sixteen-vertex model on the tree {of simple vertices}.  
 {Such $\Delta_{16}^{sv}$ predicts a linear shift of all the transition
 lines by $2e$ with respect to the phase diagram of the eight-vertex model, in favor of
 the PM phase. In addition 
 it accurately reproduces the transition lines obtained with the tree made of  
 plaquettes in the limit in which two FM vertices or two AF vertices dominate over the others. 
 However, in the general case, the critical surfaces obtained within this plaquette approximation
 present more subtle non-linearities in the vertex fugacities, as shown in Fig.~\ref{fig:phase_diagram-16vertex}.
 }
Moreover such $\Delta_{16}^{sv}$ might provide a good estimate of the  transition
parameters of the spin-ice model on the pyrochlore lattice, the properties of which are well-described
by a mean-field treatment~\cite{Jaubert2008}.

In the finite dimensional problem
the SL-FM transition becomes second order as soon as $d>0$ or $e>0$. We find that the critical 
exponents depend on the particular values of the fugacities, although we are not able to determine their
explicit dependence,
due to the difficulty of the numerical analysis. A real-space RG method could be helpful 
to deal with this question precisely. 
The fact that all transitions in the sixteen-vertex model are continuous is confirmed by
the calculations  
on the tree of single vertices and on the tree of plaquettes (cavity method).  

The results obtained with the cavity method on the Bethe lattice of single vertices 
generalize some of the calculations of Refs.~\cite{Slater1941,Jaubert2008,Yoshida2004,Jaubert-thesis,jaubert2010}
for the pyroclore system. In particular,
such treatment was used to study the so-called 
Kasteleyn transition in presence of a magnetic field~\cite{Jaubert2008}, and to discuss the
properties of multicritical points and KDP transitions found on the pyrochlore lattice~\cite{jaubert2010}. 
Moreover in~\cite{Jaubert-thesis}, disregarding vertices of type $d$, 
a critical transition temperature towards the $a$-FM phase was determined, which is in 
agreement with our $a_c^{sv}$ if one imposes the same approximation.
Notice that while in~\cite{Jaubert2008,Jaubert-thesis,jaubert2010} the original motivation 
was the study of the pyroclore lattice, ultimately the mean-field theory that is developed
is equivalent to the one that we used for the $2D$ lattice (when we considered the single vertex problem
and with no distinction among the different direction thus not giving access to staggered ordered phases).
This is because one ends up with a tree structure of vertices with connectivity four and spins
(arrows) shared by two neighboring vertices. Contrary to these papers, our approach keeps track of the four different
directions and allows us to study simultaneously all possible phases, including the AF ones 
that are relevant for artificial spin-ice samples in $2D$~\cite{Demian-thesis}. More details about such 
experimental applications will be discussed elsewhere~\cite{Levis12a}. In addition, 
our approach allows us to access the non-homogeneity brought about by certain preferential paths.
Indeed, as we discussed in Sec.~\ref{Section:fluctuations}, we found that both
the PM and the ordered phases are anisotropic (except for some special combinations of the
fugacities) and domains of arrows or vertices and thermal fluctuations of string defects
develop along some preferential axes. The same anisotropy is the one that drives
the coarsening dynamics out-of-equilibrium, as found in~\cite{Levis2012}, and should be
easy to visualize experimentally.

\section{Acknowledgements}

We thank T. Blanchard, C. Castelnovo, O. Cepas,  B. Estienne, P. Holdsworth, Y. Ikhlef, L. Jaubert, K. Kitanine, 
A. Maggs, I. P\'erez-Castillo and  R. Santachiara  for very useful discussions. We acknowledge financial 
support from ANR-BLAN-0346 (FAMOUS).

\newpage

\appendix
\setcounter{equation}{0}
\renewcommand{\theequation}{A.\arabic{equation}}

\section{The CTMC algorithm}
\label{app1}

In this Appendix we give some details on the implementation of the Continuous Time Monte Carlo algorithm 
that we used to study the phase diagram of the sixteen-vertex model.

We chose to use single spin updates.
We can easily predict the time needed to flip an arrow. Suppose that the system is in state $\mu(t_0)$  
at time $t_{0}$. The \emph{exact} probability of leaving the state $\mu$ after $\Delta t$ trials is \[
\left(W(\mu\rightarrow\mu)\right)^{\Delta t}\,\left(1-W(\mu\rightarrow\mu)\right)=
\left(W(\mu\rightarrow\mu)\right)^{\Delta t}-\left(W(\mu\rightarrow\mu)\right)^{\Delta t+1}\, .\]
In order to get an estimate for this quantity we have to generate a random number  $\xi$ uniformly distributed 
between $0$ and $1$. The latter corresponds to $\Delta t$ trials if 
$0<\xi<\left(W_{\mu\mu}\right)^{\Delta t}-\left(W_{\mu\mu}\right)^{\Delta t+1}$ then $\Delta t+1<\frac{\ln \xi}{\ln W_{\mu\mu}}<\Delta t$. 
It follows that the number of steps needed to flip an arrow should be computed by

\begin{equation}
\Delta t=\mbox{Int}\left(\frac{\ln \xi}{\ln \left(1-\sum_{I=1}^{2N}W\left(\mu\rightarrow\mu^{I}\right)\right)}\right)+1
\label{TimeUpdate}
\end{equation}

Equation (\ref{TimeUpdate})  shows that we need to know  the transition probabilities for all possible arrow-flips at each step. This is the main difficulty for the implementation of the CTMC algorithm.
A simple choice suggested by the detailed balance is 
\begin{equation}
W(\mu\rightarrow\mu^I)=g(\mu\rightarrow\mu^I)A(\mu\rightarrow\mu^{I})
\end{equation}
where we have split the transition probabilities into an edge-selection probability 
\begin{equation}
g(\mu\rightarrow\mu^{I})=1/2L^2,\,\,\forall I
\end{equation}
and a flip-acceptance probability
\begin{equation}
 A(\mu\rightarrow\mu^I)=\left\{
\begin{tabular}{cc}
$\exp\{-\beta(E(\mu^I)-E(\mu))\}$ & $\mbox{if}\,  E(\mu^{I})-E(\mu)>0$ \tabularnewline
1 & \mbox{otherwise} \tabularnewline
\end{tabular}
\right. .
\end{equation}
The transition probabilities defining the dynamics of the system can now be written as
\begin{equation}
 W(\mu\rightarrow\mu^{I})=\frac{1}{2L^2}\mbox{min}\left(1,e^{-\beta(E(\mu^{I})-E(\mu)}\right)
 \label{probaDyn}
\end{equation}
satisfying detailed balance and ergodicity. This transition probabilities are the same as for the FSMC algorithm, but we should notice 
that in this case the aim is to compute the time we have to wait before we do a flip instead of sampling the configurations of the 
system. This transition probabilities are expressed as a  function of the state of the chosen spin $I$. 
We have to compute the probability to stay in the same state $W(\mu\rightarrow\mu)=1-\sum_{I=1}^{2N}W(\mu\rightarrow{\mu^{I})}$ to 
obtain $\Delta t$ using eq. (\ref{TimeUpdate}) and then we need to know every possible energy change a single flip can 
produce. In the sixteen-vertex model there is a finite number of  such possible processes (and then a finite number of possible 
transition probabilities) independently of the system size. This procedure can then be applied by making a list of all the arrows 
classified by their state, noted from now on $l$ and defined by its neighborhood ({\it i.e.} the type of its two adjacent vertices). Since 
each vertex can take sixteen different configurations, there are $8\times8$ such states for vertical and horizontal arrows, so a total of 
$64$ states for each type of arrow. Following the original name of this method~\cite{Bortz1975,Barkema-Newman_Book} this 
algorithm is a 256-fold way. The transition probability of the process $\mu$ $\rightarrow$ $\mu^{I}$ only depends on the 
state $l$ of the $I$-th arrow 
before the flip. This can be clearly seen by rewriting \[
\exp\left[-\beta(E_{\mu^{(I)}}-E_{\mu})\right]=
\exp\left[{-\beta\left(E\left[V_{1,I}^{\mu^{(I)}}\right]+E\left[V_{2,I}^{\mu^{(I)}}\right]-E\left[V_{1,I}^{\mu}\right]-
E\left[V_{2,I}^{\mu}\right]\right)}\right] 
\]
here $E\left[V_{1,I}^{\mu^{(I)}}\right]$ is the energy of the first adjacent vertex of the $I$-th arrow after the flip from the state $\mu$. To 
know the type of the neighboring vertices $V_{1,I}$ and $V_{2,I}$ at state $\mu$ is equivalent to know the state of the concerned 
arrow before the flip (the vertex types of the neighboring vertices after the flip are determined by the vertex types before the flip): the 
energy change after a flip depends only in its initial state. We define
\[P_{l}=\frac{1}{2N}\min\left(1,e^{-\beta\,\varepsilon_{l}}\right)\]
where $\varepsilon_{l}$ is the energy difference after flipping an arrow in state $l$. It is useful for the implementation to note that we 
can compute $\Delta t$ by counting the number of arrows occupying each one of the different possible states at each step. We 
substitute the latter equation by
\begin{equation}
Q=\sum_{I=1}^{2N}W(\mu\rightarrow{\mu^{(I)})}=\sum_{l=1}^{256}g_{l}\,P_{l}\label{Q}
\end{equation}
where $g_{l}$ is the number of arrows in state $l$. We then need to keep record of the state of every arrow on a list at each step. After 
a transition this list must be updated. 

\section{{Equation for the single vertex}}
\label{Appendix_Eq_cavity}

An alternative way to study the system of equations~(\ref{psiU_8vertex}) which might be more suitable for  
an analytical treatment is to identify each direction $\alpha$ with a couple of binary indices $[i,j]$
as follows $\{d, l, r, u\} \equiv \{[-1,-1], [1,-1], [-1,1], [1,1] \}$ and define
$\psi^{[i,j]} = \frac{e^{h^{[i,j]}}}{e^{h^{[i,j]}}+e^{-h^{[i,j]}}}$.
In this way from (\ref{psiU_8vertex}) one obtains:
\beq
 \displaystyle  h^{[i_0,j_0]}   \displaystyle = \text{atanh}\, \left[
 \frac{a \, \sinh \text{F}_a  + 
b \, \sinh \text{F}_b +
c \, \sinh \text{F}_c +
d \, \sinh \text{F}_d }{
a \, \cosh \text{F}_a  + 
b \, \cosh \text{F}_b +
c \, \cosh \text{F}_c +
d \, \cosh \text{F}_d  } \right]
\eeq
wit $[i_0,j_0] \in  \{[-1,-1], [1,-1], [-1,1], [1,1] \}$ and 
\beq
 \begin{array}{lll}
\displaystyle \text{F}_a   \displaystyle =  \sum_{[i,j] \neq [-i_0,-j_0]} h^{[i,j]} & \hspace{0.2cm} &
\displaystyle  \text{F}_b   \displaystyle = \sum_{[i,j] \neq [-i_0,-j_0]} (-1)^{\frac{i+j-i_0-j_0}{2}} h^{[i,j]}  
  \\ \vspace{-0.2cm} \\
\displaystyle  \text{F}_c   \displaystyle = \sum_{[i,j] \neq [-i_0,-j_0]}  (-1)^{\frac{i-i_0+2}{2}} h^{[i,j]}  & 
\hspace{0.2cm} &
\displaystyle  \text{F}_d   \displaystyle = \sum_{[i,j] \neq [-i_0,-j_0]}  (-1)^{\frac{j-j_0+2}{2}}  h^{[i,j]} \ .
     \end{array}
\eeq
We limit ourselves to study the PM and  $a$-FM phases, recovering the full phase diagram by symmetry arguments.
In this case we take the homogeneous solution $h^{[i,j]} = h$ and the equation simplifies to:
\beq
\displaystyle  h = \text{atanh} \Big[ \tanh h \left( \frac{a - b - c - d + 2 a \cosh 2 h}{- a + b + c + d + 2 a \cosh 2 h}\right) \Big] \ .
\eeq
The paramagnetic solution $h=0$ is stable when $a<b+c+d$. At the critical point $a_c = b+ c+d$
it turns out that all values of the field $h$ are solutions of the equation and for $a>a_c$ the stable solution 
is $h=\infty$, which implies $\psi^{\alpha}=1$ $\forall \alpha$.
%
In a similar manner one can study the $a$-PM/FM transition in presence of vertices of type $e$. 
The homogeneous equation for the field $h$ associated with the system of Eqs.~(\ref{psiU_16vertex})
when $\psi^{\alpha}=\psi$ $\forall \alpha$ becomes:
\beq
h = \text{atanh}\Big[ \tanh h  \left(\frac{a - b - c - d + 2 e + 2 (a - e) \cosh 2 h}{-a + b + c + d + 2 e + 2 (a + e) \cosh 2 h} \right)\Big] \ ,
\eeq
and one recovers the results discussed in Section~\ref{16vertex_single_vertex}.

\newpage

\bibliographystyle{phjcp}
\bibliography{MC-Bethe.bib}

\end{document}